\documentclass[sn-nature]{sn-jnl_edited}% Standard Nature Portfolio 

\jyear{2023}%

\theoremstyle{thmstyleone}%
\theoremstyle{thmstyletwo}%

\theoremstyle{thmstylethree}%

\newcommand{\covid}{\textsc{covid}-19}
\newcommand{\sars}{\textsc{sars}-\textsc{c}o\textsc{v}-2}

\usepackage{bm}

\usepackage{siunitx}

\usepackage{pdflscape}
\usepackage{gensymb}

\graphicspath{{fig/}}

\usepackage{caption}

\begin{document}

\title{Assessing the impact of forced and voluntary behavioral changes on economic-epidemiological co-dynamics}
\subtitle{A comparative case study between Belgium and Sweden during the 2020 COVID-19 pandemic}


\author*[1]{\fnm{Tijs W.}
\sur{Alleman}}\email{tijs.alleman@ugent.be}

%\author[2]{\fnm{Koen}
%\sur{Schoors}}

\author[1]{\fnm{Jan M.}
\sur{Baetens}}

\affil[1]{\orgdiv{BionamiX}, \orgname{Department of Data Analysis and Mathematical Modelling, Ghent University}, \orgaddress{\street{Coupure Links 653}, \city{Ghent}, \postcode{9000}, \country{Belgium}}}

%\affil[2]{\orgdiv{Complex Systems Institute (CSI)}, \orgname{Department of Economics, Ghent University}, \orgaddress{\street{Tweekerkenstraat 2}, \city{Ghent}, \postcode{9000}, \country{Belgium}}}


\maketitle

\vspace{-0.5cm}
{\small
\noindent\textbf{Abstract} During the COVID-19 pandemic, governments faced the challenge of managing population behavior to prevent their healthcare systems from collapsing. Sweden adopted a strategy centered on voluntary sanitary recommendations while Belgium resorted to mandatory measures. Their consequences on pandemic progression and associated economic impacts remain insufficiently understood. This study leverages the divergent policies of Belgium and Sweden during the COVID-19 pandemic to relax the unrealistic -- but persistently used -- assumption that social contacts are not influenced by an epidemic's dynamics. We develop an epidemiological-economic co-simulation model where pandemic-induced behavioral changes are a superposition of voluntary actions driven by fear, prosocial behavior or social pressure, and compulsory compliance with government directives. Our findings emphasize the importance of early responses, which reduce the stringency of measures necessary to safeguard healthcare systems and minimize ensuing economic damage. Voluntary behavioral changes lead to a pattern of recurring epidemics, which should be regarded as the natural long-term course of pandemics. Governments should carefully consider prolonging lockdown longer than necessary because this leads to higher economic damage and a potentially higher second surge when measures are released. Our model can aid policymakers in the selection of an appropriate long-term strategy that minimizes economic damage.\\

%\noindent\textbf{Keywords} \covid{}, Behavioral Epidemiology, Macro-econometrics, Disease Transmission

%\noindent\textbf{Word count} 219 (summary), 1770 words (main text); 3000 words (methods). \\
}

%%%%%%%%%%%%%%%%%%%%%%%%%%%%%%%%%%%%%%%%%%%%%%%%%%%%%%%%%%%%%%%%%%%%%%%%%%%%%%
%%%%%%%%%%%%%%%%%%%%%%%%%%%%%%%%%%%%%%%%%%%%%%%%%%%%%%%%%%%%%%%%%%%%%%%%%%%%%%

\pagebreak

\section*{Summary}

During the \covid{} pandemic, governments faced the challenge of managing population behavior to prevent their healthcare systems from collapsing. Sweden adopted a strategy centered on voluntary behavioral changes based on sanitary recommendations  \citep{ludvigsson2020} while Belgium resorted to mandatory measures \citep{luyten2021}. Their consequences on pandemic progression and associated economic impacts remain insufficiently understood \citep{nigmatulina2009,buonomo2020,yan2021}. This study leverages the divergent policies of Belgium and Sweden during the COVID-19 pandemic to relax the unrealistic -- but persistently used -- assumption that social contacts are not influenced by the epidemic's dynamics. We develop an epidemiological-economic co-simulation model where pandemic-induced behavioral changes are a superposition of voluntary actions driven by fear, prosocial behavior or social pressure \cite{bavel2020}, and compulsory compliance with government directives. Our findings emphasize the importance of early responses, which reduce the stringency of measures necessary to safeguard healthcare systems and minimize ensuing economic damage. Voluntary behavioral changes lead to a pattern of recurring epidemics, which should be regarded as the natural long-term course of pandemics. Governments should carefully consider prolonging lockdown longer than necessary because this leads to higher economic damage and a potentially higher second surge when measures are released. Our model can aid policymakers in the selection of an appropriate long-term strategy that minimizes economic damage.

\section{Methods}

\subsection{Epinomic model}\label{section:epinomic_model}

\textbf{Overview} The \textit{epinomic} models of Belgium and Sweden consist of three connected submodels: 1) A spatially explicit compartmental disease transmission model for \sars{} with recurrent mobility and seasonal forcing of the transmission rate \citep{alleman2023a}. 2) A dynamic production network model based on input-output tables with a relaxed Leontief production function and hiring and firing of workers, used to quantify the impact of supply and demand shocks on employment and gross output, inspired by Pichler et al. \citep{pichler2022} and validated for Belgium \citep{alleman2023c}. 3) A collective memory feedback model to incorporate ``voluntary" behavioral changes, making the present number of social contacts and consumption patterns dependent on the history of \covid{} hospitalizations. The collective memory feedback model is inspired by previous work on the application of control theory to pandemic response \citep{alleman2020}, as well as the works of Ronan et al. \citep{ronan2021} and Nigmatulina and Larson \citep{nigmatulina2009}. The model was implemented using our in-house simulation software for $n$-dimensional pySODM \citep{alleman2023b}.\\

%%%%%%%%%%%%%%%
%% flowchart %%
%%%%%%%%%%%%%%%

\begin{figure}[h!]
    \centering
    \includegraphics[width=1.07\linewidth]{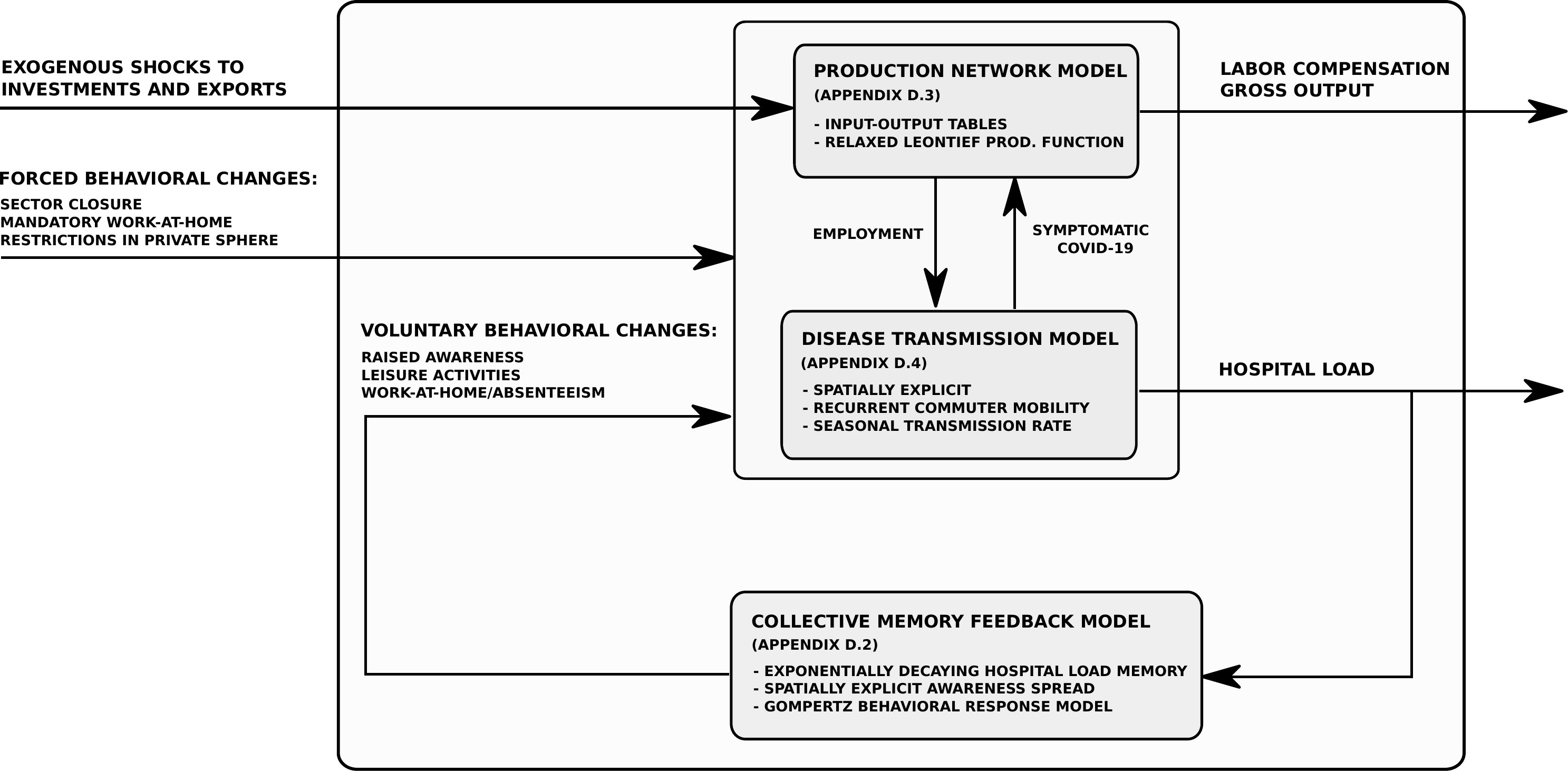}
    \caption{Schematic representation of the \textit{epinomic} model.} 
    \label{fig:epinomic_flowchart_simple}
\end{figure}

As schematically shown in Figure \ref{fig:epinomic_flowchart_simple}, the epinomic model has four inputs: 1) The shocks to investments and exports, which evolve exogenously in accordance with trade data for Belgium and Sweden (Appendix \ref{section:demand}). 2) The prohibition of an economic activity (Table \ref{tab:NACE64}), from hereon referred to as ``sector closure", 3) The obligation to work from home in a given economic activity. 4) The prohibition of leisure contacts in the private sphere. The collective memory feedback model, incorporating the ``voluntary" behavioral changes requires no external inputs, but its parameters have to be calibrated (Appendix \ref{app:calibration}). At every timestep, the production network submodel and dynamic transmission submodel exchange information. We model the impact of laid-off employees on \sars{} spread by using the (sectoral) labor compensation obtained from the production network model to compute the reduction of workplace contacts. We also model the impact of symptomatic \covid{} on labor supply and household demand, as symptomatic individuals are less likely to go to work or attend leisure activities if they are sick \citep{vankerckhove2013}. Our model keeps track of three output variables: 1) The gross economic output, 2) labor compensation, and 3) the regional hospital load. Because simulated trends in gross economic output and labor compensation are similar, we restrict our attention to labor compensation. The model is presented in full detail in Appendix \ref{app:detailed_model_description}.\\

\textbf{The definition of voluntary and forced behavioral changes} In this work, ``voluntary" behavioral changes are a black box encompassing all behavioral changes that result from awareness to \sars{}, including sanitary recommendations, fear, prosocial behavior, and social pressure \citep{bavel2020}. Awareness can be spread by individuals, scientific institutes, and governments. Opposed, ``forced" behavioral changes are the result of enforcement by law and are thus always induced by governments. Behavioral changes in Sweden were predominantly ``voluntary" according to our definition \citep{ludvigsson2020}, although we do not imply the Swedish population's mentality during the 2020 \covid{} pandemic was as a ``business-as-usual" one. We do not expect an uninformed or misinformed population to alter its behavior timely in response to an epidemic. Pressure to alter behavior increases as \sars{} incidence surges, blurring the line as ``voluntary" behavioral changes may eventually be perceived as ``forced". Opposed to Sweden, Belgian policies were predominantly ``forced", hence our choice to compare these countries in a case study.\\

\textbf{Collective memory feedback model} A collective memory feedback model is used to incorporate ``voluntary" behavioral changes by making the present number of social contacts and consumption patterns dependent on the history of \covid{} hospitalizations. As its input, the collective memory feedback model uses the hospital load obtained from the disease transmission model. To mimic the exponentially decaying nature of an individual's memory \citep{white2001}, on every timestep, we add the hospital load in every spatial patch (Swedish county or Belgian province) to a six-month-long ``memory" which we then weigh degressively using a negative exponential function with a mean lifetime of $\nu$ days \citep{ronan2021}. To account for the fact that awareness of local surges in \sars{} spreads spatially, we model the \textit{Exponential Moving Average} (EMA) hospital load in every spatial patch $g$ as a connectivity-weighted average (Fig. \ref{fig:map_popdens}) between the EMA hospital load on spatial patch $g$ and the EMA hospital load on the spatial patch with the maximum hospital load \citep{nigmatulina2009}. In this way, we allow awareness to \sars{} induced by a local epidemic to spread nationally, arguably due to coverage in national media and government communication. To take into account the detrimental impact of a healthcare system (HCS) collapse, which we assume happens when the maximum number of available IC beds has been surpassed, we normalize the EMA hospital load with the number of IC beds available in both countries. We then translate the ``perceived" hospital load in every spatial patch to a behavioral change, bound between zero and one, by means of a two-parameter Gompertz function. We model three behavioral changes: 1) A decrease in the per-contact effectivity to spread \sars{} as awareness rises with increasing hospital loads \citep{alleman2021}, 2) a voluntary reduction in leisure activities, both privately and publicly, as the hospital load increases \citep{google_mobility,alleman2021}, 3) a voluntary reduction in workplace contacts as the hospital load increases, first through workers voluntarily working from home and eventually due to absenteeism \citep{google_mobility,alleman2021}.\\

% In this way, we allow the epidemic and economic dynamics to be partly uncoupled, as individuals may still make contact at work or during leisure activities but be more prudent \citep{alleman2021}.

\begin{figure}[h!]
    \centering
    \includegraphics[width=\linewidth]{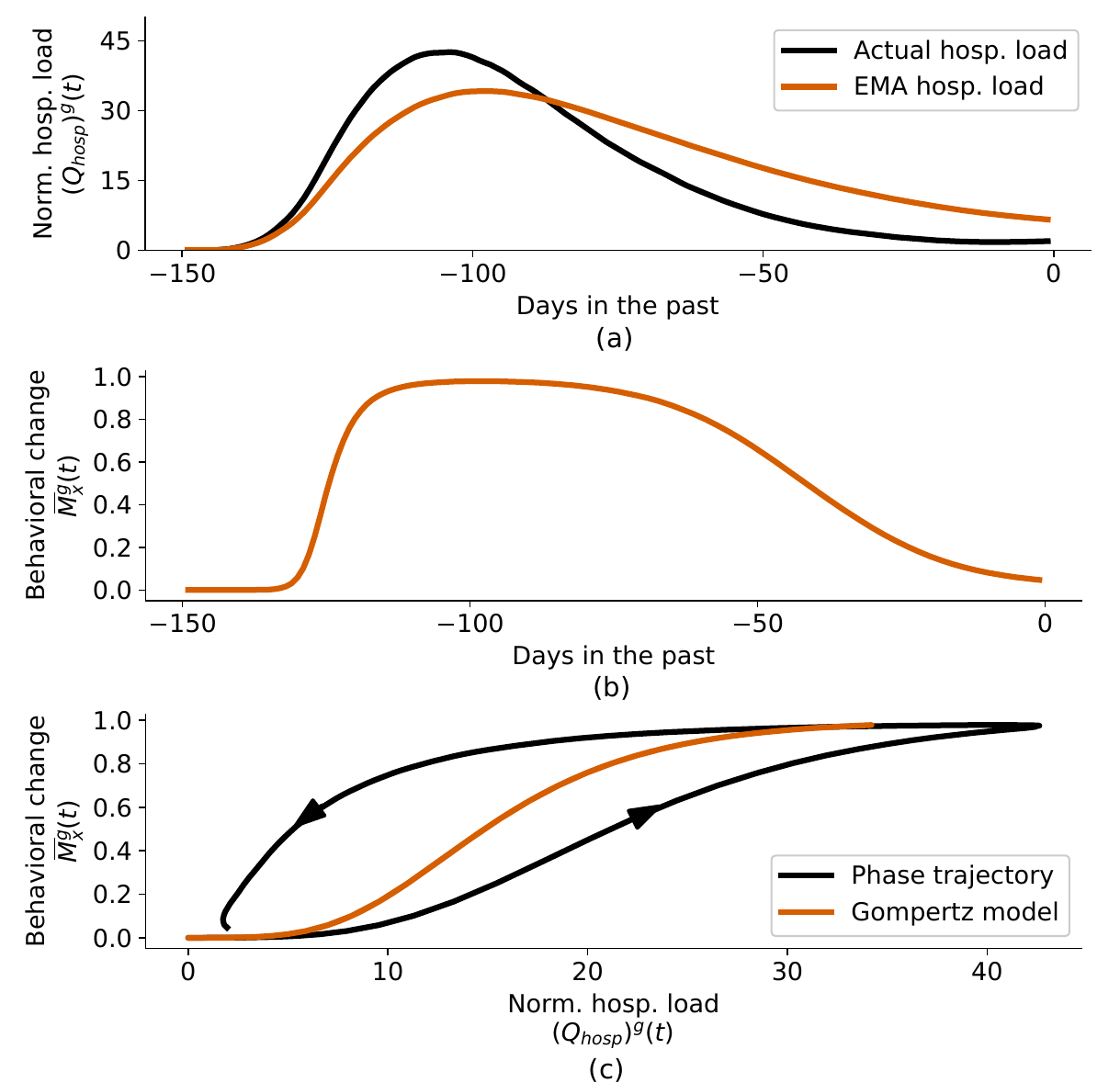}
    \caption{Time course (b) and phase trajectory (c) of voluntary behavioral changes in response to a hypothetical surge of the hospital load (a).} 
    \label{fig:memory}
\end{figure}

By using an EMA of the hospital load as the input to our Gompertz behavioral response model, a delay is introduced compared to the actual hospital load (Fig. \ref{fig:memory}a). In the upward phase of an epidemic, individuals underestimate their likelihood of contracting the disease because of ``optimism bias", and hence, the initial voluntary response will likely be too slow \citep{bavel2020}. In the downward phase of an epidemic, there is a tendency to prolong measures longer than needed \citep{alleman2023a}. The introduced delay thus results in the incorporation of favorable dynamics in the model. The time lag makes the voluntary behavior changes irreversible, introducing ``hysteresis" in the trajectory of the epidemic \citep{liu2016,lacitignola2021}. The Gompertz behavioral model and phase trajectory of the voluntary behavioral change are shown in Figure \ref{fig:memory} (b), while the time course of the voluntary behavioral change is shown in Figure \ref{fig:memory} (c). A mathematical description of the collective memory feedback model can be found in Appendix \ref{app:behavoiral_feedback}.\\

% \begin{figure}[t!]
%   \raggedright
%   \footnotesize
%   \begin{subfigure}{\textwidth}
%     \centering
%     \includegraphics[width=\linewidth]{EMA.pdf}
%     \subcaption{(a) Time course of moving average hospital load in spatial patch $g$ (hypothetical data). Actual hospital load (black), EMA of the actual hospital load with halflife $\nu=21~d.$ (orange).}
%     \label{subfig:EMA_memory}
%   \end{subfigure}
%   \vspace{0.25cm}
%     \begin{subfigure}{\textwidth}
%     \centering
%     \includegraphics[width=\linewidth]{M_time.pdf}
%     \subcaption{(b) Time course of voluntary behavioral changes $\bar{M}^g_X(t)$.}
%     \label{subfig:M_phasediagram}
%   \end{subfigure}
%   \vspace{0.25cm}
%   \begin{subfigure}{\textwidth}
%     \centering
%     \includegraphics[width=\linewidth]{M_phasediagram.pdf}
%     \subcaption{(c) A Gompertz function is used to convert the perceived hospital load into a voluntary behavioral change (orange). The phase trajectory depicts the values of the voluntary behavioral change $\bar{M}^g_X(t)$ as a function of the actual hospital load $(Q_{\text{hosp}})^g(t)$. The trajectory is not reversible, giving rise to \textit{hysteresis}.}
%     \label{subfig:M_timecourse}
%   \end{subfigure}
%   \hspace{0.5cm}
%   \caption{Time course (b) and phase trajectory (c) of voluntary behavioral changes in response to a hypothetical surge of the hospital load (a).}
% \end{figure}

\textbf{Disease transmission model} The compartmental structure and parameters of the disease transmission model for \sars{} used in this work are similar to our previously developed model \citep{alleman2023a}. A flowchart depicting the various compartments of the \sars{} model used in this work is shown in Fig. \ref{fig:flowchart_covid}. The model incorporates pre-symptomatic transmission and asymptomatic transmission of \sars{}, and accounts for different \covid{} severities, ranging from asymptomatic disease to death. Waning of antibodies (seroreversion) is included, and we have previously demonstrated the model and its parametrization adequately captures the seroprevalence in the Belgian population during the 2020-2021 \covid{} pandemic \citep{alleman2021,alleman2023a}. Every disease compartment is stratified into 17 five-year age categories to account for the fact that social contact and disease severity differ substantially with age. Inspired by Haw et al. \citep{haw2022}, we analyzed the 2012 social contact survey by B\'eraud et al. \citep{beraud2015} in metropolitan France to construct contact matrices because it is the only (large-scale) study that includes sector-specific information of respondents  (Appendix \ref{app:beraud}). The model is further stratified into 21 counties for Sweden (Fig. \ref{fig:map_SWE}) and 11 provinces for Belgium (Fig. \ref{fig:map_SWE}) to account for spatial heterogeneity across the territories. Publicly available data on the daily number of commuters residing in spatial patch $g$ and working in spatial patch $h$ were used to quantify the inter-patch connectivity. The Belgian provinces are more strongly connected than the Swedish counties implying diseases will spread more readily and homogeneously across the Belgian territory (Fig. \ref{fig:map_popdens}). We use the composition of the labor market to inform the number of social contacts in every spatial patch (Figs. \ref{fig:map_employment}, \ref{fig:map_lmc_ABC}, \ref{fig:map_lmc_GIRST}, \ref{fig:n_contacts_origin}, \ref{fig:n_contacts_destination}). We use seasonal forcing on the transmission rate to fit the model to the epidemiological data of (more than) one year \citep{gavenciak2022,alleman2023a}. We present a detailed overview of the disease transmission model in Appendix \ref{app:dynamic_transmission_model}.\\

\textbf{Production network model} The dynamic production network model is used to track the labor compensation and gross output for the 63 economic activities listed in the \textit{Nomenclature des Activit\'es \'Economiques dans la Communaut\'e Europ\'eenne} (NACE) \citep{NACE}. A detailed index of the 63 economic activities (sectors) is given in Table \ref{tab:NACE64} and an aggregation of 21 economic activities is given in Table \ref{tab:NACE21}. The model uses the $63 \times 63$ input-output matrix of Belgium and Sweden to inform the intermediate flows of services and products in the country's domestic production networks. The economy uses intermediates and labor to satisfy the demand of two end users: households and \textit{other sources} (government and non-profit consumption, investments, and exports). We model the changes in other demand exogenously using trade data for Belgium and Sweden obtained from various sources (Appendix \ref{section:demand}). Household demand and labor supply evolve in function of government restrictions and voluntary behavioral changes. The gross output of sector $k$ is the sum of the intermediate consumption of its goods and services by all other sectors, household consumption, and exogenous consumption. Prices are assumed time-invariant and capital is not explicitly modeled. One representative firm is modeled for each sector and there is one representative household \citep{pichler2022}.\\

Every firm keeps an inventory of inputs from all other firms and draws from these inventories to produce outputs. Intermediates in production are modeled explicitly as deliveries replenishing the firm's inventory. Due to shocks in household demand, exogenous demand, and labor supply caused by the epidemic, firms may run out of intermediate inputs and may need to stop production.  Depending on their ability to meet demand, firms may also hire or fire workers. By using a survey on the criticality of inputs to production, it is possible to relax the Leontief production function, which assumes every intermediate input is critical to production \citep{pichler2022}. For instance, the closure of restaurants during the \covid{} pandemic would restrain the output of construction companies, which is not realistic and leads to large exaggerations of economic damages. To the best of our knowledge, DAEDALUS by Haw et al. \citep{haw2022} does not include such a relaxed Leontief production function. Further, Haw et al. \citep{haw2022} make the limiting assumption that consumer demand does not evolve in response to the pandemic. We present a detailed overview of the production network model in Appendix \ref{app:production_network_model}, a flowchart is available in Fig. \ref{fig:EPNM_flowchart}. We have previously implemented the model of Pichler et al. \citep{pichler2022} and found its projections are in excellent agreement with data on B2B transactions, surveys on revenue and employment, as well as GDP data, made available by the Belgian National Bank \citep{alleman2023c}. We further found the model economy to be more sensitive to shocks in the supply of labor than to shocks in household demand. We present a detailed overview of the production network model in Appendix \ref{app:production_network_model}.\\

\begin{figure}[h!]
    \centering
    \includegraphics[width=\linewidth]{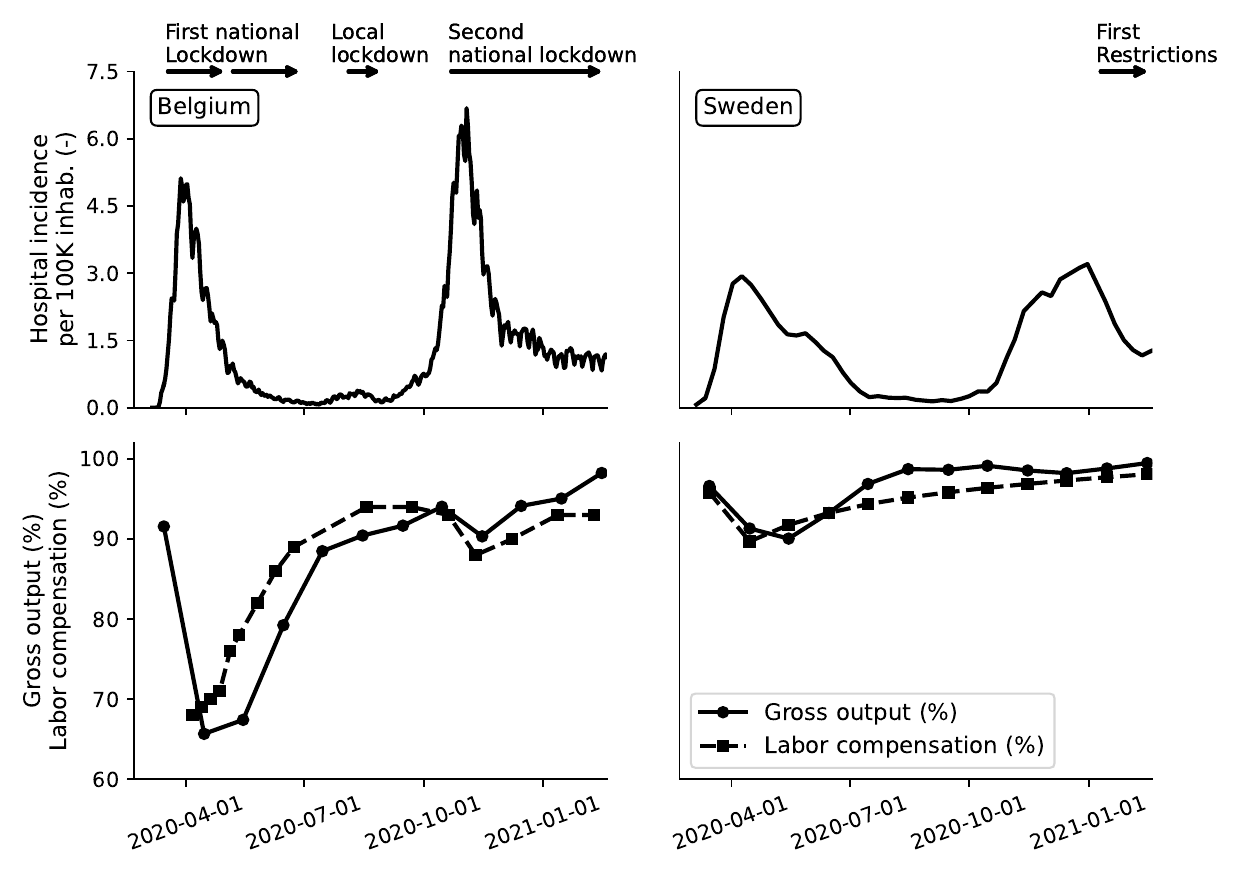}
    \caption{(top row) Daily new \covid{} hospitalizations per \num{100000} inhabitants in Belgium and Sweden. (bottom row) Percentage of gross aggregated output and labor compensation retained during the pandemic, as compared to the first quarter of 2020. The horizontal arrows denote the periods with severe social restrictions (lockdown), along with their subsequent release in Belgium.} 
    \label{fig:data_calibration}
\end{figure}

\textbf{Empirical data and policies}  For both countries, time series on the evolution of the (regional) hospitalization incidence, gross domestic product, and employment are publicly available \citep{socialstyrelsen2023,sciensano2023,scb2023b,nbb2023a,ermg2021,scb2023a}. During 2020, Belgium and Sweden faced a similar pattern of two sequential \sars{} epidemics with a period of less circulation in between them during summer (Fig. \ref{fig:data_calibration}). Belgium faced both greater surges in \covid{} hospital incidence and greater economic damage than Sweden. During 2020, Belgium faced a 13.4~\% decline in GDP while Sweden only faced a 3.1~\% decline. Especially during the first \covid{} epidemic in the second quarter of 2020, Belgium was much more severely struck, with a 29.2~\% decline in GDP while Sweden only faced an 8.5~\% decline. Labor compensation, which was partly furloughed by both governments, faced a similar decline during 2020 of 4.3~\% in Sweden and 12.2~\% in Belgium (Table \ref{tab:calibration_comparison_model_data}). The economic impact of the second \covid{} surge and lockdown in Belgium is more pronounced than in Sweden. \\

The governments of Belgium and Sweden imposed very different measures to counter the spread of \sars{}. The Belgian government forced behavioral changes by imposing two lockdowns (Fig. \ref{fig:data_calibration}) which involved the mandatory closures of many economic activities, distance learning, and restrictions on private leisure contacts \citep{luyten2021}. In contrast, Sweden responded earlier, at a hospital incidence of 0.2 instead of 0.6 patients per \num{100000} inhabitants, relied mostly on voluntary sanitary recommendations, and was able to keep most of its schools open for children up to the age of 16 \citep{ludvigsson2020}. Only on January 7, 2021, did the Swedish government impose restrictions in restaurants and commercial areas, household quarantine, and the obligation to wear facemasks on public transport \citep{ludvigsson2023}. From the start, the Swedish strategy was aimed at mitigation, while the Belgian policy was aimed at suppression.\\

\textbf{Calibration procedure} The values of 12 model parameters are \textit{a priori} unknown and are calibrated using epidemiological and economic data from Belgium and Sweden (Table \ref{tab:overview_L2_priors}). We fit the parameters to data from both countries simultaneously as fitting our model parameters to both countries separately would not yield a valid basis for comparisons. We implemented the government-mandated policies imposed during the 2020 \covid{} pandemic in Belgium and Sweden as realistically as possible (Tables \ref{tab:policies_BE_A} and \ref{tab:policies_BE_EF}, Fig. \ref{fig:data_calibration}). In addition to the 12 model parameters, the initial condition that resulted in the observed initial spatial spread of \sars{} in Belgium and Sweden was unknown. We used an iterative optimization procedure in which the initial condition and parameters were calibrated sequentially. For Sweden, the strongest seed was located in the Stockholm metropolitan area, and to a much lesser extent J\"onk\"oping. Keeping the initial condition fixed, we used the obtained estimates of the 12 parameters to start the affine-invariant ensemble sampler for Markov Chain Monte Carlo (MCMC) \citep{goodman2010,emcee2013}. The calibration procedure and its results are described in detail in Appendix \ref{app:calibration}. In Figure \ref{fig:calibration_epinomic_national}, we present a comparison between the model realizations and the empirical data. In Table \ref{tab:calibration_comparison_model_data}, we compare the observed and modeled reductions of gross aggregated output and labor compensation at the quarterly temporal level. Our model is able to capture the long-term epidemiological and economic trends in two countries using one calibrated set of parameters justifying its use in modeling counterfactual scenarios.

\subsection{Scenarios}\label{section:scenarios}
%We first examine the impact of alternative policies during the first 2020 \covid{} surge in Belgium. We then study how prolonging restrictions at low \sars{} prevalence can negatively influence a resurgence, and we illustrate the oscillatory nature of epidemiological dynamics in the absence of government forcing. Finally, we exploit the spatially-explicit nature of our epinomic model to demonstrate the difference between having a \sars{} epidemic that spreads from a single point-of-origin versus having a \sars{} epidemic that spreads from two points-of-origin.\\

\textbf{Scenario 1: How does the timing of restrictions influence the strictness needed to safeguard the Belgian healthcare system?} \\

We define four possible policy interventions for the Belgian government to impose on three dates: March 12, 2020, March 15, 2020, and March 18, 2020 (Table \ref{tab:scenarios_policies_BE}). We then gradually release the measures over a two-month period starting on May 4, 2020, as happened in reality. We set our model up to replicate the 2020 \covid{} pandemic in Belgium. We use the calibrated initial condition for Belgium, we include seasonal forcing on the transmission rate, we distinguish between regular weeks and holiday weeks, and we impose shocks to investments and exports to replicate the 2020 \covid{} pandemic in Belgium. These scenarios will allow us to assess the epidemiological and economic impact of four policy alternatives ranging from very strict to completely voluntary. Additionally, we can gauge if voluntary recommendations, as used in Sweden, would have worked in Belgium.\\

\textbf{Scenario 2: How does the timing of releasing lockdown affect the subsequent resurgence of hospitalizations?} \\

We now set up our model without seasonal forcing on the transmission rate, we do not distinguish holidays and we impose no shocks to investments and exports, thus no references to events from the 2020 \covid{} pandemic are made. We assume the first infected individuals are located in Stockholm County and Brussels in Sweden and Belgium respectively. We start the simulation with an $R_0=3$ and trigger awareness to \sars{} and a government intervention. The date of the intervention is chosen so that the peak hospital incidence of the first \sars{} epidemic is (quasi) equal to the nominal number of IC beds available in both countries. The government intervention is assumed to be equal to the lockdown measures taken by the Belgian government during the first 2020 \sars{} epidemic (Table \ref{tab:scenarios_policies_BE}, policy P1). We then release the imposed measures after 2, 3, 4, and 5 months and observe the impact on the IC load during the next year.\\

\begin{table}[h!]
\caption{Modeled government interventions at the start of the \covid{} pandemic in March 2020 for Belgium. The interventions range from most (P1) to least (P4b) strict.}
{\renewcommand{\arraystretch}{1.35}
\begin{tabular}{p{0.7cm}>{\raggedright\arraybackslash}p{3.8cm}>{\raggedright\arraybackslash}p{2.2cm}>{\raggedright\arraybackslash}p{1.4cm}>{\raggedright\arraybackslash}p{1.7cm}}
    \midrule
    \textbf{Policy} & \textbf{Economic closures} & \textbf{Schools} & \textbf{Mandated telework} & \textbf{Social restrictions} \\ \midrule
    P1 & All economic activities except utilities (D, E) & Closed during the first 14 days & Yes & Yes \\
    P2 & All economic activities involving worker-customer interactions$^1$ & Closed during the first 14 days & Yes & Yes \\
    P3 & Accommodation (I55-56), recreation (R), and membership organisations (S94) & Closed during the first 14 days & Yes & Yes  \\
    P4a & None & Remain open & Yes & No \\
    P4b & None & Remain open & No & No \\ \bottomrule
\end{tabular}
}
\label{tab:scenarios_policies_BE}
%\vspace{1.10cm}
{\raggedright{\footnotesize{$^1$Retail (G47), Transport on land, sea, and air (H49, H50, H51), Accommodation (I55-56), Real estate (L68), Rental \& leasing (N77), Travel agencies (N79), Recreation (R), Other (personal) services (S), Activities of households as employers (T).}}}
\end{table}

\textbf{Scenario 3: How does adjusting social contacts based on the history of the hospital load influence an epidemic's dynamics?} \\

We set up our epinomic model without seasonal forcing on the transmission rate, without distinguishing holidays, and without imposing shocks to investments and exports so no references to events from the 2020 \covid{} pandemic can be made. Additionally, we do not incorporate any government measures and assume the simulation starts with an $R_0=3$ and general awareness to \sars{} already triggered. We assume the first infected individuals are located in Stockholm County and Brussels in Sweden and Belgium respectively. We simulate the model while varying the mean lifetime of the population's collective memory ($\nu$; Eq. \eqref{eq:EMA_hospital_load}), which is a key parameter governing how much weight people give to past hospital loads when making decisions in the present (Fig \ref{fig:memory}a). We simulate the model for two years using three values of $\nu$: 7, 28, and 62 days. During the model's calibration, a mean lifetime of $\nu=20.8\ \text{d.}$ (Table \ref{tab:calibration_results}) was obtained.\\

\textbf{Scenario 4: What would have been the impact of multiple points-of-entry of \sars{} in Sweden and Belgium during the March 2020 \covid{} surge?} \\

%\textbf{Scenario 4: What would have been the impact of multiple points-of-origin of \sars{} in Sweden and Belgium during the March 2020 \covid{} surge?} \\

In Sweden, \sars{} seems to have spread mostly from the Stockholm metropolitan area, and to a much lesser extent from J\"onk\"oping, where the first case was detected on January 31, 2020 \citep{krisinformation2020} (Appendix \ref{app:calibration}). Our aim is to exploit the spatially explicit nature of our epinomic model to study how multiple points-of-origin of \sars{} would have impacted the course of the first \covid{} epidemic in 2020. We once again set up our epinomic model without seasonal forcing on the transmission rate, without distinguishing holidays, and without imposing shocks to investments and exports. We assume the epidemic is seeded by two individuals infected with \sars{}. We assume the first infected individual is always located in Stockholm and Brussels for Sweden and Belgium respectively. The second infected individual is placed in every other Swedish county or Belgian province, so we simulate 20 and 10 trajectories for Sweden and Belgium respectively. We start the simulation with $R_0=3$, we trigger general awareness to \sars{} at a hospital incidence of 0.2 patients per \num{100 000} inhabitants. The threshold was computed by interpolating the weekly hospital incidence data by Socialstyrelsen \citep{socialstyrelsen2023} to March 11th, 2020, corresponding to day the first sanitary recommendations were issued \citep{ludvigsson2020}. We simulate the model for 150 days, resulting in an epidemic trajectory similar to the actual first \covid{} epidemic in Sweden and Belgium  \citep{ludvigsson2020,stralin2021,sciensano2023}.\\

\section{Results and discussion}

\textbf{Scenario 1: Responding late necessitates restrictions with higher economic damages} \\

\begin{figure}[b!]
    \centering
    \includegraphics[width=\linewidth]{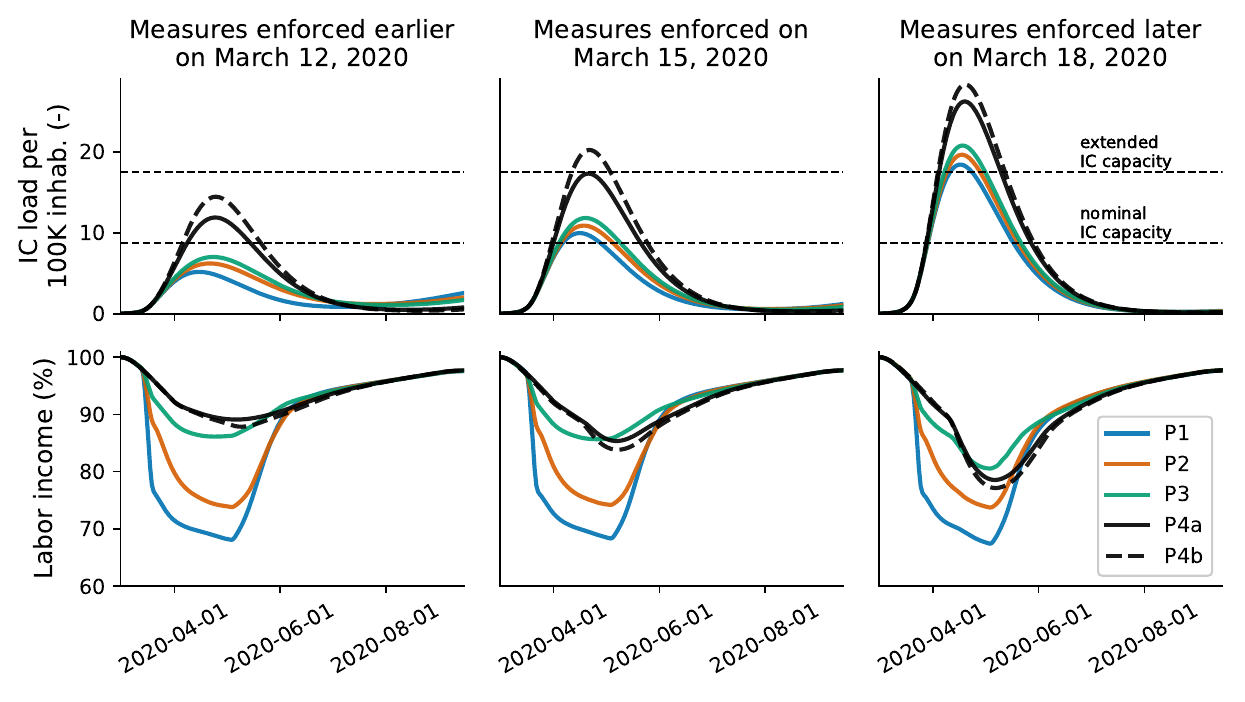}
    \caption{Simulated IC load per \num{100000} inhabitants and percentage labor income as compared to before the \covid{} pandemic, for Belgium. Policies range a strict lockdown as implemented in Belgium (P1) to voluntary recommendations as implemented in Sweden (P4b) (Table \ref{tab:scenarios_policies_BE}).} 
    \label{fig:hypothetical_scenarios_BE}
\end{figure}

An early response reduces the stringency of measures necessary to safeguard the Belgian HCS and minimizes ensuing economic damage (Figs. \ref{fig:hypothetical_scenarios_BE} and \ref{fig:hypothetical_scenarios_BE_full}, Table \ref{tab:hypothetical_scenario_results}). Regardless of the date measures are imposed, economic damages are the smallest for policy P4a (mandated telework). If policy P4a had been imposed on March 3, 2020, the cumulative number of IC patients in the second quarter of 2020 would have been identical to policy P1 implemented on March 15, 2020, but the gross aggregated output would have fallen by only 12.2~\% instead of 24.0~\% (Table \ref{tab:hypothetical_scenario_results}). Stricter measures are more effective at countering the epidemic trend than voluntary measures. Indeed, implementing policy P1 over P2 or P3 has diminishing epidemiological gains but results in much more economic damage (Table \ref{tab:hypothetical_scenario_results}). However, the scope of the epinomic model presented in this work is too limited to adequately balance public health and the economy, as reductions in labor income and the occupied number of IC beds are only two of several societal costs of the \covid{} pandemic. To address this important limitation, we should transform the model into a whole-of-society modeling framework to quantify the foremost damages to the population’s health and the economy.\\

The cost of policies P1, P2, and P3, which involve economic closures, is almost identical, regardless of the date the measures are imposed, whereas the cost of mostly voluntary policies P4a and P4b increases when measures are taken late. Opposed to policies P1, P2, and P3, policies P4a and P4b in themselves incur no direct economic shocks, but as the hospital load increases, individuals will voluntarily start avoiding activities outside the house, as was observed in Sweden \citep{yarmolMatusiak2021}. On March 18, 2020, the implementation of voluntary policies P4a and P4b can incur more damage to gross output and employment than policy P3, in which only accommodation, recreation, and activities of membership organizations (e.g. religious gatherings) are closed (Fig. \ref{fig:hypothetical_scenarios_BE}).\\

Voluntary recommendations, as given by the Swedish government, and implemented on March 15, 2020, had likely not succeeded in safeguarding the Belgian HCS. However, the Belgian government reacted later than the Swedish government, triggering measures at a hospital incidence of 0.6 and 0.2 patients per \num{100000} inhabitants respectively. Imposing the obligation to work from home (P4a), or a milder lockdown (P3) just three days earlier on March 12, 2020, could have sufficed to safeguard the Belgian HCS. As compared to scenario P1 imposed on March 15, 2020, the cost of furloughing for policies P3 and P4a implemented on March 12, 2020, would have been 14.0 billion euros and 16.3 billion euros, or 2.7~\% and 3.1~\% of Belgian GDP lower (2019 prices). The best way to minimize damages is thus by acting timely, a finding consistent with our previous review \citep{vandepitte2021}. Consequently, we recommend policymakers to implement measures earlier rather than later.\\

\textbf{Scenario 2: Prolonging lockdown increases economic damage and can lead to increased strain on the HCS when measures are released}\\

%can strain the healthcare system and increase economic damages}\\

\begin{figure}[h!]
    \centering
    \includegraphics[width=\linewidth]{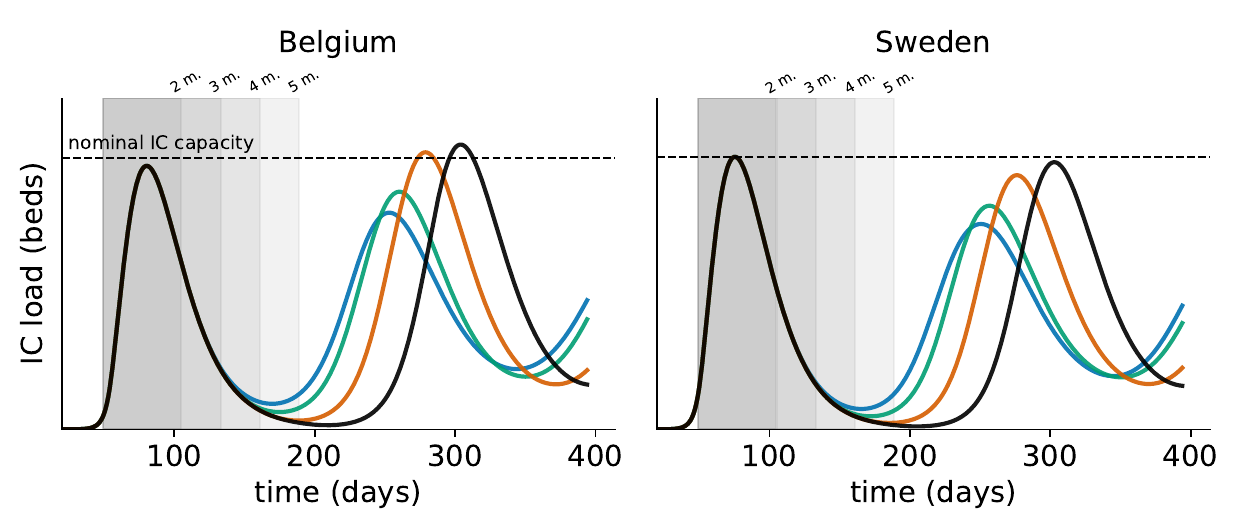}
    \caption{Simulated IC load in Belgium and Sweden during one year. Both governments impose a lockdown to counter the initial epidemic. The lockdown is then maintained for 2 (blue), 3 (green), 4 (orange) and 5 months (black). We use grey background shading to indicate the period of government intervention, though the shade does not refer to lockdown strictness. After releasing the lockdown, no measures are in place.} 
    \label{fig:prevention_paradox}
\end{figure}

In Fig. \ref{fig:prevention_paradox} we demonstrate the impact of prolonging lockdown under low \sars{} incidences after an initial \covid{} surge in Belgium and Sweden. When releasing lockdown after 2 months, we observe no divergence in the number of occupied IC beds in the first month post-relaxation compared to maintaining the lockdown. Prolonged restrictions lead to a higher peak incidence in the second \covid{} surge, potentially straining the HCS. Maintaining lockdown results in only a small decline in IC patients but significantly higher economic damages (Fig. \ref{fig:prevention_paradox_full}, Table \ref{tab:prevention_paradox_results}). Overall, sustaining restrictions appears to carry more risks and costs than benefits. However, our epinomic model's scope is currently too limited to comprehensively balance public health and the economy.

By prolonging restrictions ``forcefully" after the initial surge, \sars{} incidence gradually decreases. However, this approach risks erasing the urgency associated with the past \covid{} wave from collective memory, leading society to desire a return to 'normal' life, even if not permitted by the government. Easing restrictions suddenly results in a sharp rise in social contacts, potentially causing a worse second \covid{} surge, as happened in Belgium \citep{luyten2021}. Based on our simulations, releasing lockdown measures early and gradually is preferred over releasing measures late and suddenly. Prolonging restrictions in pursuit of a ``crush-the-curve" (suppression) strategy will inevitably result in increasing friction between the government and the population, with increasing pressure to relax measures. Given that both Belgium and Sweden never managed to eliminate \sars{} completely from their territories during the entire pandemic, and the efficacy of testing, tracing, and quarantine was insufficient to control \sars{} \citep{kucharski2020}, we have herein implicitly presented an argument against a suppression strategy. Before pursuing such a strategy, epidemiologists and policymakers should carefully consider whether the virus can be realistically controlled at low prevalence using testing, tracing, and quarantine or fully eliminated without the chance of re-importation.\\

\textbf{Scenario 3: Making social contacts dependent on the hospital load's history introduces oscillations in the epidemic dynamics} \\

\begin{figure}[h!]
    \centering
    \includegraphics[width=\linewidth]{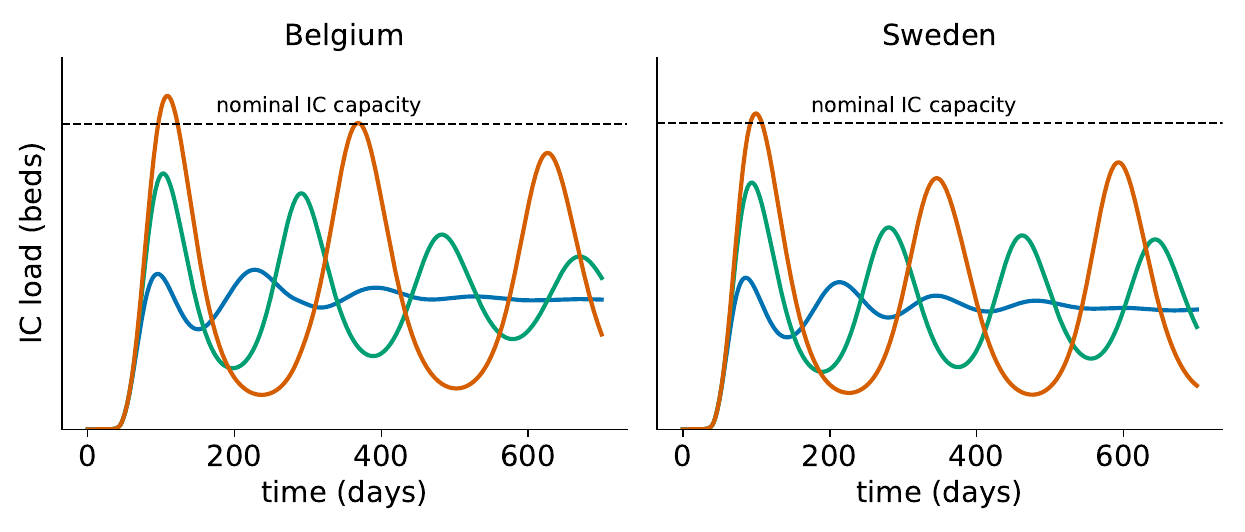}
    \caption{Simulated IC load in Belgium and Sweden during two years year under varying lengths of hospital load memory mean lifetime ($\nu$). $\nu=7~d.$ (blue), $\nu=28~d.$ (green), and $\nu=62~d.$ (orange). A longer mean lifetime implies individuals assign a high weight to past hospital loads in their present behavioral decisions, resulting in slower behavioral adjustments.} 
    \label{fig:variate_parameters-nu-epi_only}
\end{figure}

Making the number of social contacts dependent on the history of the hospital load introduces oscillations in the system's dynamics (Fig. \ref{fig:variate_parameters-nu-epi_only}). Every time the hospital load surges, individuals adjust their behavior and make fewer contacts, which in turn lowers the effective reproduction number, as such curbing the epidemic. Because we feed back an exponentially weighted average hospital load to compute the behavioral change, we introduce a time lag in an individual's response (Fig. \ref{fig:memory}a). Hence, after the epidemic is curbed, individuals will not immediately make more contacts, gradually lowering \sars{} incidence further. Eventually, the first surge will fade from the collective memory, restarting the cycle. A pattern of re-emergent hospitalization waves should be regarded as the natural long-term course of a pandemic and should be accounted for by epidemiologists and policymakers looking to manage a pandemic.

In Fig. \ref{fig:variate_parameters-nu-epi_only} we demonstrate the impact of varying the length of this collective memory, defined by the memory's mean lifetime $\nu$. Larger values of $\nu$ imply individuals assign more weight to the history of the hospital load in their present decision-making, which results in slower behavioral adjustments. Increasing the memory's mean lifetime results in low-frequency, high-amplitude oscillations while lowering the memory's mean lifetime results in high-frequency, low-amplitude oscillations. For $\nu=7~\text{days}$, we observe an endemic equilibrium at $\approx50\%$ of the nominal IC bed capacity available in both countries. The emergence of such an endemic equilibrium with oscillatory dynamics is consistent with the model of Ronan et al. \cite{ronan2021}. Changing the memory's mean lifetime does not seem to alter the dynamic equilibrium of the system, only the frequency and amplitude of the oscillations. We intend to study the properties of the dynamic equilibrium in future research. Keeping the mean lifetime short minimizes the system's oscillations, rendering it more easy to control. However, translating this shorter mean lifetime into practical advice is challenging. An important limitation of the collective memory feedback model is the absence of a temporal delay caused by the process of collecting, interpreting and communicating the daily hospital incidence, whose influence will likely destabilize the system, which will be addressed in future work.\\

\textbf{Scenario 4: Sweden may be more resilient to the spread of \sars{} than Belgium but its economy is not - Why Sweden may have been fortunate} \\

\begin{figure}[h!]
    \centering
    \includegraphics[width=\linewidth]{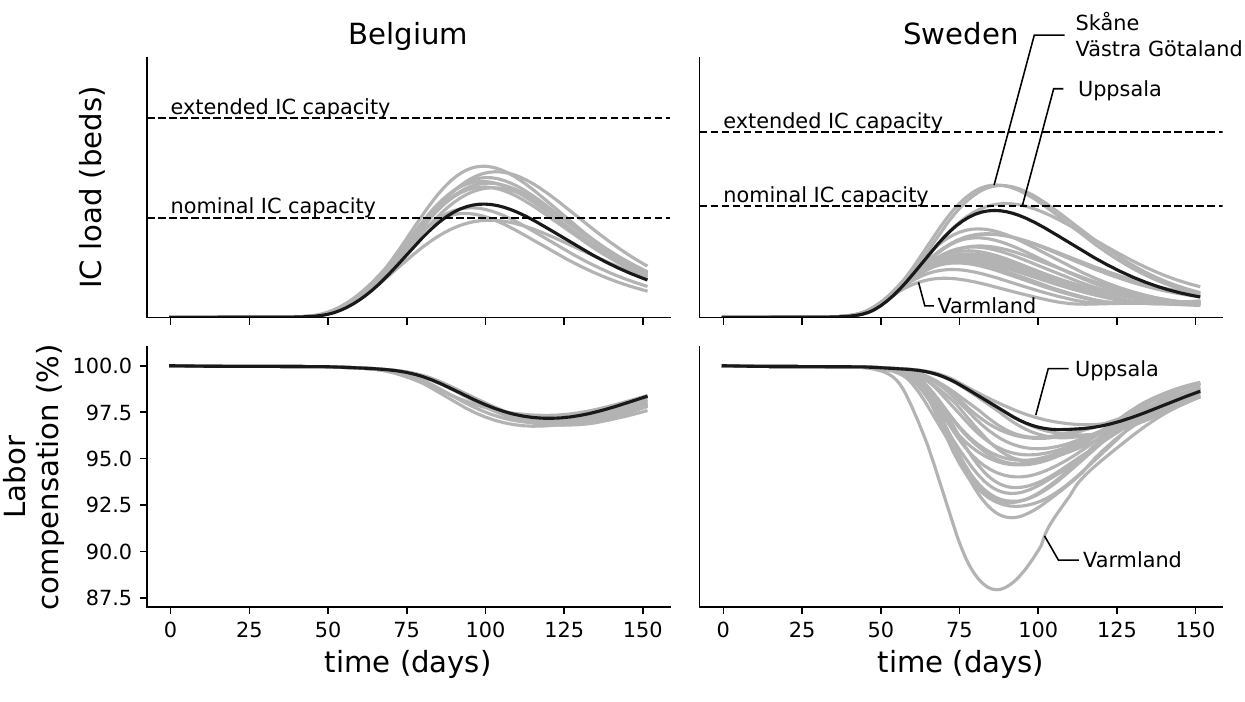}
    \caption{Impact of spatial heterogeneity in the distribution of the first two infected individuals in Belgium and Sweden on the course of the epidemic and economy. (black) The first two infected individuals are located in Stockholm (Sweden), and Brussels (Belgium). (grey) The first infected individual is located in Stockholm (Sweden), and Brussels (Belgium). The second infected individual is located in one of the 20 other counties (Sweden), or 10 provinces (Belgium).} 
    \label{fig:visualise-hypothetical_spatial_spread}
\end{figure}

In Fig. \ref{fig:visualise-hypothetical_spatial_spread}, we exploit the spatially explicit nature of our epinomic model to demonstrate the difference between having a \sars{} epidemic that spreads from a single point-of-origin (black) versus two points-of-origin (gray). In Belgium, having two points-of-origin, and thus a more spatially uniform initial spread of \sars{}, results in more occupied IC beds as compared to having a single point-of-origin for seven out of ten provinces. In contrast, in Sweden, this results in a considerably lower number of occupied IC beds in 17 out of 20 counties. From an epidemiological point-of-view, the Swedish territory thus seems more robust to the spread of \sars{} than Belgium's. \\

Most notable for Sweden is the effect of seeding one infected individual in Stockholm and the other in the counties of Skåne or Västra Götaland, which are Sweden's number two and three counties in terms of population density, as well as containing Sweden's number two and three largest cities, Malmö and Göteborg. If the epidemic had simultaneously been seeded in either county, the resulting epidemic could have surpassed the nominal IC bed capacity by 22~\%. On the other hand, if the epidemic had been seeded in the much more sparsely populated V\"armland, the epidemic would likely have been much smaller. In Belgium, no relationship exists between the population density of the second individual's seed province and the maximum impact of the resulting \sars{} epidemic on the occupied number of IC beds and the reduction in labor income (Fig. \ref{fig:hypothetical_spatial_spread_1}). In Sweden, seeding the epidemic in more sparsely populated areas results in fewer occupied IC beds but more economic damage (Fig. \ref{fig:hypothetical_spatial_spread_1}). This is caused by the way we model the spatial spread of awareness to \sars{}. Indeed, if the epidemic is seeded in a more rural area, the ensuing surge will overwhelm local HCS capacity, even though its absolute magnitude is small. Similar to the early March 2020 \covid{} surge in Bergamo, Italy, images of overcrowded hospitals will make it into (inter)national media, raising awareness to \sars{} as such resulting in voluntary behavioral changes \citep{cereda2021}. The result is a smaller overall epidemic but larger economic damage due to reductions in consumption outside the impacted county. The Swedish territory, which is much larger, more sparsely populated, and less connected than Belgium's, therefore lends itself more to a strategy of voluntary recommendations. However, Swedish policymakers should be wary of pursuing such a strategy during future epidemics. If \sars{} had simultaneously reached Sweden from Denmark at Malmö in March 2020, a counterfactual scenario that is not unimaginable at all, there could have been an acute IC bed shortage during the first \covid{} surge. Instead, based on the counterfactual scenarios for Belgium (Fig. \ref{fig:hypothetical_scenarios_BE}), mandating work-at-home (policy P4a) would result in less \sars{} circulation without inducing economic shocks, thus being the least disruptive alternative policy.

\section{Conclusions}

We developed an economic-epidemiological co-simulation model to explore the impact of voluntary and forced behavioral changes on pandemic progression and its associated economic impacts. Leveraging the divergent policies in Sweden and Belgium, alongside their respective epidemiological and economic outcomes, we calibrated our model and set up several scenarios to showcase the impact of behavioral changes in pandemic management. Our findings emphasize the importance of early responses, which reduce the stringency of measures necessary to safeguard healthcare systems thereby minimizing ensuing economic damage. Early and voluntary reactions seem better than late and forced ones. However, we demonstrated that Sweden's sparsely populated and poorly connected territory may be better suited to a strategy based on voluntary recommendations than Belgium's small, urban, and highly connected territory. Making present voluntary behavioral changes dependent on past hospitalizations introduces fluctuations in the system's dynamics. Re-emergent surges due to behavioral changes should be regarded as the natural long-term course of a pandemic. We also demonstrated that prolonging lockdown after an initial surge leads to higher economic damage and a potentially higher second surge when measures are released. Before pursuing such a strategy, consideration should be given to the possibility of controlling the virus at low incidences using testing, tracing, and quarantine, or fully eliminating the virus without the chance of re-importation. Our adaptable model offers valuable insights for policymakers aiming to minimize economic damage while managing long-term pandemics effectively.

%%%%%%%%%%%%%%%%%%%%%%%%%%%%%%%%%%%%%%%%%%%%%%%%%%%%%%%%%%%%%%%%%%%%%%%%%%%%%
%%%%%%%%%%%%%%%%%%%%%%%%%%%%%%%%%%%%%%%%%%%%%%%%%%%%%%%%%%%%%%%%%%%%%%%%%%%%%
\clearpage
\pagebreak

\backmatter % Difference between bookmark levels is greater than one

\bmhead{Availability of Data and Code}

The source code of the model is freely available on GitHub: \url{https://github.com/twallema/pyIEEM}. The model is implemented using our in-house code for simulating $n$-dimensional dynamical systems in Python 3 named \textit{pySODM} \citep{alleman2023b}. All data used are publicly available.

\bmhead{Supplementary information}

This work contains supplementary information on the (geospatial) differences in demography, recurrent mobility, and employment between Belgium and Sweden (Appendix \ref{app:demography_mobility_employment}). An index of economic activities in the \textit{Nomenclature des Activit\'es Economiques dans la Communaut\'e Europ\'eenne} (NACE) Rev. 2 (Appendix \ref{app:NACE}). The analysis of the contact survey by B\'eraud et al. \citep{beraud2015} (Appendix \ref{app:beraud}). A detailed mathematical description of the epinomic model presented in this study (Appendix \ref{app:detailed_model_description}). An overview of the model's limitations and assumptions (Appendix \ref{app:assumptions}). A detailed description of the epinomic model's calibration to empirical data (Appendix \ref{app:calibration}). A sensitivity analysis of the collective memory feedback model's parameters, in support of the detailed mathematical description of the model (Appendix \ref{app:sensitivity}). Supplementary results not included in the main text (Appendix \ref{app:supplementary_results}).

\bmhead{Author contributions}

\textbf{Tijs W. Alleman}: Conceptualization. Methodology. Formal Analysis. Software. Writing – original draft. Writing – review \& editing. Funding acquisition. \textbf{Jan M. Baetens}: Conceptualization. Writing – review \& editing. Supervision. Project administration. Funding acquisition.

\bmhead{Acknowledgements}

The authors would like to thank Ruben Savels (Ghent University), Tim Van Wesemael (Ghent University), Michiel Rollier (Ghent University), Veerle Vanlerberghe (Institute of Tropical Medicine Antwerp), and Prof. Koen Schoors (Ghent University) for proofreading the manuscript. We would like to extend our gratitude to Prof. Philip Gerlee (Chalmers University of Technology) for proofreading the manuscript and providing us with more insights on the Swedish 2020 \covid{} pandemic. This work was financially supported by \textit{Crelan}, the \textit{Ghent University Special Research Fund}, by the \textit{Research Foundation Flanders} (FWO), Belgium, project numbers G0G2920N/3G0G9820, and, by \textit{VZW 100 km Dodentocht Kadee}, through the organization of the 2020 100 km COVID-Challenge.

\bmhead{Conflict of interest} The authors declare that they have no known competing financial interests or personal relationships that could have appeared to influence the work reported in this paper. The funding sources played no role in study design; in the collection, analysis, and interpretation of data; in the writing of the report; and in the decision to submit the article for publication.

%\bmhead{Ethics approval} Not applicable.

%\bmhead{Consent to participate} Not applicable.

%\bmhead{Consent for publication} All authors have consented to publication of the manuscript in a peer-reviewed scientific journal, preceded by preprint publication in an open-access archive.

\pagebreak
\bibliography{bibliography}

\pagebreak
\begin{appendices}

\section{Demography, mobility, and employment in Belgium and Sweden}\label{app:demography_mobility_employment}

\begin{figure}[h!]
    \centering
    \includegraphics[scale=0.40]{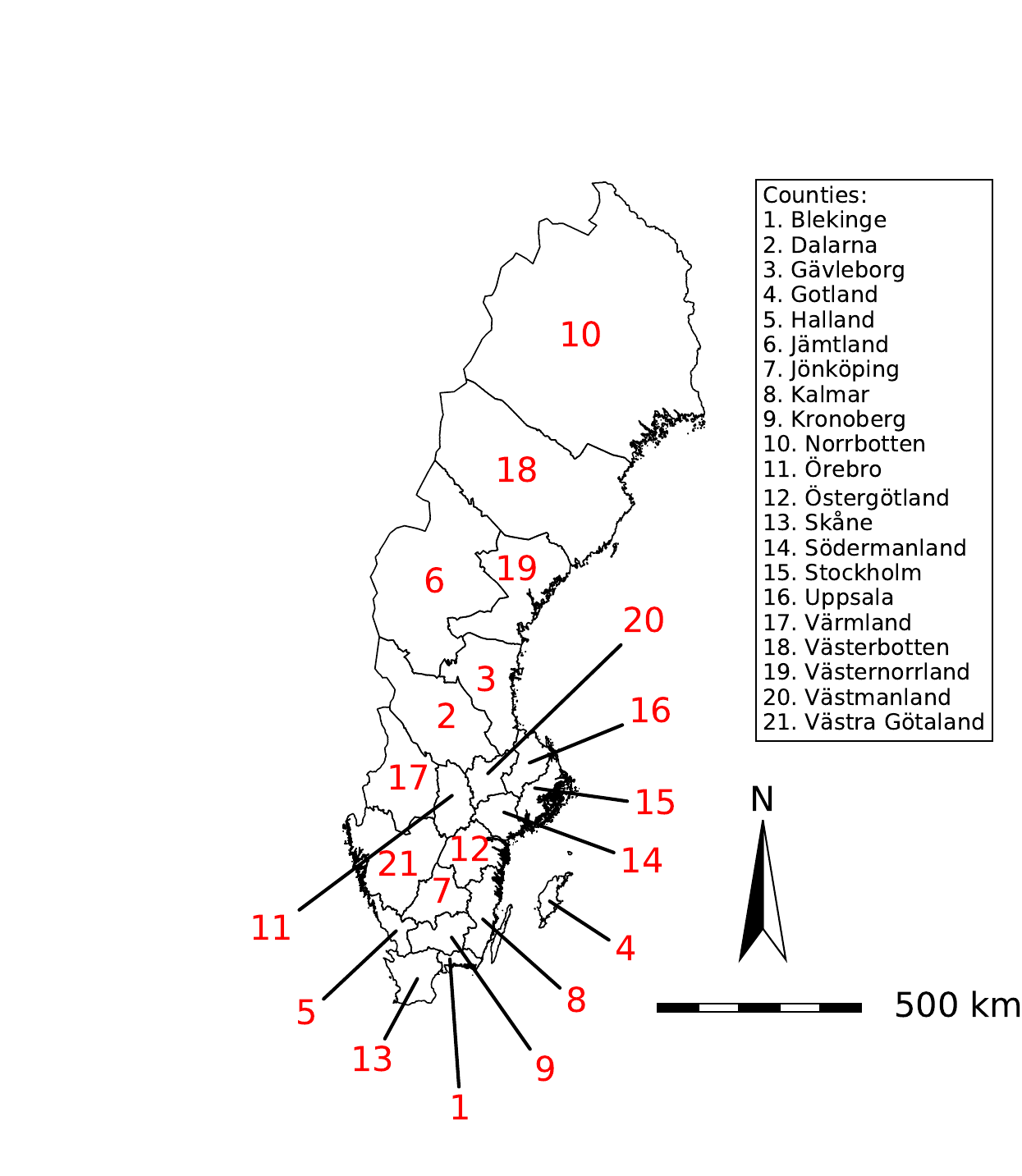}
    \caption{Sweden and its 21 counties (NUTS 2 level).} 
    \label{fig:map_SWE}
\end{figure}

\begin{figure}[h!]
    \centering
    \includegraphics[scale=0.40]{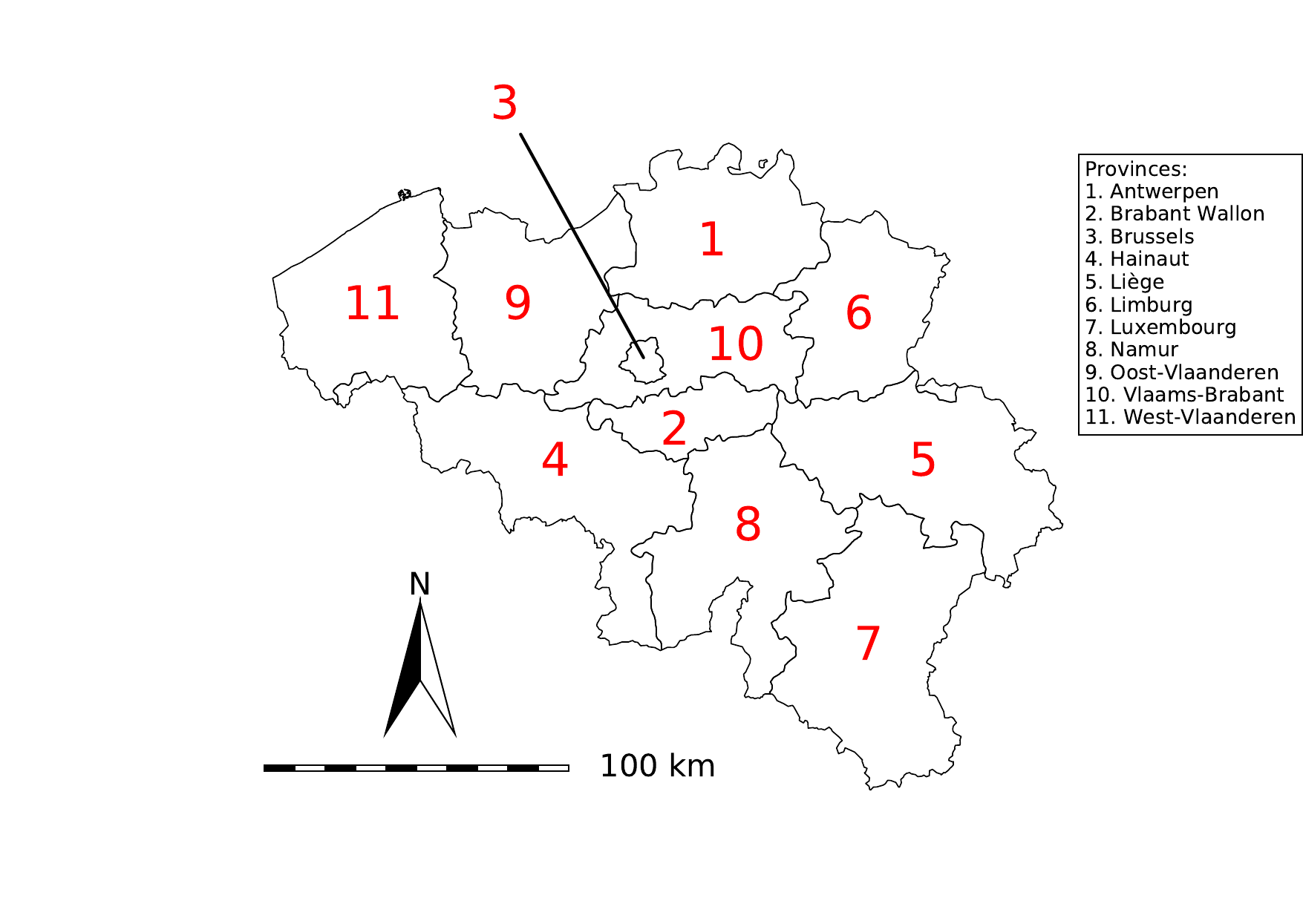}
    \caption{Belgium, its 10 provinces and the Brussels capital region (NUTS 2 level).} 
    \label{fig:map_BE}
\end{figure}

\textbf{Territorial division} Our spatially explicit compartmental disease transmission models are set up at the \textit{Nomenclature of territorial units for statistics} (NUTS) 2 level \citep{NUTS}. This corresponds to 11 spatial units for Belgium and 21 spatial units for Sweden (Figs. \ref{fig:map_SWE} and \ref{fig:map_BE}). \\

\begin{figure}[h!]
    \centering
    \includegraphics[width=\linewidth]{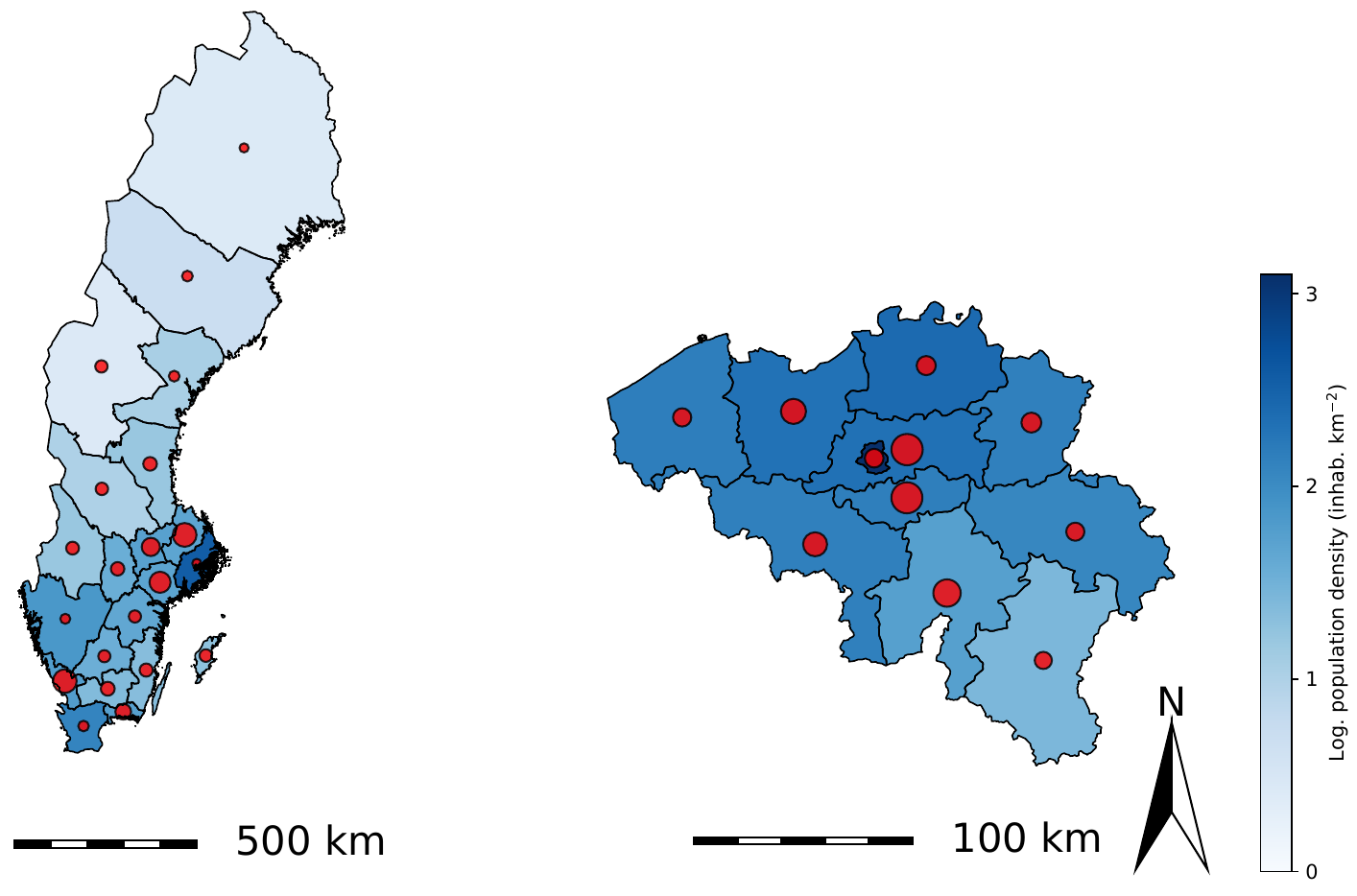}
    \caption{Population density and fraction of the population commuting outbound for Sweden and Belgium at the NUTS 2 level. Population density is indicated by background shading, and outbound commuting mobility is indicated by the size of the red marker (with an identical scale in both countries).} 
    \label{fig:map_popdens}
\end{figure}

\textbf{Demography and mobility} Belgium has an overall population density of 386 $\text{inhab.}\text{km}^{\text{-}2}$, while Sweden has a much lower overall population density of 26 $\text{inhab.}\text{km}^{\text{-}2}$. Population density is more homogeneously distributed in Belgium than in Sweden (Tables \ref{tab:demography_SWE} and \ref{tab:demography_BE}, Fig. \ref{fig:map_popdens}). In Sweden, the counties Stockholm, Skåne (Malm\"o), and V\"astra G\"otaland (G\"oteborg), which contain Sweden's three largest cities, have an elevated population density compared to the other counties (Table \ref{tab:demography_SWE}). The fraction of the active population (16-65 years old) crossing a county/province border for their daily work commute is much lower in Sweden (5.2~\%) than in Belgium (16.2~\%) (Tables \ref{tab:demography_SWE}, \ref{tab:demography_BE}). The Belgian provinces are thus more strongly connected than the Swedish counties implying diseases will spread more readily and homogeneously across the Belgian territory.\\

\begin{figure}[h!]
    \centering
    \includegraphics[width=\linewidth]{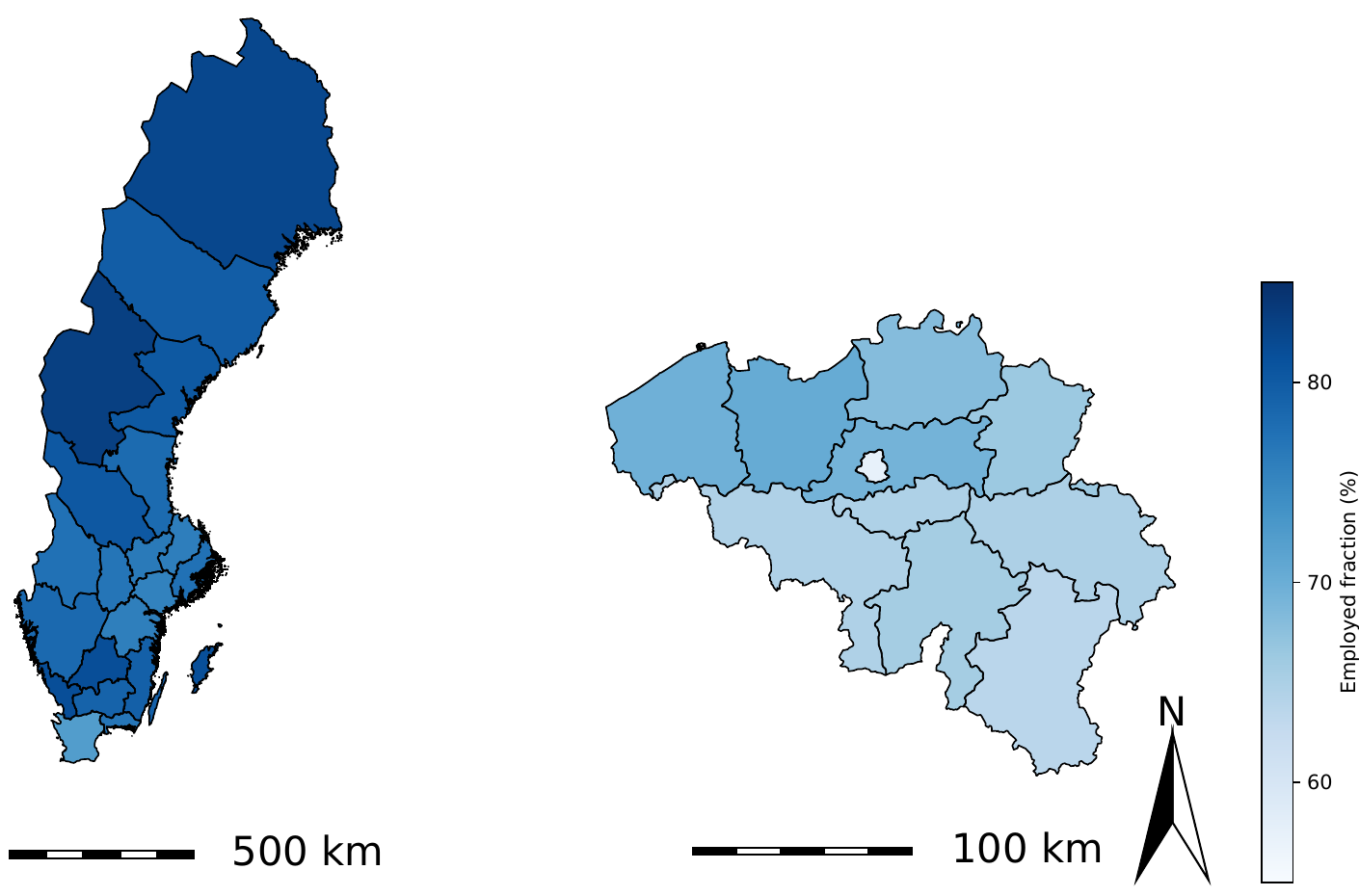}
    \caption{Employed fraction of the active population (16-65 years old) for Sweden and Belgium at the NUTS 2 level.} 
    \label{fig:map_employment}
\end{figure}

\begin{figure}[h!]
    \centering
    \includegraphics[width=\linewidth]{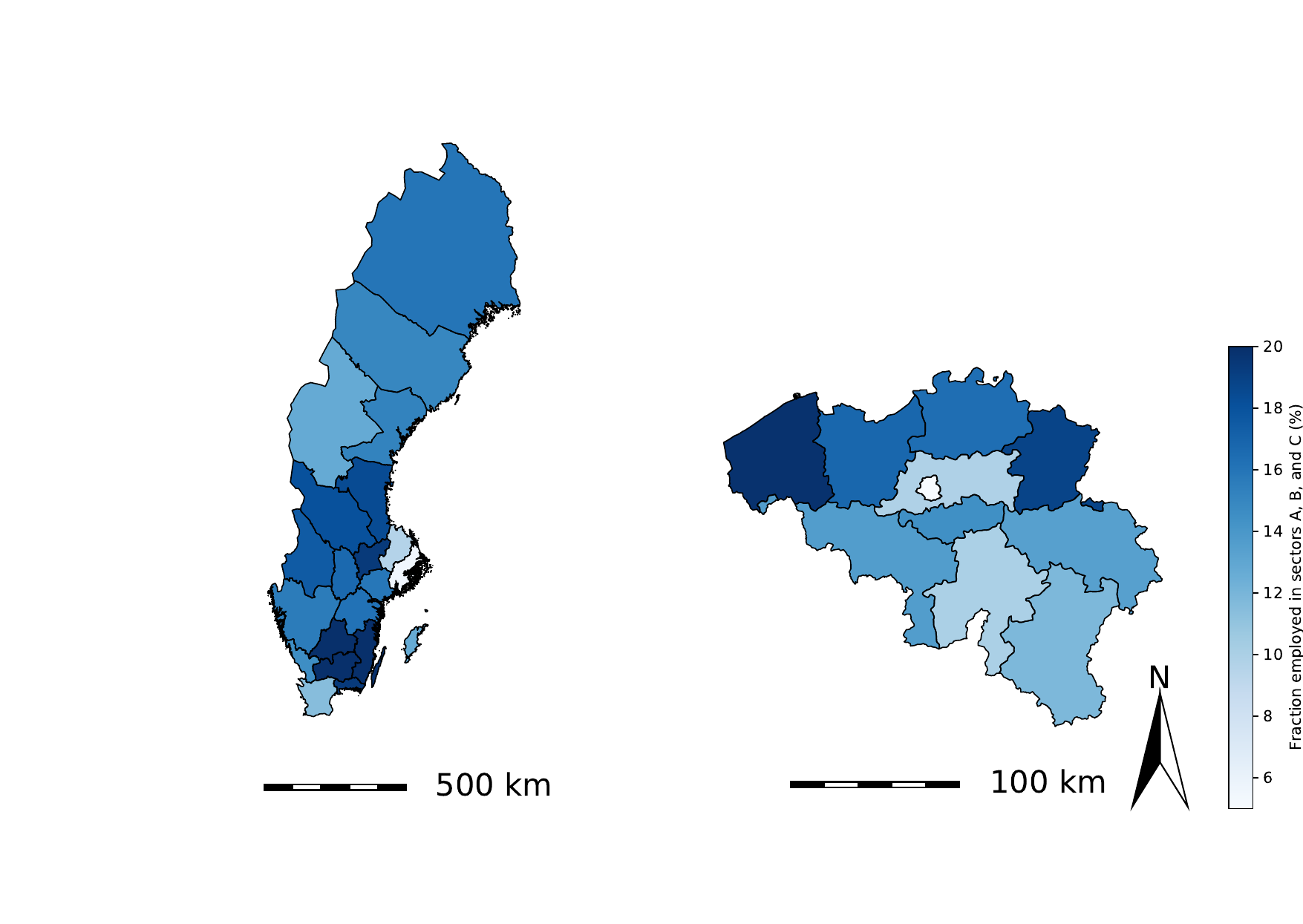}
    \caption{Employees active in sectors A (Agriculture, Forestry, Fishing), B (Mining and Quarrying), and C (Manufacturing) for Sweden and Belgium at the NUTS 2 level.} 
    \label{fig:map_lmc_ABC}
\end{figure}

\begin{figure}[h!]
    \centering
    \includegraphics[width=\linewidth]{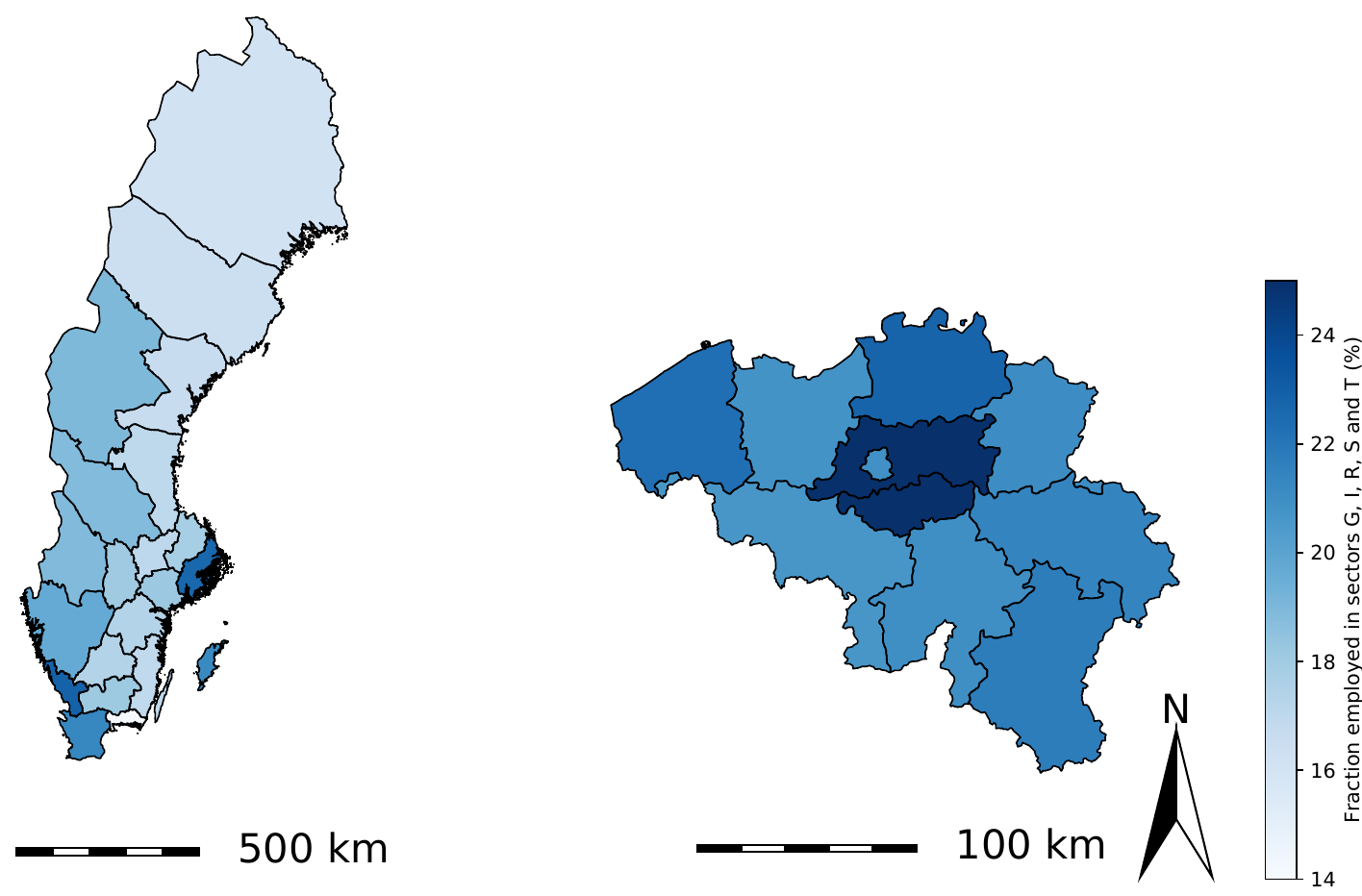}
    \caption{Employees active in sectors G (Retail), I (Accommodation), R (Recreation), S (Personal services), and T (households as employees) for Sweden and Belgium at the NUTS 2 level.} 
    \label{fig:map_lmc_GIRST}
\end{figure}

\textbf{Employment and economic activity} Overall, Belgium has a lower employment rate than Sweden (66.4~\% vs. 77.5~\%) (Tables \ref{tab:demography_SWE}, \ref{tab:demography_BE}). Within both countries, regional differences in employment are small (Fig. \ref{fig:map_employment}). In terms of labor market composition, the overall fraction of employees working in sectors A (Agriculture, Forestry, Fishing), B (Mining and Quarrying), and C (Manufacturing), which produce goods, are similar in both countries (13.4~\% vs. 13.9~\%) (Tables \ref{tab:demography_SWE} and \ref{tab:demography_BE}, Fig. \ref{fig:map_lmc_ABC}). Sectors A, B, and C, are not likely to experience restrictions and/or voluntary reductions in demand during an epidemic. The overall fraction of employees working in sectors G (Retail), I (Accommodation), R (Recreation), S (Personal services), and T (households as employers)  is 19.7~\% in Sweden and 22.0~\% in Belgium. Spatial heterogeneity in Sweden is higher than in Belgium. The latter sectors are most likely to experience restrictions and/or voluntary reductions in demand during an epidemic (Tables \ref{tab:demography_SWE} and \ref{tab:demography_BE}, Fig. \ref{fig:map_lmc_GIRST}). Belgium has more outbound commuters compared to Sweden but a smaller fraction of its population is employed (Figs. \ref{fig:map_popdens} and \ref{fig:map_employment}). The population commuting outbound is not correlated to population density or the employed fraction of the population.\\

\begin{landscape}
\thispagestyle{empty}
\begin{table}
    \centering
    \caption{Number of inhabitants, population density, active population (16-65 years old) commuting outbound, active population (16-65 years old) employed, employed population active in sectors A, B, and C, and, employed population active in sectors G, I, R, S, and T for Sweden \citep{scb2023c,scb2023d,scb2023e}.}
    \begin{tabular}{p{2.2cm}p{2.2cm}p{2.2cm}p{2.2cm}p{2.2cm}p{2.8cm}p{2.8cm}}
        \toprule
        \multicolumn{7}{l}{\textbf{Sweden}} \\
        \midrule
        \textbf{County} & \textbf{Inhabitants ($\bm{10^3}$ inhab.)} & \textbf{Density (inhab. km}\bm{$^{\text{-}2}$}) & \textbf{Active pop. commuting outbound (\%)} & \textbf{Active pop. employed (\%)} & \textbf{Employees in sectors A, B, C (\%)} & \textbf{Employees in sectors G, I, R, S, T (\%)}  \\ \midrule
        Blekinge & 160 & 54 & 7.8 & 76.9 & 19.4 & 14.2 \\
        Dalarma & 288 & 10 & 4.9 & 80.3 & 18.1 & 18.9 \\
        G\"avleborg & 287 & 16 & 5.9 & 78.1 & 18.4 & 17.0 \\
        Gotland & 60 & 19 & 5.0 & 81.4 & 12.7 & 21.2 \\
        Halland & 334 & 62 & 17.5 & 81.5 & 14.6 & 22.9 \\
        J\"amtland & 131 & 3 & 4.8 & 83.2 & 12.8 & 19.0 \\
        J\"onk\"oping & 364 & 35 & 4.6 & 81.5 & 25.3 & 17.4 \\
        Kalmar & 245 & 22 & 5.5 & 79.4 & 22.1 & 16.9 \\
        Kronoberg & 202 & 24 & 6.1 & 79.1 & 21.6 & 18.1 \\
        Norrbotten & 250 & 3 & 2.7 & 82.3 & 16.0 & 16.1 \\
        \"Orebro & 305 & 36 & 5.7 & 77.0 & 16.7 & 18.1 \\
        \"Osterg\"otland & 466 & 44 & 4.7 & 75.9 & 16.2 & 17.4 \\
        Skåne & 1378 & 126 & 3.6 & 72.3 & 11.5 & 21.3 \\
        S\"odermanland & 298 & 49 & 13.9 & 75.6 & 15.9 & 18.2 \\
        Stockholm & 2377 & 365 & 2.6 & 77.5 & 5.5 & 22.6 \\
        Uppsala & 384 & 47 & 17.7 & 76.1 & 9.6 & 17.8 \\
        V\"armland & 282 & 16 & 5.2 & 77.5 & 17.5 & 18.9 \\
        V\"asterbotten & 272 & 5 & 3.7 & 79.8 & 15.0 & 16.4 \\
        V\"asternorrland & 245 & 11 & 3.7 & 80.2 & 15.2 & 16.6 \\
        V\"astmanland & 276 & 54 & 10.4 & 76.5 & 19.4 & 17.1 \\
        V\"astra G\"otaland & 1726 & 73 & 3.2 & 78.3 & 15.6 & 19.7 \\ \midrule
        Total & \num{10328} & 26 & 5.2 & 77.5 & 13.4 & 19.7\\ \bottomrule
    \end{tabular}
    \label{tab:demography_SWE}
\end{table}
\end{landscape}

\begin{landscape}
\thispagestyle{empty}
\begin{table}
    \centering
    \caption{Number of inhabitants, population density, active population (16-65 years old) commuting outbound, active population (16-65 years old) employed, employed population active in sectors A, B, and C, and, employed population active in sectors G, I, R, S, and T for Belgium. \citep{statbel2023a,census2011a,census2011b}.}
    \begin{tabular}{p{2.5cm}p{2.2cm}p{2.2cm}p{2.2cm}p{2.2cm}p{2.8cm}p{2.8cm}}
        \toprule
        \multicolumn{7}{l}{\textbf{Belgium}} \\
        \midrule
        \textbf{Province} & \textbf{Inhabitants ($\bm{10^3}$ inhab.)} & \textbf{Density (inhab. km}\bm{$^{\text{-}2}$}) & \textbf{Active pop. commuting outbound (\%)} & \textbf{Active pop. employed (\%)} & \textbf{Employees in sectors A, B, C (\%)} & \textbf{Employees in sectors G, I, R, S, T (\%)}  \\ \midrule
        Antwerpen & 1858 & 255 & 11.5 & 68.1 & 16.4 & 22.8 \\
        Brabant Wallon & 404 & 147 & 30.2 & 64.8 & 14.5 & 25.0 \\
        Brussels & 1208 & 3088 & 11.0 & 57.5 & 3.1 & 20.9 \\
        Hainaut & 1344 & 142 & 18.0 & 64.6 & 13.6 & 20.7 \\
        Li\`ege & 1107 & 117 & 10.3 & 64.9 & 13.4 & 21.5 \\
        Limburg & 874 & 142 & 12.4 & 66.4 & 18.8 & 21.1 \\
        Luxembourg & 285 & 27 & 9.3 & 63.6 & 11.8 & 21.8 \\
        Namur & 494 & 56 & 23.5 & 65.5 & 10.0 & 21.0 \\
        Oost-Vlaanderen & 1515 & 200 & 19.8 & 70.4 & 16.8 & 20.8 \\
        Vlaams-Brabant & 1146 & 207 & 30.7 & 69.4 & 9.8 & 25.9 \\
        West-Vlaanderen & 1196 & 148 & 10.5 & 69.7 & 19.9 & 22.3 \\ \midrule
        Total & \num{11431} & 386 & 16.2 & 66.4 & 13.9 & 22.0 \\ \bottomrule
    \end{tabular}
    \label{tab:demography_BE}
\end{table}
\end{landscape}

\clearpage
\pagebreak

\section{NACE Rev. 2 classification}\label{app:NACE}

\begin{table}[ht!]
    \caption{Aggregation of the \textit{Nomenclature des Activit\'es \'Economiques dans la Communaut\'e Europ\'eenne} (NACE) Rev. 2 in 21 economic activities \citep{NACE}.}
    \centering
    \begin{tabular}{>{\raggedright\arraybackslash}p{1.1cm}>{\raggedright\arraybackslash}p{6.5cm}}
        \toprule
        \textbf{Code} & \textbf{Name} \\
        \midrule
        A & Agriculture, forestry and fishing \\ 
        B & Mining and quarrying \\ 
        C & Manufacturing \\ 
        D & Electricity, gas, steam, and air conditioning supply \\ 
        E & Water supply, sewerage, waste management and remediation \\ 
        F & Construction \\ 
        G & Wholesale and retail trade \\ 
        H & Transport and storage \\ 
        I & Accommodation and food service \\ 
        J & Information and communication \\ 
        K & Finance and insurance \\ 
        L & Real estate \\ 
        M & Professional, scientific and technical activities \\ 
        N & Administration and support services \\ 
        O & Public administration and defense \\ 
        P & Education \\ 
        Q & Human health and social work \\ 
        R & Arts, entertainment and recreation \\ 
        S & Other service activities \\ 
        T & Activities of households as employers \\ 
        \bottomrule
    \end{tabular}
    \label{tab:NACE21}
\end{table}

\clearpage
\begin{table}[!ht]
    \caption{Aggregation of the \textit{Nomenclature des Activit\'es \'Economiques dans la Communaut\'e Europ\'eenne} (NACE) Rev. 2 in 64 economic activities \citep{NACE}.}
    \centering
    \tiny
    \begin{tabular}{>{\raggedright\arraybackslash}p{1.1cm}>{\raggedright\arraybackslash}p{8.5cm}}
        \toprule
        \textbf{Code} & \textbf{Name} \\
        \midrule
        A01 & Agriculture \\ 
        A02 & Forestry and logging \\ 
        A03 & Fishing and aquaculture \\ 
        B05-09 & Mining and quarrying \\ 
        C10-12 & Manufacture of food, beverages and tobacco products \\ 
        C13-15 & Manufacture of textiles, wearing apparel and leather \\ 
        C16 & Manufacture of wood and of products of wood and cork, except furniture \\ 
        C17 & Manufacture of paper and paper products \\ 
        C18 & Printing and reproduction of recorded media \\ 
        C19 & Manufacture of coke and refined petroleum products \\ 
        C20 & Manufacture of chemicals and chemical products \\ 
        C21 & Manufacture of basic pharmaceutical products and pharmaceutical preparations \\ 
        C22 & Manufacture of rubber and plastic products \\ 
        C23 & Manufacture of other non-metallic mineral products \\ 
        C24 & Manufacture of basic metals \\ 
        C25 & Manufacture of fabricated metal products, except machinery and equipment \\ 
        C26 & Manufacture of computer, electronic and optical products \\ 
        C27 & Manufacture of electrical equipment \\ 
        C28 & Manufacture of machinery and equipment \\ 
        C29 & Manufacture of motor vehicles, trailers and semi-trailers \\ 
        C30 & Manufacture of other transport equipment \\ 
        C31-32 & Manufacture of furniture and other manufacturing \\ 
        C33 & Repair and installation of machinery and equipment \\ 
        D35 & Electricity, gas, steam and air conditioning supply \\ 
        E36 & Water collection, treatment and supply \\ 
        E37-39 & Sewerage; Waste collection, treatment and disposal activities; material recovery; remediation activities \\ 
        F41-43 & Construction of buildings; Civil engineering; Specialised construction activities \\ 
        G45 & Wholesale and retail trade and repair of motor vehicles and motorcycles \\ 
        G46 & Wholesale trade, except of motor vehicles and motorcycles \\ 
        G47 & Retail trade, except of motor vehicles and motorcycles \\ 
        H49 & Land transport and transport via pipelines \\ 
        H50 & Water transport \\ 
        H51 & Air transport \\ 
        H52 & Warehousing and support activities \\ 
        H53 & Postal and courier activities \\ 
        I55-56 & Accommodation and food services \\ 
        J58 & Publishing activities \\ 
        J59-60 & Motion picture, video and television programme production, sound recording and music publishing; Programming and broadcasting activities \\ 
        J61 & Telecommunications \\ 
        J62-63 & Computer programming, consultancy, information services \\ 
        K64 & Financial services, except insurances and pension funding \\ 
        K65 & Insurance, reinsurance, and pension funding, except compulsory social security \\ 
        K66 & Activities auxiliary to financial services and insurance activities \\ 
        L68 & Real estate \\ 
        M69-70 & Legal and accounting \\ 
        M71 & Activities of head offices; management consultancy \\ 
        M72 & Scientific research and development \\ 
        M73 & Advertising and market research \\ 
        M74-75 & Other professional, scientific and technical activities; veterinary activities \\ 
        N77 & Rental and leasing activities \\ 
        N78 & Employment activities \\ 
        N79 & Travel agencies, tour operators, and other reservation services \\ 
        N80-82 & Security and investigation activities; Services to buildings and landscape activities; Office administrative, office support and other business support activities \\ 
        O84 & Public administration and defense; compulsory social security \\ 
        P85 & Education \\ 
        Q86 & Human health activities \\ 
        Q87-88 & Residential care activities; Social work activities without accommodation \\ 
        R90-92 & Creative, arts and entertainment; Libraries, archives, museums and other cultural activities; Gambling and betting \\ 
        R93 & Sports activities and amusement and recreation activities \\ 
        S94 & Activities of membership organizations \\ 
        S95 & Repair of computers and personal and household goods \\ 
        S96 & Other personal service activities \\ 
        T97-98 & Activities of households as employers of domestic personnel; Undifferentiated goods- and services-producing activities of private households for own use \\ 
        \bottomrule
    \end{tabular}
    \label{tab:NACE64}
\end{table}

\clearpage

\pagebreak
\section{Social contact data}\label{app:beraud}

\textbf{Introduction} Many respiratory infectious diseases, such as influenza and \sars{}, are transmitted through close person-to-person contact and thus many population-based surveys of social contacts targeted at understanding the spread of respiratory infections have been conducted in different populations \citep{eames2012, leung2017}. The number of social contacts in disease transmission models is typically informed by a square matrix $\bm{N}$, where an element $N_{ij}$ is the number of social contacts made by an individual in age group $i$ with an individual in age group $j$. To couple our disease transmission model with the production network model, data on contact structures with respect to economic activity are needed. In this work, synthetic contact matrices for public leisure contacts and work contacts, dependent on economic activity, are used. However, few contact studies have estimated contact structures with respect to economic activity. Similar to Haw et al. \citep{haw2022}, we re-analyze the 2012 COMES-F contact survey by B\'eraud et al. \citep{beraud2015} because it is the only (large-scale) study that includes sector-specific information of respondents. \\

\textbf{Survey design} B\'eraud et al. \citep{beraud2015} contacted \num{24250} individuals in mainland France according to quota for age, gender, days of the week, and school holidays by random digit dialing of both landlines and mobile numbers. Contact diaries were sent to \num{3977} people who accepted to participate, among whom \num{2029} ended up in the final dataset. These participants provided personal information, information on their environment, socio-professional background, and all their contacts for 2 consecutive days on a paper diary. Each contact was described by means of its age, gender, location, frequency, type (skin contact or not), and duration. If a correspondent had more than 20 professional contacts, an estimated number of professional contacts and the age groups the contacts belonged to were provided (Supplementary Professional Contacts; SPC). Of special interest to our work are the location of the contact, and, if the correspondent was employed, the economic activity he/she was employed in. Contacts were reported for the following locations: homes, schools, private leisure activities, public leisure activities, public transport, indoor workplaces, and outdoor work and/or leisure activities.\\

\textbf{Alignment of economic activities} Employed individuals could choose their economic activity of employment from 11 options, which we then aligned with the economic activities of the NACE 21 classification (Table \ref{tab:aligment_economic_activities}). Economic activities not included in the survey were aligned based on strong similarities between the activities, or similar incidences of \sars{} infections prior to the second \covid{} wave in Belgium \citep{verbeeck2021}. For education (P), we truncated the reported workplace contacts for individuals employed in sectors P and Q above 20 years old. Social contacts are extrapolated from 21 economic activities (Table \ref{tab:NACE21}) to 64 economic activities (Table \ref{tab:NACE64}) by assuming social contact is the same for all sub-activities in a given group. \\

\begin{table}[h]
    \caption{Alignment of economic activities in the contact survey of B\'eraud et al. \citep{beraud2015} with the NACE Rev. 2 classification (Table \ref{tab:NACE21}).}
    \centering
    \begin{tabular}{p{4.5cm}p{2.7cm}p{3cm}}
        \toprule
        \textbf{Economic activity in survey} & \textbf{Closest economic activity in NACE} & \textbf{Used as proxy for NACE activities} \\
        \midrule
        Agriculture, Forestry, Fishing & A & A \\
        Other industry & C & B, C, H \\
        Energy & D & D \\
        Construction & F & E, F \\
        Commerce & G & G, I, R \\
        Finance and Insurance & K & K \\
        Services to companies & M & J, M \\
        Services to individuals & S, T & L, S, T \\
        Education, Health, Social work & P, Q & P, Q \\
        Administration & N, O & N, O \\
        \bottomrule
    \end{tabular}
    \label{tab:aligment_economic_activities}
\end{table}

\textbf{Extraction of the contact matrices} First, age was transformed into 17 five-year age intervals, and days of the week were transformed into a binary weekday/weekend variable to avoid data sparseness. Then, for the SPC contacts, no information was available on: the duration of the contacts, the type of day (week/weekend, vacation), and, only four large age categories for the age of the contactees were available, Thus, the SPC contacts were imputed to match the survey's contact durations, types of day, and, five-year age categories. For every economic activity, the distribution of contact durations, as well as the ratios of the number of reported work contacts on weekdays, weekend days, and holidays for correspondents reporting between 10-20 contacts at work was derived. We then distributed the SPC contacts over the possible contact durations and types of day using these distributions \citep{beraud2015}. Demographic weighting was used to distribute the contacts over the age categories. Second, the number of contacts in a given location was modeled using a Generalized Estimation Equation (GEE) with a negative binomial distribution. Feature selection was performed by starting with all available independent variables and dropping those that were statistically insignificant. Because of sparseness, we assumed only workplace contacts depend on the economic activity of the participant. Third, the contacts made on public transport, which comprise only a small fraction of the total number of contacts, were incorporated in the indoor work contacts, school contacts, or public leisure contacts depending on the age of the correspondent. The contacts made during outdoor work and/or leisure activities were aggregated with the private leisure contacts for individuals younger than 20 years or older than 60 years. For working-aged individuals, contacts made during outdoor work and/or leisure activities were aggregated with the indoor work contacts on weekdays, and with private leisure contacts on weekend days. Fourth, reciprocity was enforced for contacts at home and during public and private leisure activities. Finally, the contacts matrices for workplaces were smoothed using a Generalized Additive Model (GAM) with a negative binomial distribution \citep{beraud2015}.\\

\begin{figure}[h!]
    \centering
    \includegraphics[width=1.02\textwidth]{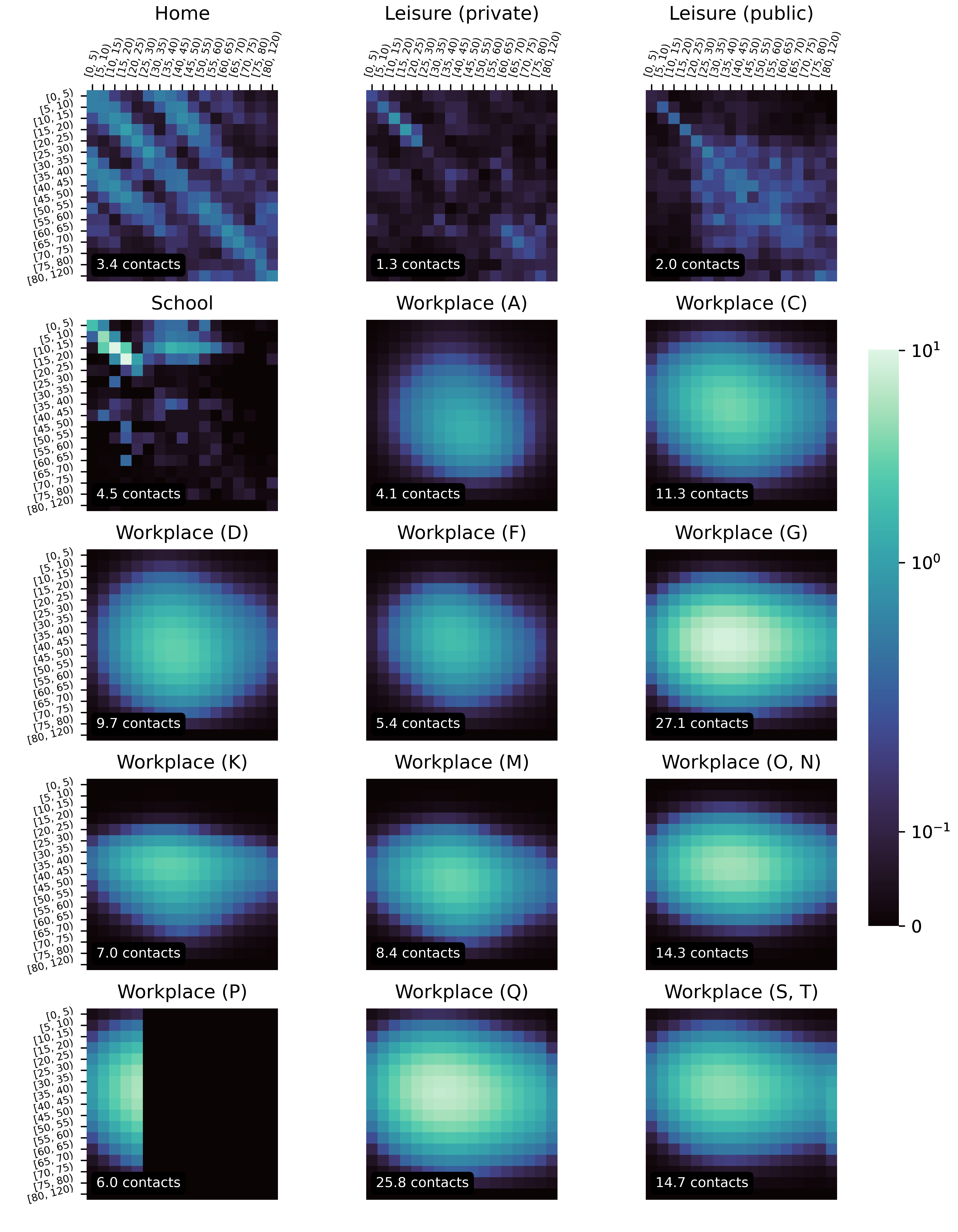}
    \caption{Daily number of social contacts at home, during public and private leisure activities, at school, and in workplaces, in 5-year age groups. An overview of economic activities in NACE Rev. 2 can be found in Table \ref{tab:NACE21}. The average number of social contacts is shown on each contact matrix. Inferred from the social contact survey of B\'eraud \citep{beraud2015}. } 
    \label{fig:contacts}
\end{figure}

\textbf{Results} The daily number of social contacts at home, during public and private leisure activities, at schools, and in workplaces (for the different economic activities) are shown in Figure \ref{fig:contacts}. Without the inclusion of workplace contacts, youths have the greatest number of social contacts, and, thus the biggest potential impact on the spread of infectious diseases, which is consistent with B\'eraud et al. \citep{beraud2015}. At home, the characteristic on-off diagonal mixing pattern of three generations is clearly visible \citep{mossong2008}. Further, our estimate of 3.4 contacts at home per day corresponds exactly to those reported by Verelst et al. \citep{verelst2021}. Employees in retail \& wholesale (G), and human health and social Work (Q) make the most contacts at work (27.1, 25.8), while employees in agriculture, forestry \& farming (A) and construction (F) make the least contacts at work (4.1, 5.4). Prior to the second COVID-19 wave in Belgium, Verbeeck et al. \citep{verbeeck2021} reported the lowest incidences in agriculture (A)($25~\%$)\footnote{Estimates obtained by normalizing the absolute incidence with the mean incidence over the study period, and weighing with the number of employees per economic activity.} and construction (F) ($78~\%$) while the highest incidences were reported for accommodation and food services (I) ($122~\%$), human health (Q) ($139~\%$), and recreation (R) ($147~\%$). The reported incidence in Retail (G47) (105~\%) was only slightly above average. However, mandatory preventive measures, such as face masks, may have had a higher impact on the transmission of \sars{} in retail (G47) than in accommodation, human health (Q), and recreation (R). Individuals working in administration (O, N) have an above-average number of social contacts (14.3), as well as an above-average \sars{} incidence prior to the second COVID-19 wave in Belgium ($112-116~\%$). Overall, the contact matrices at the sectoral level seem to capture the observed trends in \sars{} incidence reported by Verbeeck et al. \citep{verbeeck2021}, justifying their use to couple social contacts and economic activity in our epidemiological-economical co-simulation.\\

\clearpage

\pagebreak
\section{Detailed model description}\label{app:detailed_model_description}

\subsection{Overview}

\textbf{Overview} The \textit{epinomic} models of Belgium and Sweden consist of three connected submodels: 1) A spatially explicit compartmental disease transmission model for \sars{} with recurrent mobility and seasonal forcing of the transmission rate \citep{alleman2023a}. 2) A dynamic production network model based on input-output tables with a relaxed Leontief production function and hiring and firing of workers, used to quantify the impact of supply and demand shocks on employment and gross output, inspired by Pichler et al. \citep{pichler2022} and validated for Belgium \citep{alleman2023c}. 3) A collective memory feedback model to incorporate ``voluntary" behavioral changes, making the present number of social contacts and consumption patterns dependent on the history of \covid{} hospitalizations. In this work, ``voluntary" behavioral changes are a black box encompassing all behavioral changes that result from awareness to \sars{}, including fear, prosocial behavior, and social pressure \citep{bavel2020}. Awareness can be spread by individuals, scientific institutes, and governments. Opposed, ``forced" behavioral changes are the result of enforcement by law and are thus always induced by governments. During the 2020 \covid{} pandemic, Belgian policies were predominantly ``forced" while Swedish policies were predominantly ``voluntary", hence our choice to compare these countries in a case study. The collective memory feedback model is inspired by previous work on the application of control theory to pandemic response \citep{alleman2020}, as well as the works of Ronan et al. \citep{ronan2021} and Nigmatulina and Larson \citep{nigmatulina2009}. The model was implemented using pySODM \citep{alleman2023b}.\\

At every timestep, the production network submodel and dynamic transmission submodel exchange information. We model the impact of laid-off employees on \sars{} spread by using the (sectoral) labor compensation obtained from the production network model to compute the reduction of workplace contacts. We also model the impact of symptomatic \covid{} on labor supply and household demand, as symptomatic individuals are less likely to go to work or attend leisure activities if they are sick \citep{vankerckhove2013}. \\

\begin{figure}[h!]
    \centering
    \includegraphics[width=1.08\linewidth]{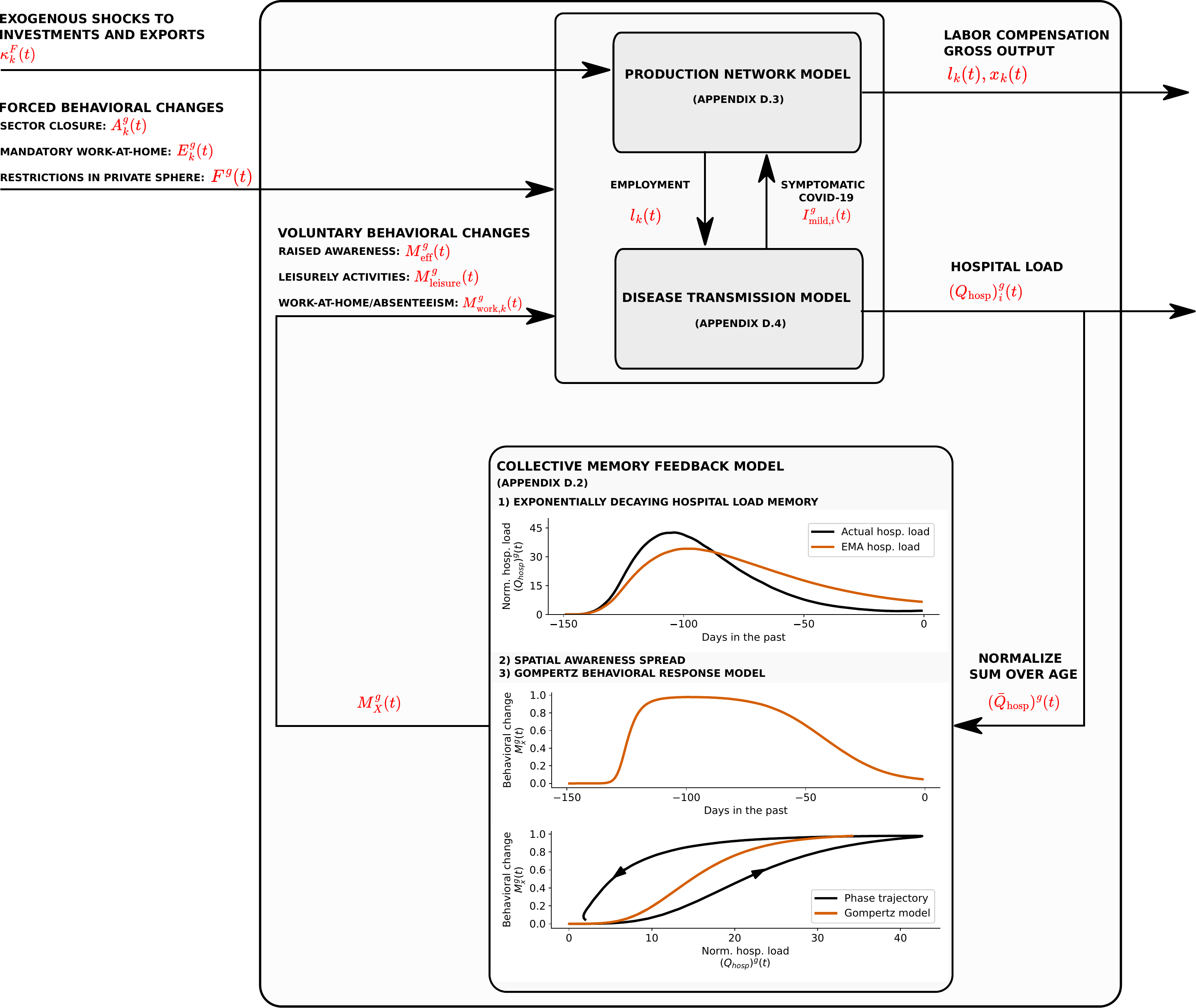}
    \caption{Schematic representation of the \textit{epinomic} model. Subscripts $g$ refer to the spatial patches of the disease transmission model, subscripts $i$ refer to an individual's age in the disease transmission model, and, subscripts $k$ refer to economic activity (Table \ref{tab:NACE64}) in the production network model.} 
    \label{fig:epinomic_flowchart}
\end{figure}

As shown schematically in Figure \ref{fig:epinomic_flowchart}, the epinomic model has four inputs: 1) $\kappa_k^F(t)$, quantifying the shocks to investments and exports in economic activity $k$, which we model exogenously in accordance with trade data for Belgium and Sweden obtained from various sources (Appendix \ref{section:demand}). 2) $A^g_k(t)$, the prohibition of economic activity $k$ in spatial patch $g$ (Table \ref{tab:NACE64}), 3) $E^g_k(t)$, the obligation for employees in economic activity $k$ in spatial patch $g$ to work from home. 4) $F^g(t)$, the prohibition of leisure contacts in the private sphere in spatial patch $g$. The collective memory feedback model, incorporating the ``voluntary" behavioral changes requires no external inputs, but its parameters have to be calibrated (Appendix \ref{app:calibration}). At every timestep, the production network submodel and dynamic transmission submodel exchange information. We model the impact of laid-off employees on \sars{} spread by using the (sectoral) labor compensation obtained from the production network model to compute the reduction of workplace contacts. We also model the impact of symptomatic \covid{} on labor supply and household demand, as symptomatic individuals are less likely to go to work or attend leisure activities if they are sick \citep{vankerckhove2013}. In the absence of government intervention, surging \sars{} cases will lead to a (tardy) reduction of both workplace and leisure contacts, which in turn induces labor supply and household demand shocks to the economy. Our model keeps track of three output variables 1) $x_k(t)$, the gross economic output, 2) $l_k(t)$, labor compensation as a proxy for employment, and 3) $(Q_{\text{hosp}})^g_i(t)$, the regional hospital load. In sections \ref{app:behavoiral_feedback}, \ref{app:dynamic_transmission_model}, and \ref{app:production_network_model} we discuss each of the submodels in detail.\\

\clearpage
\pagebreak
\subsection{Collective memory feedback model}\label{app:behavoiral_feedback}

The collective memory feedback model uses the absolute hospital load per age group $i$ and spatial patch $g$, $(Q_{\text{hosp}})_i^g$, computed using the disease transmission model (Section \ref{app:dynamic_transmission_model}) as its sole input, and thus does not require any user input. First, we eliminate the age groups and normalise the hospital load to \num{100000} inhabitants,
\begin{equation}
(\bar{Q}_{\text{hosp}})^g(t) = 10^5 \frac{\sum_i (Q_{\text{hosp}})^g_i (t)}{T^g},
\end{equation}
where $T^g$ is the total population in spatial patch $g$. On every new day, we add the hospital load in every spatial patch $g$ to a ``memory"  of the hospital load of length $l=6~\text{months}$.
We then degressively weigh the past hospital load using a negative exponential function with a mean lifetime of $\nu$ days.
The impact of $\nu$ on the model's trajectories is presented in Fig. \ref{fig:sensitivity_nu}.
Mathematically, the exponential moving average (EMA) hospital load of spatial patch $g$ and day $t$ is computed as,
\begin{equation}\label{eq:EMA_hospital_load}
\text{EMA}\big[ (\bar{Q}_{\text{hosp}})^g(t) \big] = \frac{1}{N} \sum_{\tau=0}^l  e^{-\tau \nu^{-1}} (\bar{Q}_{\text{hosp}})^g(t-\tau),
\end{equation}
with normalisation factor,
$$N=\sum_{\tau=0}^l  e^{\tau \nu^{-1}}.$$
A comparison between the time course of the actual and EMA hospital load is visualized in Fig. \ref{fig:memory} (a). Next, we account for the spatial spread of awareness by assuming the ``perceived" hospital load in every spatial patch $g$ is computed as a connectivity-weighted average (Fig. \ref{fig:map_popdens}) between the previously computed EMA hospital load on spatial patch $g$ and the EMA hospital load on the spatial patch with the maximum hospital load. To take into account the detrimental impact of a healthcare system (HCS) collapse, which we assume happens when the maximum number of available IC beds has been surpassed\footnote{At this point, hospitals will be forced to ration IC beds.}, we normalize the EMA hospital load with the ratio of IC beds available in both countries. Mathematically,
\begin{equation}\label{eq:EMA_spatial_average}
    (\widetilde{Q}_{\text{hosp}})^g(t) = r~\frac{\text{EMA}\big[ (\bar{Q}_{\text{hosp}})^g(t) \big] + \mu C^{gh} \text{EMA}\big[ (\bar{Q}_{\text{hosp}})^h(t) \big]}{1 + \mu C^{gh}},
\end{equation}
where $h$ is the index corresponding to the spatial patch with the maximum hospital load,
\begin{equation}
h = \arg \max_g \big\{ \text{EMA}\big[ (\bar{Q}_{\text{hosp}})^g(t) \big] \big\},
\end{equation}
and $C^{gh}$ represents the (normalized) connectivity of spatial patch $g$ to spatial patch $h$. For every spatial patch $g$, the average connectivity to the other spatial patches $h$ is computed. We the normalize the connectivity of all spatial patches $h$ connected to spatial patch $g$ with the average connectivity. We refer to $(\widetilde{Q}_{\text{hosp}})^g(t)$ as the ``perceived" hospital load in spatial patch $g$ on day $t$. The spatial connectivity of the hospital load awareness network is governed by $\mu$, where low values of $\mu$ result in low spatial spread of awareness.
The impact of $\mu$ on the model's trajectories is presented in Fig. \ref{fig:sensitivity_mu}.
$r$ is the ratio of IC beds in Belgium (nominally 1000 beds) and Sweden (nominally 600 beds).
The perceived hospital load in every spatial patch is then translated to a fraction bounded between zero and one by means of a simple two-parameter Gompertz model. Mathematically,
\begin{equation}
\bar{M}^g_X(t) = \exp \big( -\xi_X e^{-\pi_X (\widetilde{Q}_{\text{hosp}})^g(t)} \big)~\text{for } X \in \{\text{eff},\text{leisure},\text{work}\}.
\end{equation}
For ease of notation in the subsequent sections,
\begin{equation}\label{eq:gompertz}
M^g_X(t) = 1 - \bar{M}^g_X(t),
\end{equation}
where $\xi_X$ governs the value of $M^g_X(t)$ in the absence of hospitalisations and $\pi_X$ governs the steepness of the curve. We compute voluntary behavioral changes for three $X$: 1) $M^g_{\text{eff}}(t)$, the effectivity of social contacts to spread \sars{}, as people may still have contacts but be more prudent or take preventive measures, inspired by previous work \citep{alleman2021}, 2) $M^g_{\text{leisure}}(t)$, governing the reduction of public and private leisure contacts (Sections \ref{app:dynamic_transmission_model} and \ref{app:production_network_model}), and 3) $M^g_{\text{work},k}(t)$, governing the reduction of work contacts (Sections \ref{app:dynamic_transmission_model} and \ref{app:production_network_model}).\\

We assume that in the absence of hospitalizations, there is no voluntary reduction of leisure and work contacts, so we set $\xi_{\text{leisure}} = \xi_{\text{work}} = 10$. The remaining four parameters, $\xi_{\text{eff}}$, $\pi_{\text{eff}}$, $\pi_{\text{work}}$ and, $\pi_{\text{leisure}}$ are calibrated (Appendix \ref{app:calibration}). The impact of these parameters on the model's trajectories is presented in Figs. \ref{fig:sensitivity_xi_eff}, \ref{fig:sensitivity_pi_eff}, \ref{fig:sensitivity_pi_work} and \ref{fig:sensitivity_pi_leisure}. The steepness of the voluntary reduction of work contacts, $\pi_{\text{work}}$, will depend on the economic activity of employment $k$. We assume the ``willingness" to voluntarily reduce workplace contacts depends on the (normalized) product of the fraction of achievable telework and on the physical proximity with clients or other workers in economic activity $k$. Employees in economic activities where physical proximity is high and working from home is possible are thus assumed to reduce their workplace contacts more quickly. Mathematically, we multiply $\pi_{\text{work}}$ with the ``willingness" to lower workplace contacts $W_k$,
\begin{equation}
    \pi_{\text{work}, k} = W_k \pi_{\text{work}},
\end{equation}
where,
\begin{equation}\label{eq:hesitancy}
W_k = \frac{\text{FP}_k f_{\text{telework},k}}{\sum_k \text{FP}_k f_{\text{telework},k} \bigg( \dfrac{n_{k}}{\sum_l n_{l}} \bigg)},
\end{equation}
with $f_{\text{telework},k}$ the fraction of employees working from home during the first 2020 \covid{} epidemic in Belgium (Table \ref{tab:ERMG_lav}). $\text{FP}_k$ is the physical proximity in the workplace, retrieved from Pichler et al. \citep{pichler2020}, and $n_{k}$ is the number of employees employed in economic activity $k$. The values of $W_k$ are presented in Table \ref{tab:hesitancy}.\\

We present an overview of the key components of the collective memory feedback model in Fig. \ref{fig:memory}. The top panel contains, for an imaginary epidemic, the time courses of the (normalized) hospital load, $(\bar{Q}_{\text{hosp}})^g(t)$, and the (normalized) EMA hospital load, $\text{EMA}\big[ (\bar{Q}_{\text{hosp}})^g(t) \big]$. Taking the EMA of the actual hospital load clearly introduces a time delay between the curves. In the middle panel, an example of a Gompertz function to convert the perceived hospital load, $(\widetilde{Q}_{\text{hosp}})^g(t)$, into a voluntary behavioral change, $M^g_X(t)$, is shown in orange. In black, for our imaginary epidemic, the phase trajectory of the voluntary behavioral change, $M^g_X(t)$, is shown in function of the actual hospital load, $(\bar{Q}_{\text{hosp}})^g(t)$. In the upward phase of the epidemic, the perceived hospital load is lower than the actual hospital load and thus the voluntary response is (too) slow. In this way, we incorporate ``optimism bias" in the model \citep{bavel2020}. In the downward phase, the perceived hospital load is higher than the actual hospital load, and the behavioral return to ``normal" is more gradual. In the bottom panel, we show the time course of $M_X^g(t)$. The time lag introduced by using the EMA of the hospital load makes the behavioral changes $M^g_X(t)$ irreversible, introducing ``hysteresis" in the trajectory of the epidemic \citep{liu2016,lacitignola2021}. We argue that the incorporation of voluntary behavioral dynamics in the epinomic model is beneficial. Moreover, this approach aligns with our previous work on disease transmission models for \sars{}. We incorporated a phenomenological reduction in the effectivity of social contacts to spread \sars{} when lockdown measures were taken with a very similar time course to $M^g_X(t)$ (see Fig. B6, Alleman et al. \citep{alleman2023a}). \\

% \begin{figure}
%   \raggedright
%   \footnotesize
%   \begin{subfigure}{\textwidth}
%     \centering
%     \includegraphics[width=\linewidth]{EMA.pdf}
%     \subcaption{(a) Time course of moving average hospital load in spatial patch $g$. Actual hospital load (black), EMA of the actual hospital load with halflife $\nu=21~d.$ (red). Taking the EMA of the actual hospital load introduces a time lag.}
%     \label{subfig:EMA_memory}
%   \end{subfigure}
%   \vspace{0.25cm}
%   \begin{subfigure}{\textwidth}
%     \centering
%     \includegraphics[width=\linewidth]{M_phasediagram.pdf}
%     \subcaption{(b) A Gompertz function is used to convert the perceived hospital load into a voluntary behavioral change (red). The phase trajectory depicts the values of the voluntary behavioral change $M^g_x(t)$ as a function of the actual hospital load. The trajectory is not reversible, giving rise to \textit{hysteresis}.}
%     \label{subfig:M_timecourse}
%   \end{subfigure}
%   \vspace{0.25cm}
%   \begin{subfigure}{\textwidth}
%     \centering
%     \includegraphics[width=\linewidth]{M_time.pdf}
%     \subcaption{(c) Time course of $M^g_x(t)$.}
%     \label{subfig:M_phasediagram}
%   \end{subfigure}
%   \hspace{0.5cm}
%   \caption{Time course and phase trajectory of voluntary behavioral changes resulting from the collective memory feedback model.}
%   \label{fig:memory}
% \end{figure}

\clearpage
\begin{table}[!ht]
\tiny
\centering
\caption{Fraction of employees employed in Belgium, retrieved from the national accounts. Physical proximity index \citep{pichler2020}. Fraction of employees working from home during the first Belgian 2020 COVID-19 lockdown. Computed as the average of four weekly surveys performed during April 2020 by the Economical Risk Management Group (ERMG) \citep{ermg2021}. ``Willingness" to voluntarily reduce work contacts, computed using Eq. \eqref{eq:hesitancy}.}
\begin{tabular}{>{\raggedright\arraybackslash}p{1.2cm}>{\centering\arraybackslash}p{2cm}>{\centering\arraybackslash}p{2cm}>{\centering\arraybackslash}p{2.2cm}>{\centering\arraybackslash}p{2.4cm}}
\toprule
\textbf{Economic activity} & \textbf{Fraction of employees (\%)} & \textbf{Physical proximity index (-)} & \textbf{Work from home (\%)} & \textbf{Willingness (-)} \\
\midrule

    A01 & 1.2 & 0.35 & 5.5 & 0.08 \\ 
    A02 & $<$ 0.1 & 0.28 & 5.5 & 0.06 \\ 
    A03 & $<$ 0.1 & 0.63 & 5.5 & 0.13 \\ 
    B05-09 & 0.1 & 0.57 & 22.0 & 0.49 \\ 
    C10-12 & 2.1 & 0.61 & 20.0 & 0.47 \\ 
    C13-15 & 0.4 & 0.56 & 20.0 & 0.44 \\ 
    C16 & 0.3 & 0.54 & 27.0 & 0.56 \\ 
    C17 & 0.2 & 0.55 & 16.8 & 0.35 \\ 
    C18 & 0.3 & 0.52 & 39.0 & 0.78 \\ 
    C19 & 0.1 & 0.55 & 30.3 & 0.65 \\ 
    C20 & 0.9 & 0.54 & 30.3 & 0.63 \\ 
    C21 & 0.6 & 0.53 & 30.3 & 0.62 \\ 
    C22 & 0.5 & 0.56 & 17.0 & 0.37 \\ 
    C23 & 0.6 & 0.57 & 17.0 & 0.37 \\ 
    C24 & 0.5 & 0.56 & 15.8 & 0.34 \\ 
    C25 & 1.1 & 0.55 & 15.8 & 0.33 \\ 
    C26 & 0.2 & 0.50 & 59.8 & 1.16 \\ 
    C27 & 0.3 & 0.55 & 30.0 & 0.64 \\ 
    C28 & 0.6 & 0.54 & 30.0 & 0.62 \\ 
    C29 & 0.6 & 0.55 & 18.4 & 0.39 \\ 
    C30 & 0.1 & 0.54 & 6.8 & 0.14 \\ 
    C31-32 & 0.4 & 0.55 & 8.5 & 0.18 \\ 
    C33 & 0.5 & 0.61 & 39.0 & 0.93 \\ 
    D35 & 0.4 & 0.58 & 42.0 & 0.94 \\ 
    E36 & 0.1 & 0.55 & 33.0 & 0.70 \\ 
    E37-39 & 0.5 & 0.52 & 30.0 & 0.61 \\ 
    F41-43 & 5.9 & 0.64 & 21.3 & 0.53 \\ 
    G45 & 1.6 & 0.60 & 18.9 & 0.44 \\ 
    G46 & 4.2 & 0.55 & 25.3 & 0.54 \\ 
    G47 & 6.4 & 0.67 & 7.1 & 0.18 \\ 
    H49 & 2.5 & 0.59 & 20.9 & 0.47 \\ 
    H50 & 0.1 & 0.67 & 35.0 & 0.90 \\ 
    H51 & 0.1 & 0.76 & 21.3 & 0.62 \\ 
    H52 & 1.9 & 0.57 & 30.8 & 0.67 \\ 
    H53 & 0.7 & 0.60 & 36.0 & 0.84 \\ 
    I55-56 & 3.2 & 0.75 & 2.5 & 0.07 \\ 
    J58 & 0.2 & 0.49 & 73.0 & 1.39 \\ 
    J59-60 & 0.3 & 0.58 & 73.0 & 1.64 \\ 
    J61 & 0.5 & 0.55 & 73.0 & 1.55 \\ 
    J62-63 & 1.7 & 0.48 & 73.0 & 1.37 \\ 
    K64 & 1.1 & 0.50 & 81.0 & 1.58 \\ 
    K65 & 0.5 & 0.50 & 81.0 & 1.56 \\ 
    K66 & 0.7 & 0.50 & 81.0 & 1.58 \\ 
    L68 & 0.6 & 0.58 & 41.8 & 0.93 \\ 
    M69-70 & 8.2 & 0.46 & 62.8 & 1.12 \\ 
    M71 & 1.3 & 0.50 & 62.8 & 1.22 \\ 
    M72 & 0.2 & 0.47 & 62.8 & 1.15 \\ 
    M73 & 0.4 & 0.56 & 62.8 & 1.36 \\ 
    M74-75 & 0.5 & 0.54 & 62.8 & 1.30 \\ 
    N77 & 0.3 & 0.61 & 26.8 & 0.63 \\ 
    N78 & 4.3 & 0.61 & 59.8 & 1.40 \\ 
    N79 & 0.2 & 0.61 & 26.8 & 0.63 \\ 
    N80-82 & 4.7 & 0.61 & 40.1 & 0.94 \\ 
    O84 & 9.0 & 0.60 & 76.4 & 1.76 \\ 
    P85 & 8.4 & 0.66 & 100.0 & 2.57 \\ 
    Q86 & 6.8 & 0.75 & 36.0 & 1.04 \\ 
    Q87-88 & 6.5 & 0.75 & 36.0 & 1.04 \\ 
    R90-92 & 0.5 & 0.75 & 15.5 & 0.45 \\ 
    R93 & 0.4 & 0.75 & 15.5 & 0.45 \\ 
    S94 & 0.9 & 0.61 & 15.5 & 0.36 \\ 
    S95 & 0.1 & 0.61 & 7.1 & 0.17 \\ 
    S96 & 1.4 & 0.64 & 2.5 & 0.06 \\ 
    T97-98 & 0.8 & 0.61 & 2.5 & 0.06 \\ 
\bottomrule
\end{tabular}
\label{tab:hesitancy}
\end{table}

\clearpage
\pagebreak
\subsection{Disease transmission model}\label{app:dynamic_transmission_model}

\begin{figure}[h!]
    \centering
    \includegraphics[width=\textwidth]{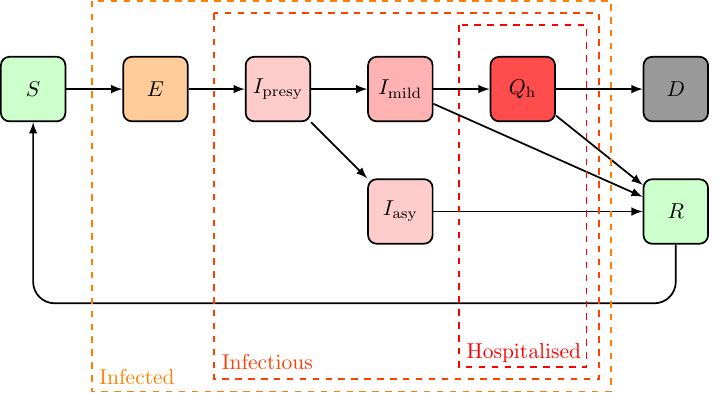}
    \caption{Flowchart of the SEIQRD model used in this work. Here, $S$ stands for susceptible, and $E$ for exposed, $I_\text{presy}$ for presymptomatic infectious, $I_\text{asy}$ for asymptomatic infectious, $I_\text{mild}$ for mildly symptomatic and infectious. $Q_{\text{h}}$ stands for hospitalized, where $Q$ refers to the simplifying assumption that hospitalized individuals are unable to spread the infection to healthcare workers. After infection, individuals are either deceased ($D$) or recovered ($R$). To model seroreversion, recovered individuals may again become susceptible.} 
    \label{fig:flowchart_covid}
\end{figure}

\textbf{Disease compartments and stratifications} The disease transmission model for \sars{} used in this work is similar to our previously established models \citep{alleman2021,alleman2023a}. A flowchart depicting the various compartments of the \sars{} model used in this work is shown in Fig. \ref{fig:flowchart_covid}. The model accounts for pre-symptomatic and asymptomatic transmission of \sars{}, and for different COVID-19 severities, ranging from mild disease to hospitalization. Opposed to the previously developed models, we omit the inclusion of detailed hospital dynamics. Waning of antibodies (seroreversion) is included, as re-susceptibility to SARS-CoV-2 is likely to influence disease dynamics over the time horizon of this work. Every disease compartment is first stratified into 17 five-year age categories: 0–5, 5–10, ..., 75-85, 85-120 years of age, to account for the fact that social contact and disease severity differs substantially between individuals of different ages (Fig. \ref{fig:contacts}). The model is then further stratified into 21 spatial patches for Sweden (Fig. \ref{fig:map_SWE}) and 11 spatial patches for Belgium (Fig. \ref{fig:map_SWE}), resulting in a total of 2856 and 1496 model states. \\

\pagebreak
\textbf{Model equations and parameters} \\

In this work, we implemented the dynamics shown in Fig. \ref{fig:flowchart_covid} using ordinary differential equations (ODEs),
\begin{eqnarray}\label{eq:DTM}
\lambda_i^g(t) &=& s_i \bar{\beta}(t) \sum_{j=1}^N \Bigg[ \bigg(N_{ij}^g(t) - (N_{\text{work}})_{ij}^g(t) \bigg) S^g_j \frac{(I_{\text{presy}})_j^g + (I_{\text{asy}})^g_j  + (I_{\text{mild}})^g_j}{T_j^g} + \nonumber \\
&& (N_{\text{work}})_{ij}^h(t) \sum_{h=1}^G  \bar{P}^{gh} S_j^h \frac{ (I_{\text{presy}})_j^h + (I_{\text{asy}})^h_j + (I_{\text{mild}})^h_j}{T_j^h} \Bigg], \nonumber \\ 
\dot{S}^g_i &=&  \zeta^{\text{-}1} R^g_i - \lambda_i^g(t), \nonumber \\ 
\dot{E}^g_i &=&  \lambda_i^g(t) - \alpha^{\text{-}1} E_i^g, \nonumber \\ 
(\dot{I}_{\text{presy}})^g_i &=& \alpha^{\text{-}1} E_i^g - \gamma^{\text{-}1} (I_{\text{presy}})_i^g,   \nonumber \\ 
(\dot{I}_{\text{asy}})^g_i &=& a_i \gamma^{\text{-}1} (I_{\text{presy}})^g_i - \delta^{\text{-}1} (I_{\text{asy}})^g_i, \nonumber \\  
(\dot{I}_{\text{mild}})^g_i &=& (1-a_i) \gamma^{\text{-}1} (I_{\text{presy}})^g_i - \delta^{\text{-}1} (I_{\text{mild}})^g_i, \nonumber \\ 
(\dot{Q}_{\text{hosp}})^g_i &=& h_i \delta^{\text{-}1} (I_{\text{mild}})^g_i - \epsilon^{\text{-}1} (Q_{\text{hosp}})^g_i, \nonumber \\ 
\dot{R}^g_i &=& \delta^{\text{-}1} (I_{\text{asy}})^g_i + (1-h_i) \delta^{\text{-}1} (I_{\text{mild}})^g_i (1-m_i) \epsilon^{\text{-}1} +  (Q_{\text{hosp}})^g_i - \zeta^{\text{-}1} R_i^g, \nonumber \\ 
\dot{D}^g_i &=& m_i \epsilon^{\text{-}1} (Q_{\text{hosp}})^g_i. \nonumber \\ 
\end{eqnarray}
Here, $S$ stands for susceptible, $E$ for exposed, $I_\text{presy}$ for presymptomatic infectious, $I_\text{asy}$ for asymptomatic infectious, $I_\text{mild}$ for mildly symptomatic and infectious. $Q_{\text{h}}$ stands for hospitalized, $R$ for recovered and $D$ for deceased. Every model state is $N \times G$ dimensional, where $N$ is the number of age groups and $G$ is the number of spatial patches. We summarise the disease transmission model's parameters in Table \ref{tab:DTM_parameters}.

\begin{landscape}
\thispagestyle{empty}
\begin{table}[!h]
    \centering
    \caption{Parameters used to simulate the dynamics between the various disease transmission model compartments shown in Fig. \ref{fig:flowchart_covid}.}
    {\renewcommand{\arraystretch}{1.10}
    \begin{tabular}{p{1cm}p{8cm}p{1.8cm}p{6cm}}
        \toprule
        \textbf{Symbol} & \textbf{Meaning} & \textbf{Value(s)} & \textbf{Source}  \\ \midrule
        $\alpha$ & Length of latent period & $4.5~\text{d}.$ & Computed so the incubation period $\alpha + \gamma = 5.2~\text{d}.$ \citep{li2020} \\
        $\bar{\beta}(t)$ &  Transmission coefficient at day $t$. Time-dependent due to seasonal forcing. & 0.034 (SWE), 0.031 (BE) & Computed so $R_0=3$.\\
        $\gamma$ &  Length of presymptomatic infectiousness & $0.7~\text{d}.$ & \citep{wei2020,he2020} \\
        $\delta$ &  Length of asymptomatic and symptomatic infection & $7.0~\text{d}$ & \citep{he2020} \\
        $\epsilon$ &  Average length of hospital stay & $11.4~\text{d}.$ & \citep{alleman2021} \\
        $\zeta$ &  Average time to seroreversion & $9.2~\text{m}.$ & \citep{alleman2021} \\
        \midrule
        $\bm{a}$ & Asymptomatic fraction. Elements $a_i$ denote the fraction of infected in age group $i$ that don't develop symptoms. & Table \ref{tab:DTM_agedist_parameters} & \citep{poletti2021} \\
        $\bm{s}$ &  Susceptibility. Elements $s_i$ denote the relative susceptibility to infection of an individual in age group $i$. & Table \ref{tab:DTM_agedist_parameters} & \citep{viner2021} \\
        $\bm{h}$ & Case hospitalisation ratio. Elements $h_i$ denote the fraction of mildly symptomatic patients in age group $i$ requiring hospitalisation. & Table \ref{tab:DTM_agedist_parameters} & \citep{alleman2021} \\
        $\bm{m}$ &  Hospital fatality ratio. Elements $m_i$ denote the mortality of hospitalised patients in age group $i$. & Table \ref{tab:DTM_agedist_parameters} & \citep{alleman2021} \\
        \midrule
        $\bm{\bar{P}}$ &  Mobility matrix. Elements $\bar{P}^{gh}$ denote the fraction of the active population of spatial patch $g$ commuting to spatial patch $h$. & Fig. \ref{fig:map_popdens} & \citep{scb2023c} (SWE) and \citep{census2011a} (BE) \\
        $\bm{N}(t)$ &  Social contact matrix. Elements $N^g_{ij}$ denote the number of social contacts made by an individual in age group $i$ with an individual in age group $j$ in spatial patch $g$ on day $t$. & Fig. \ref{fig:contacts} & Appendix \ref{app:beraud}, \citep{beraud2015} \\
      \bottomrule
    \end{tabular}
    }
    \label{tab:DTM_parameters}
\end{table}
\end{landscape}

\begin{table}[!h]
\centering
\caption{Fraction of infected individuals remaining asymptomatic ($a_i$), fraction of mildly symptomatic individuals requiring hospitalization ($h_i$), mortality of patients in the hospital ($m_i$), and, susceptibility to infection ($s_i$).}
\begin{tabular}{ p{3cm} p{1.5cm} p{1.5cm} p{1.5cm} p{1.5cm}} 
\toprule
\textbf{Age class $i$ (years)} & $a_i$ (\%) & $h_i$ (\%) & $m_i$ (\%) & $s_i$ (\%)\\ \midrule
$[0,5[$ & 82.0 & 1.0 & 0.0 & 56.0\\
$[5,10[$ & 82.0 & 1.0 & 0.0 & 56.0\\
$[10,15[$ & 82.0 & 1.0 & 1.2 & 82.0\\
$[15,20[$ & 82.0 & 1.2 & 1.2 & 100 \\
$[20,25[$ & 78.0 & 1.5 & 1.5 & 100 \\
$[25,30[$ & 78.0 & 2.5 & 1.5 & 100 \\
$[30,35[$ & 78.0 & 2.5 & 2.7 & 100 \\
$[35,40[$ & 78.0 & 3.0 & 2.7 & 100 \\
$[40,45[$ & 70.0 & 3.0 & 4.1 & 100 \\
$[45,50[$ & 70.0 & 6.0 & 4.1 & 100 \\
$[50,55[$ & 70.0 & 6.0 & 8.0 & 100 \\
$[55,60[$ & 70.0 & 12.0 & 8.0 & 100 \\
$[60,65[$ & 65.0 & 12.0 & 16.4 & 100 \\
$[65,70[$ & 65.0 & 45.0 & 16.4 & 100 \\
$[70,75[$ & 65.0 & 45.0 & 26.6 & 100 \\
$[75,80[$ & 65.0 & 95.0 & 26.6 & 100 \\
$[80,\infty[$ & 17.8 & 97.0 & 40.4 & 100 \\ \midrule
\textbf{Population average} & 71.4 & 14.7 & 21.4 & 94.1 \\ \bottomrule
\end{tabular}
\label{tab:DTM_agedist_parameters}
\end{table}

\textbf{Recurrent mobility} To quantify the inter-patch connectivity, a square origin-destination matrix $\bm{P}$, whose elements $P^{gh}$ represent the number of daily commuters residing in patch $g$ and working in patch $h$ was retrieved from the Central Bureau of Statistics \citep{scb2023d} for Sweden and from Statbel \citep{census2011a} for Belgium. The mobility matrix was normalized with the size of the active population (16-65 years),
\begin{equation}
\bar{P}^{gh} =\frac{{P^{gh}}}{T_{\text{active}}^g} ,
\end{equation}
where $\bar{P}^{gh}$ is the fraction of the active population of spatial patch $g$ commuting to spatial patch $h$, and, $T_{\text{active}}^g$ is the size of the active population of spatial patch $g$. By normalizing, every row $g$ of the mobility matrix sums to the employed fraction of the active population in spatial patch $g$. The mobility matrix $\bm{\bar{P}}$ is used to determine the number of susceptible and infectious people from province $g$ that visit spatial patch $h$ for work. Revisiting Eq. \eqref{eq:DTM}, we assume spatial patches are connected by commuters having social contact at work. A map containing the fraction of the active population of spatial patch $g$ commuting outbound is shown in Fig. \ref{fig:map_popdens}. \\

\begin{figure}[h!]
    \centering
    \includegraphics[width=\linewidth]{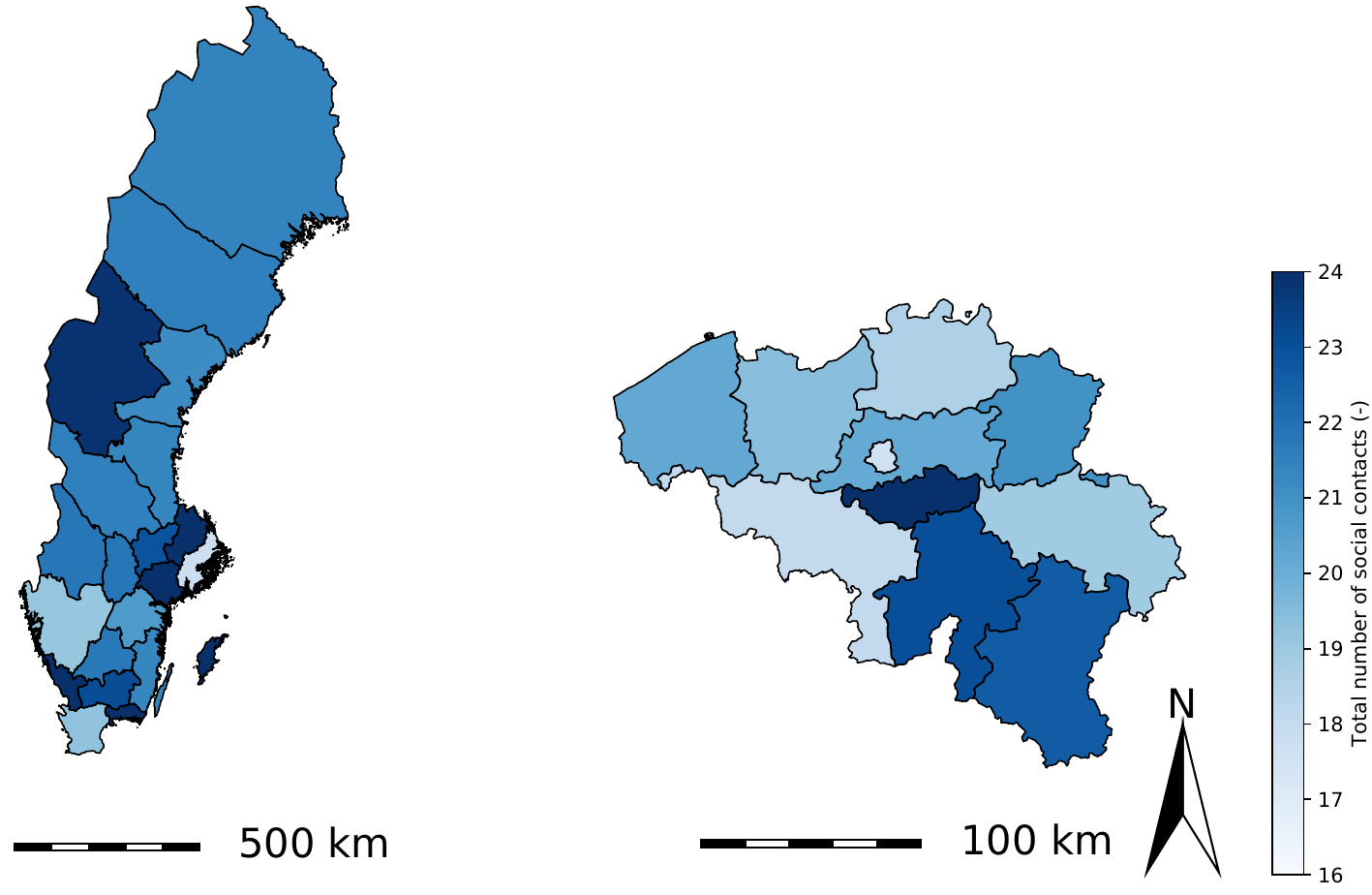}
    \caption{Origin-based number of daily social contacts in Sweden and Belgium at the NUTS 2 level. To be interpreted as the number of daily social contacts of an individual residing in spatial patch $g$.} 
    \label{fig:n_contacts_origin}
\end{figure}

\begin{figure}[h!]
    \centering
    \includegraphics[width=\linewidth]{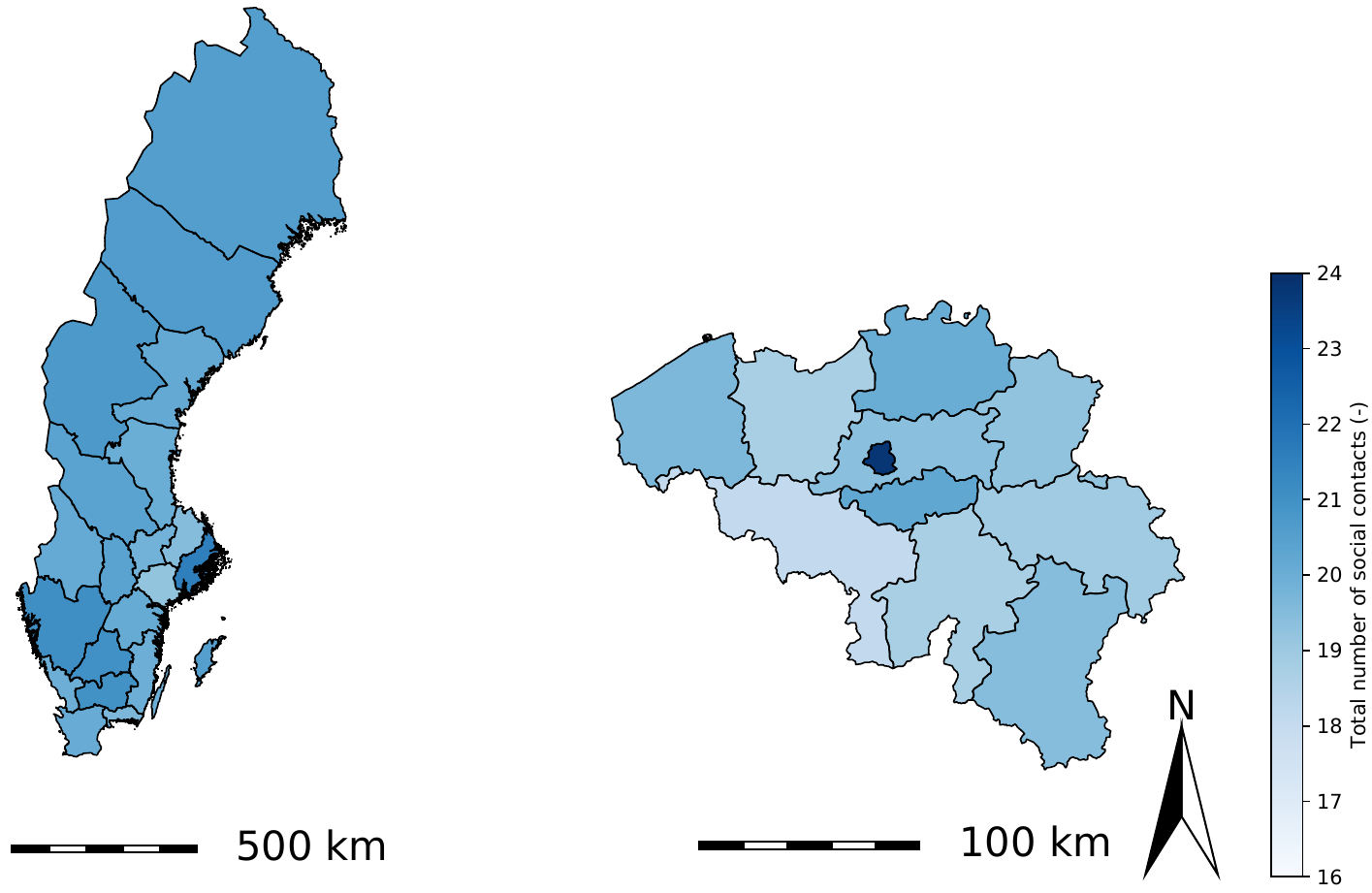}
    \caption{Destination-based number of daily social contacts in Sweden and Belgium at the NUTS 2 level. To be interpreted as the number of daily social contacts happening on the territory of spatial patch $g$. Social contacts take place in metropolitan areas (destination-based contacts) by people residing outside these metropolitan areas (origin-based contacts).} 
    \label{fig:n_contacts_destination}
\end{figure}

\textbf{Social contacts} Social behavior during the pandemic must be translated into a number of social contacts. To this end, we rescale the prepandemic contact matrices obtained by reanalyzing the 2012 social contact study of B\'eraud et al. \citep{beraud2015} (Appendix \ref{app:beraud}). The linear combination of prepandemic interaction matrices used to model pandemic social contact is given by,
\begin{multline}
    \bm{N}^g(t) = \alpha^g(t) \bm{N}^\text{home} + \beta^g(t) \bm{N}^\text{schools} + \sum_k \text{LMC}^g_k \gamma_k^g(t)\bm{N}_k^\text{work}  \\
    + \delta^g(t)\bm{N}^\text{leisure pub.} + \epsilon(t)^g\bm{N}^\text{leisure priv.},
\end{multline}
with,
\begin{equation}
    \left\{
    \renewcommand*{\arraystretch}{1.4}
        \begin{array}{rl}
            \alpha^g(t) &= 1,\\
            \beta^g(t) &= A^{g}_{\text{P85}}(t)M^g_{\text{eff}}(t) ,\\
            \gamma_k^g(t) &= B_k^g(t) M^g_{\text{eff}}(t) ,\\
            \delta^g(t) &= C^g(t) M^g_{\text{eff}}(t),\\
            \epsilon^g(t) &= D^g(t) M^g_{\text{eff}}(t).\\
        \end{array}
    \right.
    \label{eq:coefficients_matrices}
\end{equation}\\

where $\bm{N}^g(t)$ is the square social contact matrix in spatial patch $g$ at time $t$. $\bm{N}^{\text{home}}$, $\bm{N}^{\text{schools}}$, $\bm{N}^{\text{leisure pub.}}$, and $\bm{N}^{\text{leisure priv.}}$ are the prepandemic social contact matrices for homes, schools and leisure activities. $\bm{N}^{\text{work}}_k$ is the prepandemic social contact matrix for employees in economic activity $k$ (Appendix \ref{app:beraud}). Our model does not distinguish between weekdays and weekends but does distinguish holidays. $\text{LMC}^g_k$ (labor market composition) is the fraction of employees of spatial patch $g$ employed in economic activity $k$ (Table \ref{tab:NACE21}). Significant spatial differences in labor market composition exist (Figs. \ref{fig:map_lmc_ABC} and \ref{fig:map_lmc_GIRST}), resulting in a different number of social contacts on every spatial patch $g$ (Figs. \ref{fig:n_contacts_origin} and \ref{fig:n_contacts_destination}). $\bm{A}(t)$, with elements $A_k^g(t)$, represents the degree to which the economic activity $k$ (Table \ref{tab:NACE64}) is allowed in spatial patch $g$ at time $t$, and $A^g_{\text{P85}}(t)$ represents the degree to which schools are open in spatial patch $g$ at time $t$. $\bm{B}(t)$, with elements $B_k^g(t)$, denotes the remaining fraction of work contacts for economic activity $k$ in spatial patch $g$ at time $t$. Its elements are computed as the minimum of mandatory sector closures, mandatory telework, voluntary telework/absenteeism, sickness, and unemployment. Mathematically,
\begin{equation}\label{eq:contact_work}
B_k^g(t) = \min \bigg(\bar{A}_k^g(t),\ \bar{E}_k^g(t),\ M_{\text{work}, k}^g(t),\ \bar{I}_{\text{mild, active}}^g(t), \frac{l_k(t)}{l_k(0)} \bigg)
\end{equation}
where
\begin{equation}\label{eq:bar_A}
\bar{A}_k^g(t) = f_{\text{workplace},k} + A_k^g(t) \big( 1 - f_{\text{workplace},k} \big),
\end{equation}
\begin{equation}\label{eq:bar_E}
\bar{E}_k^g(t) = f_{\text{telework},k} + E_k^g(t) \big( 1 - f_{\text{telework},k} \big),
\end{equation}
and
\begin{equation}\label{eq:mildactive_I}
\bar{I}_{\text{mild, active}}^g(t) = \sum_{i=16}^{65} \Bigg( \frac{I_{\text{mild}, i}^g(t)}{T^g_i(t)} \Bigg).
\end{equation}
In Eqs. \eqref{eq:bar_A} and \eqref{eq:bar_E}, $f_{\text{workplace},k}$ and $f_{\text{telework},k}$ are the fraction of employees in the workplace and working from home during the first 2020 COVID-19 epidemic in Belgium (Table \ref{tab:ERMG_lav}). $\bm{E}(t)$, with elements $E_k^g(t)$, represents the degree to which telework is mandatory in economic activity $k$, spatial patch $g$, and at time $t$. $M^g_{\text{work},k}(t)$ is the ``voluntary" reduction of work contacts (either through telework or absenteeism) in spatial patch $g$ at time $t$ obtained from the collective memory feedback model (Section \ref{app:behavoiral_feedback}). $\bar{I}_{\text{mild, active}}^g(t)$ is the fraction of the active population with mild symptoms in spatial patch $g$ at time $t$. $l_k(t)$ is the labor compensation to employees in economic activity $k$ at time $t$, obtained from the production network model. Equation \eqref{eq:bar_A} dictates that government-mandated economic closures reduce the number of work contacts to $f_{\text{workplace},k}$. During the firs 2020 \covid{} surge, employees were only allowed to go work in cases where working remotely was not possible, but only if distancing could be guaranteed. Given the lack of protective equipment and heavy fines in the event of non-compliance, in practice, these measures were a \textit{de facto} closure of the Belgian economy. \\

$\bm{C}(t)$, with elements $C^g(t)$, denotes the remaining fraction of public leisure contacts in spatial patch $g$ at time $t$, and is computed as the smallest denominator of mandatory sector closures and voluntary reductions of leisure contacts.
\begin{equation}
C^g(t) = \min \Bigg\{ \sum_k \bigg( A^g_k(t) \frac{\text{LAV}^C_k}{\sum_j \text{LAV}^C_j} \bigg),\  M^g_{\text{leisure}}(t),\ \bar{I}_{\text{mild}}^g(t) \Bigg\},
\end{equation}
where $A^g_k(t)$, represents the degree to which economic activity $k$ is allowed in spatial patch $g$ at time $t$, $M^g_{\text{leisure}}(t)$ is the ``voluntary" reduction of leisure contacts in spatial patch $g$ at time $t$ obtained from the collective memory feedback model (Section \ref{app:behavoiral_feedback}). Finally, $\bar{I}_{\text{mild}}^g(t)$ is the fraction of the total population with mild symptoms in spatial patch $g$ at time $t$,
\begin{equation}\label{eq:Imild}
\bar{I}_{\text{mild}}^g(t) = \sum_i \Bigg(\frac{I_{\text{mild}, i}^g(t)}{T^g_i(t)}\Bigg).
\end{equation}
$\text{LAV}^C_k$ is the ``Leisure Association Vector - Contacts", which denotes to what extent an economic activity $k$ is related to leisure contacts (Table \ref{tab:ERMG_lav}). Since, to the best of our knowledge, no quantitative data on the relationship between economic activities and leisure contacts is available, the values of $\text{LAV}$ are based on assumptions. These likely won't alter the results presented in this work for two reasons. First, the number of public leisure contacts is low compared to all other contacts (Fig. \ref{fig:contacts}). Second, economic activities associated with leisure contacts are typically closed together, eliminating the impact of any assumptions made regarding the $\text{LAV}$ altogether. More research on the association between economic activity and leisure contacts is needed to refine the model presented here.\\

$\bm{D}$, with elements $D^g(t)$, denotes the remaining fraction of private leisure contacts in spatial patch $g$ at time $t$. Its entries are computed as the smallest denominator between mandatory social restrictions in the private sphere and voluntary reductions of leisure contacts. Mathematically,
\begin{equation}
D^g(t) = \min \big( F^g(t),\  M^g_{\text{leisure}}(t),\ \bar{I}_{\text{mild}}^g(t) \big),
\end{equation}
where $F^g(t)$ is the degree to which leisurely interactions in the private sphere are prohibited in spatial patch $g$ at time $t$, and, $M^g_{\text{leisure}}(t)$ is the ``voluntary" reduction of leisure contacts in spatial patch $g$ at time $t$ obtained from the collective memory feedback model (Section \ref{app:behavoiral_feedback}). $\bar{I}_{\text{mild}}^g(t)$ is the fraction of the total population with mild symptoms in spatial patch $g$ at time $t$ (Eq. \eqref{eq:Imild}).\\

%Summarising for the sake of clarity, the social contact model has seven inputs, three depend on non-voluntary policies imposed by governments, three depend on voluntary behavioral changes, and the last is $l(t)$, the number of employed individuals obtained from the production network model. The non-voluntary policies are $\bm{A}$, which represents the degree to which the economic activity $k$ is allowed in spatial patch $g$ at time $t$, $\bm{E}$, the degree to which telework is mandated for economic activity $k$ in spatial patch $g$ at time $t$, and, $\bm{F}$, the degree to which leisurely interactions in the private sphere are prohibited in spatial patch $g$ at time $t$. The ``voluntary" reductions are governed by the parameters $\bm{M}$, which in turn depend on the moving average hospital load per \num{100000} inhabitants in spatial patch $g$ (see Section \textbf{X}).\\

\textbf{Seasonal forcing of transmission rate} Seasonal changes have been recognized to play a role in the spread of many viral diseases amongst humans, notably influenza \citep{martinez2018}. Evidence for \sars{} is emerging \citep{gavenciak2022} and seasonal changes played a critical role in describing long-term trends in the number of \covid{} in Belgium \citep{alleman2023a}. We therefore scale the transmission coefficient of \sars{} with a cosine function \citep{liu2021}. Its period is one year, and its amplitude is denoted by $A$, i.e.
\begin{equation}
    \bar{\beta}(t) = \beta(t)\left[ 1 + A \cos\left(2\pi \frac{t - \Delta t}{365 \text{ d.}}\right) \right],
    \label{eq:seasonality}
\end{equation}
where $A$ is the seasonal amplitude, $t$ is expressed in days since January 1st, so that, for $\Delta t=0$, we assume the $\bar{\beta}(t)$ values are maximal on January 1st. $\Delta t$ represents the possible temporal shift of the seasonal effect.\\

\begin{table}[!ht]
\tiny
\centering
\caption{Fraction of employees working from home and at the workplace during the first Belgian 2020 COVID-19 lockdown. Computed as the average of four weekly surveys performed during April 2020 by the Economical Risk Management Group (ERMG) \citep{ermg2021}. Association of economic activity to leisure contacts, $\textstyle \text{LAV}_i^C$ (assumed). For Retail (G47), Rental \& Leasing (N77), Travel agencies and tour operators (N79), and Other Personal Services (S96) we assumed 10\% of contacts associated with these economic activities are of leisurely nature. Maximum shock to household consumption $\textstyle \text{LAV}_i^D$, inspired by household consumption shocks first used by the Congressional Budget Office \citep{congressional_budget_office2006}, then used by Pichler et al. \citep{pichler2022} in his production network model, and, refined in our validation effort \citep{alleman2023c}.}
\begin{tabular}{>{\raggedright\arraybackslash}p{1.2cm}>{\centering\arraybackslash}p{2cm}>{\centering\arraybackslash}p{2cm}>{\centering\arraybackslash}p{2.2cm}>{\centering\arraybackslash}p{2.4cm}}
\toprule
\textbf{Economic activity} & \textbf{Work from home (\%)} & \textbf{At workplace (\%)} & \textbf{$\textstyle \text{LAV}_i^C$ (\%)} & \textbf{$\textstyle \text{LAV}_i^D$ (\%)} \\
\midrule
A01 & 5.5 & 88.0 & 0.0 & 10.0 \\
A02 & 5.5 & 88.0 & 0.0 & 10.0 \\
A03 & 5.5 & 88.0 & 0.0 & 10.0 \\
B05-09 & 22.0 & 71.5 & 0.0 & 10.0 \\
C10-12 & 20.0 & 61.5 & 0.0 & 10.0 \\
C13-15 & 20.0 & 21.2 & 0.0 & 10.0 \\
C16 & 27.0 & 43.0 & 0.0 & 10.0 \\
C17 & 16.8 & 48.5 & 0.0 & 10.0 \\
C18 & 39.0 & 43.0 & 0.0 & 10.0 \\
C19 & 30.2 & 49.0 & 0.0 & 10.0 \\
C20 & 30.2 & 49.0 & 0.0 & 10.0 \\
C21 & 30.2 & 49.0 & 0.0 & 10.0 \\
C22 & 17.0 & 55.8 & 0.0 & 10.0 \\
C23 & 17.0 & 55.8 & 0.0 & 10.0 \\
C24 & 15.8 & 56.5 & 0.0 & 10.0 \\
C25 & 15.8 & 56.5 & 0.0 & 10.0 \\
C26 & 59.8 & 13.5 & 0.0 & 10.0 \\
C27 & 30.0 & 38.5 & 0.0 & 10.0 \\
C28 & 30.0 & 38.5 & 0.0 & 10.0 \\
C29 & 18.4 & 33.3 & 0.0 & 10.0 \\
C30 & 6.8 & 28.0 & 0.0 & 10.0 \\
C31-32 & 8.5 & 27.0 & 0.0 & 10.0 \\
C33 & 39.0 & 43.0 & 0.0 & 10.0 \\
D35 & 42.0 & 58.0 & 0.0 & 0.0 \\
E36 & 33.0 & 67.0 & 0.0 & 0.0 \\
E37-39 & 30.0 & 70.0 & 0.0 & 0.0 \\
F41-43 & 21.2 & 36.0 & 0.0 & 0.0 \\
G45 & 18.9 & 19.8 & 0.0 & 0.0 \\
G46 & 25.3 & 28.3 & 0.0 & 0.0 \\
G47 & 7.1 & 47.4 & 10.0 & 10.0 \\
H49 & 20.9 & 24.5 & 0.0 & 67.0 \\
H50 & 35.0 & 24.5 & 0.0 & 67.0 \\
H51 & 21.2 & 22.2 & 0.0 & 67.0 \\
H52 & 30.8 & 48.5 & 0.0 & 0.0 \\
H53 & 36.0 & 64.0 & 0.0 & 0.0 \\
I55-56 & 2.5 & 6.5 & 100.0 & 100.0 \\
J58 & 73.0 & 8.3 & 0.0 & 0.0 \\
J59-60 & 73.0 & 8.3 & 0.0 & 0.0 \\
J61 & 73.0 & 8.3 & 0.0 & 0.0 \\
J62-63 & 73.0 & 8.3 & 0.0 & 0.0 \\
K64 & 81.0 & 10.8 & 0.0 & 0.0 \\
K65 & 81.0 & 10.8 & 0.0 & 0.0 \\
K66 & 81.0 & 10.8 & 0.0 & 0.0 \\
L68 & 41.8 & 23.5 & 0.0 & 0.0 \\
M69-70 & 62.8 & 10.3 & 0.0 & 0.0 \\
M71 & 62.8 & 10.3 & 0.0 & 0.0 \\
M72 & 62.8 & 10.3 & 0.0 & 0.0 \\
M73 & 62.8 & 10.3 & 0.0 & 0.0 \\
M74-75 & 62.8 & 10.3 & 0.0 & 0.0 \\
N77 & 26.8 & 3.5 & 10.0 & 100.0 \\
N78 & 59.8 & 11.3 & 0.0 & 0.0 \\
N79 & 26.8 & 3.5 & 10.0 & 100.0 \\
N80-82 & 40.1 & 29.4 & 0.0 & 0.0 \\
O84 & 76.4 & 23.6 & 0.0 & 0.0 \\
P85 & 100.0 & 0.0 & 0.0 & 0.0 \\
Q86 & 36.0 & 64.0 & 0.0 & 0.0 \\
Q87-88 & 36.0 & 64.0 & 0.0 & 0.0 \\
R90-92 & 15.5 & 4.8 & 100.0 & 100.0 \\
R93 & 15.5 & 4.8 & 100.0 & 100.0 \\
S94 & 15.5 & 4.8 & 100.0 & 100.0 \\
S95 & 7.1 & 47.4 & 0.0 & 10.0 \\
S96 & 2.5 & 0.8 & 10.0 & 100.0 \\
T97-98 & 2.5 & 0.8 & 0.0 & 100.0 \\ \bottomrule
\end{tabular}
\label{tab:ERMG_lav}
\end{table}

\clearpage
\pagebreak
\subsection{Production network model}\label{app:production_network_model}

\subsubsection{Introduction}

The production network model (Fig. \ref{fig:EPNM_flowchart}) used in this study is based on the work of Pichler et al. \citep{pichler2022}. The model’s defining feature is the relaxation of the stringent Leontief production function, which assumes every input is critical for production. In reality, the closure of restaurants during the \covid{} pandemic is not likely to prevent construction companies from building. Using a Leontief production function results in a (drastic) overestimation of economic damage \citep{pichler2022}. To the best of our knowledge, the closest relative to our model, DAEDALUS by Haw et al. \citep{haw2022}, does not include such relaxation of the Leontief assumption. Before attempting to couple the production network model of Pichler et al. \citep{pichler2022} to our disease transmission models for \sars{} \citep{alleman2023a}, we have validated the production network model to sectoral data on B2B transactions, and sectoral and aggregated surveys on revenue and employment, as well as GDP data, made available by the Belgian National Bank \citep{alleman2023c}. We found the production network model of Dr. Pichler and its parameters results in excellent agreement with all relevant timeseries of Belgian economic indicators.\\

\subsubsection{Model overview and parameters}

Economic activity is classified in $N=64$ sectors corresponding to the \textit{Nomenclature des Activit\'es \'Economiques dans la Communaut\'e Europ\'eenne} (NACE) \citep{NACE}. A detailed index of the 63 economic activities (sectors) is given in Table \ref{tab:NACE64} and an aggregation to 21 economic activities is given in Table \ref{tab:NACE21}. The model uses the $63 \times 63$ input-output matrix of Belgium and Sweden to inform the intermediate flows of services and products in the country's domestic production networks. The economy produces services and products for two end users: households and \textit{other sources} (government and non-profit consumption, investments, and exports). The gross output of sector $k$ is the sum of the intermediate consumption of its goods by all other sectors, household consumption, and exogenous consumption. Mathematically, its basic accounting structure is represented as follows,
\begin{equation}
    x_i(t) = \sum_{l=1}^N Z_{kl}(t) + c_k(t) + f_k(t)
\end{equation}
where $x_k(t)$ is the gross output of sector $k$, $Z_{kl}(t)$ is the input-output matrix containing the intermediate consumption of good $k$ by industry $l$, $c_k(t)$ is the household consumption of good $k$, and $f_k(t)$ is the exogenous consumption of good $k$.\\

We adopt the standard convention that in the input-output matrix columns represent demand while rows represent supply. Prices are assumed time-invariant and capital is not explicitly modeled. One representative firm is modeled for each sector and there is one representative household. Every firm keeps an inventory of inputs from all other firms and draws from these inventories to produce outputs. Intermediates in production are modeled as deliveries replenishing the firm's inventory. The model tracks the dynamics of seven relevant variables such as gross output and labor compensation (Table \ref{tab:overview_states}). Prior to the \covid{} pandemic, the economy was assumed to be in equilibrium and supply equals demand. The pandemic imbalances the model economy through a combination of shocks in consumer demand, exogenous demand, and labor supply. Further, firms may run out of intermediate inputs and may need to stop production. However, as opposed to a traditional Leontief production function, not every intermediate input may be critical to production \citep{pichler2022}.\\

\begin{table}[!h]
    \centering
    \caption{Overview of the production network model's states. Their initial values are listed in Tables \ref{tab:EPNM_initial_states_SWE} and \ref{tab:EPNM_initial_states_BE}.}
    {\renewcommand{\arraystretch}{1.35}
    \begin{tabular}{p{1.2cm}p{8cm}}
        \toprule
        \textbf{Symbol} & \textbf{Name} \\ \midrule
        $x_k(t)$ & Gross output of sector $k$ at time $t$\\
        $d_k(t)$ & Total demand of sector $k$ at time $t$\\
        $l_k(t)$ & Labor compensation to workers in sector $k$ at time $t$\\
        $c_k(t)$ & Realised household consumption of good $k$ at time $t$\\
        $f_k(t)$ & Realised exogenous consumption of good $k$ at time $t$\\
        $O_{kl}(t)$ & Realised B2B demand by sector $k$ of good $l$ at time $t$\\
        $S_{kl}(t)$ & Stock of material $k$ held in the inventory of sector $l$ at time $t$\\
      \bottomrule
    \end{tabular}
    }
    \label{tab:overview_states}
\end{table}

A schematic overview of the model is shown in Figure \ref{fig:EPNM_flowchart}, while its parameters and their values are listed in Table \ref{tab:EPNM_parameters}. At each timestep $t$ the model loops through the following steps.
\begin{enumerate}\itemsep0.5em 
    \item The value of the consumer demand shock, exogenous demand shock, and labor supply shock are computed using the model's four inputs: $\kappa^F_k(t)$ (shock to exogenous demand), $A^g_k(t)$ (mandated sector closures), $M^g_{\text{leisure}}(t)$ (voluntary leisure reduction) and $M^g_{\text{work}, k}(t)$ (voluntary telework or absenteeism) (Section \ref{section:shocks}).
    \item Total demand, desired consumer demand, desired exogenous demand and desired business-to-business demand are computed subject to the aforementioned shocks (Section \ref{section:demand}).
    \item Firms will produce as much as they can to satisfy demand, thus the maximum productive capacity under constrained labor availability and under available inputs is computed. Input bottlenecks are treated in five different ways depending on the criticality of the inputs (Section \ref{section:supply}).
    \item The realized output is computed. If it does not meet demand, then industries ration their output proportionally across households, exogenous agents, and businesses (Section \ref{section:rationing}).
    \item The inventories of each firm are updated using the realized B2B demand (Section \ref{section:inventory_updating}).
    \item Firms hire or fire workers depending on their ability to meet demand (Section \ref{section:hiring_firing}).
\end{enumerate}

%%%%%%%%%%%%%%%%%%%%%%
%% model parameters %%
%%%%%%%%%%%%%%%%%%%%%%

\begin{table}[!h]
    \centering
    \caption{Overview of economic production network model parameters.}
    {\renewcommand{\arraystretch}{1.35}
    \begin{tabular}{p{1.1cm}p{6cm}>{\raggedright\arraybackslash}p{3.5cm}}
        \toprule
        \textbf{Symbol} & \textbf{Name} & \textbf{Value} \\ \midrule
        $\bm{\kappa^D}(t)$ & Elements $\kappa_k^D(t)$. Household consumption shock to sector $k$ & Evolves dynamically  \\
        $\bm{\kappa^F}(t)$ & Elements $\kappa_k^F(t)$. Exogeneous consumption shock to sector $k$ & Evolves dynamically \\
        $\bm{\kappa^S}(t)$ & Elements $\kappa_k^S(t)$. Labor supply shock to sector $k$ & Evolves dynamically \\ \midrule
        $\bm{Z}$ & Elements $Z_{kl}$. Intermediate consumption by sector $k$ of good $l$. Input-Output matrix. & BE: \citep{fpb2018}, SWE: \citep{scb2023f} \\
        $\bm{\mathcal{A}}$ & Elements $\mathcal{A}_{kl}$. Technical coefficients. Payment to sector $k$ per unit produced of $l$. & $\mathcal{A}_{kl} = Z_{kl}/x_l(0)$\\
        $\bm{\mathcal{C}}$ & Elements $\mathcal{C}_k$. Critical inputs of sector $k$ & Survey \citep{pichler2022} \\
        $\bm{\mathcal{I}}$ & Elements $\mathcal{I}_k$. Important inputs of sector $k$ & Survey \citep{pichler2022} \\
        $\bm{n}$ & Elements $n_k$. Inventory kept by sector $k$ so that production can go on for $n_k$ days & Table \ref{tab:EPNM_stock}, \citep{pichler2022} \\
        $\mathcal{F}$ & Production function & Half critical, \citep{alleman2023c}\\
        $\tau$ & Speed of inventory restocking & $14~d.$, \citep{alleman2023c}\\
        $\iota_F$ & Speed of firing & Calibrated, Table \ref{tab:calibration_results} \\
        $\iota_H$ & Speed of hiring & Calibrated, Table \ref{tab:calibration_results} \\
        $\Delta s $ & Changes in the savings rate & $0.75$, \citep{basselier2021}\\
      \bottomrule
    \end{tabular}
    }
    \label{tab:EPNM_parameters}
\end{table}

%%%%%%%%%%%%%%%
%% flowchart %%
%%%%%%%%%%%%%%%

\begin{landscape}
\thispagestyle{empty}
\begin{figure}[h!]
    \centering
    \includegraphics[width=1.08\linewidth]{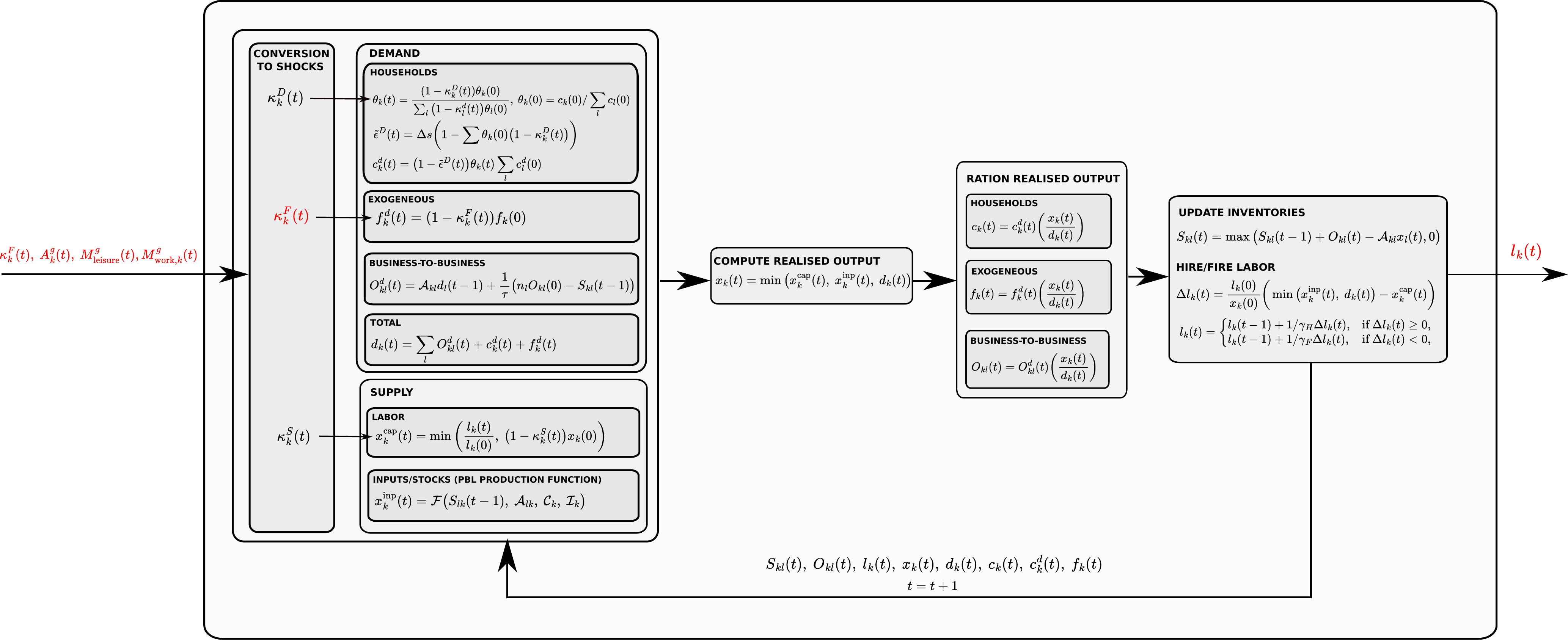}
    \caption{Schematic representation of the production network model. } 
    \label{fig:EPNM_flowchart}
\end{figure}
\end{landscape}

%%%%%%%%%%%%%%%%%%%%%%%%
%% targeted inventory %%
%%%%%%%%%%%%%%%%%%%%%%%%

\begin{table}[!ht]
    \tiny
    \centering
    \caption{Targeted number of days inventory. Inventory kept by sector $k$ so that production can go on for $n_k$ days. Retrieved from Pichler et al. \citep{pichler2022}.}
    \begin{tabular}{>{\raggedright\arraybackslash}p{1.1cm}>{\raggedright\arraybackslash}p{7.5cm}p{1cm}}
        \toprule
        \textbf{NACE 64} & \textbf{Name} & $n_l~(\text{days})$ \\
        \midrule
        A01 & Agriculture & 32.2 \\ 
        A02 & Forestry and logging & 39.2 \\ 
        A03 & Fishing and aquaculture & 73.4 \\ 
        B05-09 & Mining and quarrying & 16.8 \\ 
        C10-12 & Manufacture of food, beverages, and tobacco products & 38.5 \\ 
        C13-15 & Manufacture of textiles, wearing apparel and leather & 50.6 \\ 
        C16 & Manufacture of wood and of products of wood and cork, except furniture & 32.2 \\ 
        C17 & Manufacture of paper and paper products & 28.8 \\ 
        C18 & Printing and reproduction of recorded media & 16.8 \\ 
        C19 & Manufacture of coke and refined petroleum products & 21.5 \\ 
        C20 & Manufacture of chemicals and chemical products & 39.9 \\ 
        C21 & Manufacture of basic pharmaceutical products and pharmaceutical preparations & 47.6 \\ 
        C22 & Manufacture of rubber and plastic products & 32.8 \\ 
        C23 & Manufacture of other non-metallic mineral products & 36.5 \\ 
        C24 & Manufacture of basic metals & 49.6 \\ 
        C25 & Manufacture of fabricated metal products, except machinery and equipment & 38.5 \\ 
        C26 & Manufacture of computer, electronic and optical products & 52 \\ 
        C27 & Manufacture of electrical equipment & 46.3 \\ 
        C28 & Manufacture of machinery and equipment & 44.2 \\ 
        C29 & Manufacture of motor vehicles, trailers, and semi-trailers & 24.5 \\ 
        C30 & Manufacture of other transport equipment & 64.4 \\ 
        C31-32 & Manufacture of furniture and other manufacturing & 39.2 \\ 
        C33 & Repair and installation of machinery and equipment & 37.5 \\ 
        D35 & Electricity, gas, steam, and air conditioning supply & 13.1 \\ 
        E36 & Water collection, treatment and supply & 5.7 \\ 
        E37-39 & Sewerage; Waste collection, treatment and disposal activities; materials recovery; remediation activities & 11.7 \\ 
        F41-43 & Construction of buildings; Civil engineering; Specialised construction activities & 64.4 \\ 
        G45 & Wholesale and retail trade and repair of motor vehicles and motorcycles & 43.6 \\ 
        G46 & Wholesale trade, except of motor vehicles and motorcycles & 18.4 \\ 
        G47 & Retail trade, except of motor vehicles and motorcycles & 31.8 \\ 
        H49 & Land transport and transport via pipelines & 1.7 \\ 
        H50 & Water transport & 2 \\ 
        H51 & Air transport & 1.7 \\ 
        H52 & Warehousing and support activities & 25.8 \\ 
        H53 & Postal and courier activities & 1.3 \\ 
        I55-56 & Accommodation and food services & 7.4 \\ 
        J58 & Publishing activities & 7 \\ 
        J59-60 & Motion picture, video, and television program production, sound recording and music publishing; Programming and broadcasting activities & 11.4 \\ 
        J61 & Telecommunications & 6 \\ 
        J62-63 & Computer programming, consultancy, information services & 6.4 \\ 
        K64 & Financial services, except insurances and pension funding & 9.4 \\ 
        K65 & Insurance, reinsurance, and pension funding, except compulsory social security & 9.7 \\ 
        K66 & Activities auxiliary to financial services and insurance activities & 9.4 \\ 
        L68 & Real estate & 34.2 \\ 
        M69-70 & Legal and accounting & 21.8 \\ 
        M71 & Activities of head offices; management consultancy & 14.7 \\ 
        M72 & Scientific research and development & 8.4 \\ 
        M73 & Advertising and market research & 3.4 \\ 
        M74-75 & Other professional, scientific and technical activities; veterinary activities & 8.4 \\ 
        N77 & Rental and leasing activities & 3.4 \\ 
        N78 & Employment activities & 3.4 \\ 
        N79 & Travel agencies, tour operators and other reservation services & 3.4 \\ 
        N80-82 & Security and investigation activities; Services to buildings and landscape activities; Office administrative, office support and other business support activities & 3.4 \\ 
        O84 & Public administration and defense; compulsory social security & 9.4 \\ 
        P85 & Education & 4 \\ 
        Q86 & Human health activities & 3 \\ 
        Q87-88 & Residential care activities; Social work activities without accommodation & 3 \\ 
        R90-92 & Creative, arts and entertainment; Libraries, archives, museums and other cultural activities; Gambling and betting & 2.3 \\ 
        R93 & Sports activities and amusement and recreation activities & 2.3 \\ 
        S94 & Activities of membership organizations & 2.3 \\ 
        S95 & Repair of computers and personal and household goods & 2.3 \\ 
        S96 & Other personal service activities & 2.3 \\ 
        T97-98 & Activities of households as employers of domestic personnel; Undifferentiated goods- and services-producing activities of private households for own use & 9.4 \\ 
        \bottomrule
    \end{tabular}
    \label{tab:EPNM_stock}
\end{table}

%%%%%%%%%%%%%%%%%%%%%%%%
%% initial states SWE %%
%%%%%%%%%%%%%%%%%%%%%%%%

\begin{table}[!ht]
    \tiny
    \centering
    \caption{Overview of initial states for Sweden (in $10^6~\text{SEK}/\text{y}$). At macro-economic equilibrium, gross output and total demand are equal, $d_k(0) = x_k(0)$. The B2B demand by sector $k$ of good $l$ is equal to the intermediate consumption listed in the input-output matrix and hence $O_{kl}(0) = Z_{kl}$. The initial stock of material $k$ held in the inventory of sector $l$, $S_{kl}(0)$ is computed as $S_{kl}(0) = n_j Z_{kl}(0)$, where $n_l$ is the targeted number of days inventory of material $k$ by sector $l$ (Table \ref{tab:EPNM_stock}). Data were retrieved from the Central Bureau of Statistics \citep{scb2023f}.}
    \begin{tabular}{>{\raggedright\arraybackslash}p{1.1cm}>{\raggedright\arraybackslash}p{7cm}p{0.6cm}p{0.5cm}p{0.5cm}p{0.5cm}}
        \toprule
        \textbf{NACE 64} & \textbf{Name} & $x_{k,0}$ &  $c_{k,0}$ & $f_{k,0}$ & $l_{k,0}$ \\
        \midrule
        NACE 64 & Name & x & c & f & l \\ 
        A01 & Agriculture & 93789 & 23201 & 5752 & 9952 \\ 
        A02 & Forestry and logging & 115741 & 1690 & 11997 & 9236 \\ 
        A03 & Fishing and aquaculture & 37022 & 2589 & 30740 & 514 \\ 
        B05-09 & Mining and quarrying & 164185 & 256 & 35120 & 8477 \\ 
        C10-12 & Manufacture of food, beverages and tobacco products & 286916 & 126407 & 56312 & 25220 \\ 
        C13-15 & Manufacture of textiles, wearing apparel and leather & 91384 & 36675 & 32296 & 2637 \\ 
        C16 & Manufacture of wood and of products of wood and cork, except furniture & 114694 & 2411 & 37076 & 13630 \\ 
        C17 & Manufacture of paper and paper products & 150460 & 2082 & 107977 & 13763 \\ 
        C18 & Printing and reproduction of recorded media & 16821 & 0 & -2071 & 4535 \\ 
        C19 & Manufacture of coke and refined petroleum products & 175897 & 12314 & 84436 & 2251 \\ 
        C20 & Manufacture of chemicals and chemical products & 185922 & 7133 & 80279 & 10609 \\ 
        C21 & Manufacture of basic pharmaceutical products and pharmaceutical preparations & 121729 & 8726 & 96267 & 6528 \\ 
        C22 & Manufacture of rubber and plastic products & 94354 & 5056 & 35885 & 9059 \\ 
        C23 & Manufacture of other non-metallic mineral products & 62834 & 1997 & 9939 & 7336 \\ 
        C24 & Manufacture of basic metals & 214042 & 20 & 95464 & 14083 \\ 
        C25 & Manufacture of fabricated metal products, except machinery and equipment & 194231 & 2673 & 78091 & 31945 \\ 
        C26 & Manufacture of computer, electronic and optical products & 207467 & 20466 & 149688 & 10989 \\ 
        C27 & Manufacture of electrical equipment & 140538 & 7828 & 88233 & 10758 \\ 
        C28 & Manufacture of machinery and equipment & 343186 & 2577 & 281823 & 35492 \\ 
        C29 & Manufacture of motor vehicles, trailers and semi-trailers & 474562 & 36728 & 318121 & 32213 \\ 
        C30 & Manufacture of other transport equipment & 63570 & 4043 & 41280 & 6876 \\ 
        C31-32 & Manufacture of furniture and other manufacturing & 103774 & 19118 & 60869 & 10784 \\ 
        C33 & Repair and installation of machinery and equipment & 78811 & 542 & 10963 & 19008 \\ 
        D35 & Electricity, gas, steam and air conditioning supply & 149229 & 66988 & 13336 & 14639 \\ 
        E36 & Water collection, treatment and supply & 9669 & 0 & 191 & 2111 \\ 
        E37-39 & Sewerage; Waste collection, treatment and disposal activities; materials recovery; remediation activities & 69088 & 0 & 11129 & 11379 \\ 
        F41-43 & Construction of buildings; Civil engineering; Specialised construction activities & 678001 & 195 & 462051 & 184665 \\ 
        G45 & Wholesale and retail trade and repair of motor vehicles and motorcycles & 133425 & 64014 & 20151 & 37932 \\ 
        G46 & Wholesale trade, except of motor vehicles and motorcycles & 613250 & 145215 & 281116 & 152110 \\ 
        G47 & Retail trade, except of motor vehicles and motorcycles & 268850 & 213983 & 46561 & 94785 \\ 
        H49 & Land transport and transport via pipelines & 281691 & 67557 & 11553 & 60741 \\ 
        H50 & Water transport & 50174 & 2385 & 17727 & 5151 \\ 
        H51 & Air transport & 53841 & 15173 & 9508 & 3761 \\ 
        H52 & Warehousing and support activities & 278844 & 8003 & 113396 & 37549 \\ 
        H53 & Postal and courier activities & 31886 & 2258 & 2809 & 10294 \\ 
        I55-56 & Accommodation and food services & 187048 & 129171 & 62 & 62967 \\ 
        J58 & Publishing activities & 125313 & 14129 & 54828 & 20810 \\ 
        J59-60 & Motion picture, video and television program production, sound recording and music publishing; Programming and broadcasting activities & 93834 & 2107 & 16328 & 10553 \\ 
        J61 & Telecommunications & 142032 & 40263 & 15072 & 12920 \\ 
        J62-63 & Computer programming, consultancy, information services & 520390 & 14874 & 299828 & 112901 \\ 
        K64 & Financial services, except insurances and pension funding & 156352 & 47403 & 30850 & 44746 \\ 
        K65 & Insurance, reinsurance, and pension funding, except compulsory social security & 58844 & 40376 & 3524 & 15496 \\ 
        K66 & Activities auxiliary to financial services and insurance activities & 25161 & 8302 & 3483 & 11798 \\ 
        L68 & Real estate & 530809 & 209831 & 11551 & 63479 \\ 
        M69-70 & Legal and accounting & 364472 & 469 & 72630 & 91492 \\ 
        M71 & Activities of head offices; management consultancy & 215914 & 1141 & 25900 & 63969 \\ 
        M72 & Scientific research and development & 228033 & 0 & 219577 & 51733 \\ 
        M73 & Advertising and market research & 95924 & 272 & 17194 & 14506 \\ 
        M74-75 & Other professional, scientific and technical activities; veterinary activities & 78847 & 6899 & 4947 & 15492 \\ 
        N77 & Rental and leasing activities & 202998 & 26593 & 80944 & 31203 \\ 
        N78 & Employment activities & 89369 & 0 & 762 & 42331 \\ 
        N79 & Travel agencies, tour operators and other reservation services & 52526 & 21045 & 14125 & 4947 \\ 
        N80-82 & Security and investigation activities; Services to buildings and landscape activities; Office administrative, office support and other business support activities & 145754 & 5333 & 5149 & 57413 \\ 
        O84 & Public administration and defense; compulsory social security & 315282 & 3183 & 268193 & 120317 \\ 
        P85 & Education & 347151 & 18113 & 313563 & 189417 \\ 
        Q86 & Human health activities & 356865 & 34696 & 298995 & 177354 \\ 
        Q87-88 & Residential care activities; Social work activities without accommodation & 291510 & 47198 & 244088 & 191628 \\ 
        R90-92 & Creative, arts and entertainment; Libraries, archives, museums and other cultural activities; Gambling and betting & 64489 & 29055 & 22895 & 15502 \\ 
        R93 & Sports activities and amusement and recreation activities & 53074 & 27680 & 19604 & 17145 \\ 
        S94 & Activities of membership organizations & 51368 & 203 & 38986 & 25662 \\ 
        S95 & Repair of computers and personal and household goods & 9005 & 330 & 0 & 2705 \\ 
        S96 & Other personal service activities & 40956 & 34504 & 20 & 12956 \\ 
        T97-98 & Activities of households as employers of domestic personnel; Undifferentiated goods- and services-producing activities of private households for own use & 2517 & 2517 & 0 & 2217 \\ 
        \bottomrule
    \end{tabular}
\label{tab:EPNM_initial_states_SWE}    
\end{table}

%%%%%%%%%%%%%%%%%%%%%%%
%% initial states BE %%
%%%%%%%%%%%%%%%%%%%%%%%

\begin{table}[!ht]
    \tiny
    \centering
    \caption{Overview of initial states for Belgium (in $10^6~\text{EUR}/\text{y}$). At macro-economic equilibrium, gross output and total demand are equal, $d_k(0) = x_k(0)$. The B2B demand by sector $k$ of good $l$ is equal to the intermediate consumption listed in the input-output matrix and hence $O_{kl}(0) = Z_{kl}$. The initial stock of material $k$ held in the inventory of sector $l$, $S_{kl}(0)$ is computed as $S_{kl}(0) = n_j Z_{kl}(0)$, where $n_l$ is the targeted number of days inventory of material $k$ by sector $l$ (Table \ref{tab:EPNM_stock}). Data were retrieved from the Federal Planning Bureau \citep{fpb2018}.}
    \begin{tabular}{>{\raggedright\arraybackslash}p{1.1cm}>{\raggedright\arraybackslash}p{7.5cm}p{0.6cm}p{0.5cm}p{0.5cm}p{0.5cm}}
        \toprule
        \textbf{NACE 64} & \textbf{Name} & $x_{k,0}$ &  $c_{k,0}$ & $f_{k,0}$ & $l_{k,0}$ \\
        \midrule
        A01 & Agriculture & 16782 & 2489 & 3363 & 491 \\ 
        A02 & Forestry and logging & 648 & 93 & 163 & 23 \\ 
        A03 & Fishing and aquaculture & 429 & 206 & 77 & 28 \\ 
        B05-09 & Mining and quarrying & 24251 & 25 & 9447 & 266 \\ 
        C10-12 & Manufacture of food, beverages and tobacco products & 56386 & 14792 & 23096 & 4324 \\ 
        C13-15 & Manufacture of textiles, wearing apparel and leather & 12802 & 3979 & 6544 & 880 \\ 
        C16 & Manufacture of wood and of products of wood and cork, except furniture & 4890 & 228 & 1672 & 487 \\ 
        C17 & Manufacture of paper and paper products & 7857 & 377 & 3138 & 642 \\ 
        C18 & Printing and reproduction of recorded media & 3193 & 63 & 640 & 674 \\ 
        C19 & Manufacture of coke and refined petroleum products & 32573 & 3435 & 15024 & 233 \\ 
        C20 & Manufacture of chemicals and chemical products & 62834 & 1310 & 39364 & 3669 \\ 
        C21 & Manufacture of basic pharmaceutical products and pharmaceutical preparations & 21378 & 1304 & 14566 & 1548 \\ 
        C22 & Manufacture of rubber and plastic products & 14087 & 559 & 7425 & 1410 \\ 
        C23 & Manufacture of other non-metallic mineral products & 8864 & 351 & 3015 & 1491 \\ 
        C24 & Manufacture of basic metals & 29681 & 49 & 17145 & 1975 \\ 
        C25 & Manufacture of fabricated metal products, except machinery and equipment & 14444 & 246 & 7317 & 2274 \\ 
        C26 & Manufacture of computer, electronic and optical products & 15089 & 803 & 11006 & 672 \\ 
        C27 & Manufacture of electrical equipment & 10145 & 1134 & 6066 & 1007 \\ 
        C28 & Manufacture of machinery and equipment & 23306 & 223 & 18380 & 1812 \\ 
        C29 & Manufacture of motor vehicles, trailers and semi-trailers & 42488 & 3705 & 31168 & 1602 \\ 
        C30 & Manufacture of other transport equipment & 4474 & 474 & 3014 & 425 \\ 
        C31-32 & Manufacture of furniture and other manufacturing & 17193 & 2909 & 12780 & 721 \\ 
        C33 & Repair and installation of machinery and equipment & 8468 & 214 & 1920 & 2325 \\ 
        D35 & Electricity, gas, steam and air conditioning supply & 19084 & 5719 & 4627 & 2008 \\ 
        E36 & Water collection, treatment and supply & 1233 & 762 & 0 & 433 \\ 
        E37-39 & Sewerage; Waste collection, treatment and disposal activities; materials recovery; remediation activities & 14748 & 1255 & 3681 & 1581 \\ 
        F41-43 & Construction of buildings; Civil engineering; Specialised construction activities & 68328 & 609 & 37917 & 9383 \\ 
        G45 & Wholesale and retail trade and repair of motor vehicles and motorcycles & 11646 & 4234 & 4285 & 3127 \\ 
        G46 & Wholesale trade, except of motor vehicles and motorcycles & 56373 & 6329 & 25408 & 14907 \\ 
        G47 & Retail trade, except of motor vehicles and motorcycles & 23611 & 22494 & 1117 & 8209 \\ 
        H49 & Land transport and transport via pipelines & 27054 & 2217 & 8686 & 5460 \\ 
        H50 & Water transport & 5171 & 10 & 3027 & 208 \\ 
        H51 & Air transport & 7891 & 572 & 2638 & 477 \\ 
        H52 & Warehousing and support activities & 31465 & 263 & 13833 & 5370 \\ 
        H53 & Postal and courier activities & 4405 & 191 & 715 & 1487 \\ 
        I55-56 & Accommodation and food services & 19527 & 11300 & 1693 & 4036 \\ 
        J58 & Publishing activities & 6022 & 1151 & 1714 & 802 \\ 
        J59-60 & Motion picture, video, and television program production, sound recording and music publishing; Programming and broadcasting activities & 5167 & 868 & 1500 & 755 \\ 
        J61 & Telecommunications & 14003 & 4335 & 3237 & 1797 \\ 
        J62-63 & Computer programming, consultancy, information services & 21334 & 0 & 9662 & 5509 \\ 
        K64 & Financial services, except insurances and pension funding & 20798 & 3302 & 2537 & 3898 \\ 
        K65 & Insurance, reinsurance, and pension funding, except compulsory social security & 9448 & 4150 & 944 & 2021 \\ 
        K66 & Activities auxiliary to financial services and insurance activities & 20464 & 2691 & 6250 & 3569 \\ 
        L68 & Real estate & 46378 & 33438 & 214 & 1166 \\ 
        M69-70 & Legal and accounting & 20233 & 546 & 19687 & 6691 \\ 
        M71 & Activities of head offices; management consultancy & 13253 & 94 & 5477 & 2433 \\ 
        M72 & Scientific research and development & 20054 & 0 & 18169 & 4925 \\ 
        M73 & Advertising and market research & 9887 & 3 & 3821 & 798 \\ 
        M74-75 & Other professional, scientific and technical activities; veterinary activities & 2779 & 397 & 322 & 241 \\ 
        N77 & Rental and leasing activities & 17691 & 2292 & 4093 & 1214 \\ 
        N78 & Employment activities & 7661 & 0 & 50 & 6943 \\ 
        N79 & Travel agencies, tour operators and other reservation services & 3225 & 2770 & 17 & 365 \\ 
        N80-82 & Security and investigation activities; Services to buildings and landscape activities; Office administrative, office support and other business support activities & 13999 & 1714 & 2375 & 5543 \\ 
        O84 & Public administration and defense; compulsory social security & 33807 & 2729 & 30292 & 23682 \\ 
        P85 & Education & 27168 & 1212 & 24225 & 21167 \\ 
        Q86 & Human health activities & 32665 & 6366 & 22548 & 10263 \\ 
        Q87-88 & Residential care activities; Social work activities without accommodation & 15209 & 6653 & 8557 & 11628 \\ 
        R90-92 & Creative, arts and entertainment; Libraries, archives, museums and other cultural activities; Gambling and betting & 4914 & 1983 & 1886 & 1279 \\ 
        R93 & Sports activities and amusement and recreation activities & 2869 & 961 & 786 & 622 \\ 
        S94 & Activities of membership organizations & 6231 & 115 & 3090 & 2437 \\ 
        S95 & Repair of computers and personal and household goods & 1057 & 582 & 28 & 106 \\ 
        S96 & Other personal service activities & 3640 & 3241 & 7 & 620 \\ 
        T97-98 & Activities of households as employers of domestic personnel; Undifferentiated goods- and services-producing activities of private households for own use & 425 & 425 & 0 & 425 \\ 
        \bottomrule
    \end{tabular}
\label{tab:EPNM_initial_states_BE}    
\end{table}

\clearpage
\pagebreak

\subsubsection{Conversion of inputs to economic shocks}\label{section:shocks}

The economic production network has four inputs (Fig. \ref{fig:EPNM_flowchart}): 1) $\kappa_k^F(t)$, the demand shock to investments, exports, government and non-profit spending of economic activity $k$ at time $t$, which evolves exogenously in accordance with data on investments and trade data (Section \ref{section:demand}), 2) $A^g_k(t)$ represents the degree to which the economic activity $k$ (Table \ref{tab:NACE64}) is allowed in spatial patch $g$ at time $t$, 3) $M^g_{\text{work}, k}(t)$ is the ``voluntary" reduction of work contacts in economic activity $k$ in response to the EMA hospital load in spatial patch $g$ at time $t$, 4) $M^g_{\text{leisure}}(t)$ is the ``voluntary" reduction of leisure contacts in response to the EMA hospital load in spatial patch $g$ at time $t$. $\kappa_k^F(t)$ evolves exogenously, while the three remaining inputs, $A^g_k(t)$, $M^g_{\text{work}, k}(t)$, evolve endogenously. These inputs must be converted to a shock to household demand, $\kappa_k^D(t)$ and a shock to labor supply, $\kappa_k^S(t)$.\\

\textbf{Household demand shock} The reduction in household demand is computed as follows,
\begin{equation}
\kappa_k^D(t) = \bar{I}_{\text{mild}}(t) (1 - \text{LAV}^D_k) + (1-\bar{I}_{\text{mild}}(t)) \bar{M}_{\text{leisure}}(t) (1 - \text{LAV}^D_k),
\end{equation}
where $\textstyle \bar{I}_{\text{mild}}(t)$ is the fraction of the total population experiencing mild symptoms,
\begin{equation}
\bar{I}_{\text{mild}}(t) = \sum_i \sum_g \Bigg( \frac{I_{\text{mild}, i}^g(t)}{T^g_i(t)} \Bigg),
\end{equation}
and $\bar{M}_{\text{leisure}}(t)$ is computed as the demographic mean of $M^g_{\text{leisure}}(t)$, using the number of inhabitants per spatial patch found in Tables \ref{tab:demography_SWE} and \ref{tab:demography_BE}. $\text{LAV}^D_k$ is the ``Leisure Association Vector - Household demand", which denotes the maximum household demand shock to an economic activity $k$ out of fear of infection (Table \ref{tab:ERMG_lav}). The reduction of household demand in response to \covid{} can thus be summarized as follows. Those who develop symptomatic \covid{} stop going to activities associated with leisure altogether, and a fraction of those unaffected by \covid{} stop going to activities associated with leisure voluntarily, resulting in a reduction of demand.\\

\textbf{Labor supply shock} Employed individuals experiencing symptomatic \covid{} stay home from work resulting in a labor supply shock. Those unaffected by \covid{} infection and not able to work remotely may stay home from work out of fear of infection (absenteeism), or, the government may place restrictions on economic activity. Mathematically, the reduction in available labor is given by,
\begin{equation}
\kappa^{S,g}_k(t) = \widetilde{I}^g_{\text{mild, active}}(t) + \big( 1-\widetilde{I}^g_{\text{mild, active}}(t) \big) \max \big( \bar{A}^g_k(t), \bar{M}^g_{\text{work},k} (t)\big),
\end{equation}
where $\widetilde{I}^g_{\text{mild, active}}(t)$ is the fraction of the active and employed population experiencing symptomatic \covid{}, i.e.,
\begin{equation}
\widetilde{I}^g_{\text{mild, active}}(t) = \bar{I}^g_{\text{mild, active}} \sum_h \bar{P}^{gh},
\end{equation}
where $\bar{I}_{\text{mild, active}}(t)$ is the fraction of the active population experiencing mild symptoms (Eq. \eqref{eq:mildactive_I}) and $\bm{\bar{P}}$ is the normalized mobility matrix (Section \ref{app:dynamic_transmission_model}), whose rows sum to the employed fraction of the active population. Further,
\begin{equation}
\bar{A}^g_k(t) = A^g_k(t) \big(1 - f_{\text{workplace}, k} - f_{\text{telework}, k} \big),
\end{equation}
implying that, similar to the social contacts in the workplace (Eq. \eqref{eq:contact_work}), mandated economic closure results in a labor supply shock equal to the fraction of employees on temporary employment, $\textstyle (1 - f_{\text{workplace}, k} - f_{\text{telework}, k})$, as observed during the 2020 \covid{} pandemic in Belgium. Finally,
\begin{equation}
\bar{M}^g_{\text{work}, k} (t) =
    \begin{cases}
        0, & \text{if } M^g_{\text{work}, k} < f_{\text{telework}, k},\\
        M^g_{\text{work}, k} - f_{\text{telework}, k}, & \text{else}. \\
    \end{cases}
\end{equation}
We use the labor market composition in every spatial patch $g$ to convert the labor supply shock to the national level.

\subsubsection{Demand}\label{section:demand}

\textbf{Total demand} The total demand of industry $k$ at time $t$, denoted $d_k(t)$, is the sum of the demand from all its customers,
\begin{equation}
    d_k(t) = \sum_{j=1}^{N} O_{kl}^d(t) + c_k^d(t) + f_k^d(t),
\end{equation}
where $O_{kl}^d(t)$ is the intermediate demand from industry $k$ to industry $l$, $c_k^d(t)$ is the total demand from households and $f_k^d(t)$ denotes exogenous demand. The superscript $d$ refers to \textit{desired}, as each customer's demand may or may not be met under the imposed shocks.\\

\textbf{Household demand} The household demand for good $k$ is,
\begin{equation}
    c_k^d(t) = \big(1 - \tilde{\kappa}^D(t)\big) \theta_k(t) \tilde{c}^d(0),
\end{equation}
where $\theta_k(t)$ is the household preference coefficient, denoting the share of good $k$ in the aggregate household demand $\textstyle \tilde{c}^d(0) = \sum_j c_j^d(0)$. Before the pandemic, the share of good $k$ in total household consumption can be computed using the available data $\theta_k(0) = \textstyle c_k(0) / \sum_{j} c_{j}(0)$ (Tables \ref{tab:EPNM_initial_states_BE}, \ref{tab:EPNM_initial_states_SWE}). As household demand for good $k$ changes under the demand shocks induced by the pandemic $\kappa_k^D(t)$, the consumption preference evolves dynamically according to,
\begin{equation}
    \theta_k(t) = \frac{\big(1-\kappa_k^D(t)\big)\theta_k(0)}{\textstyle \sum_l \big(1-\kappa_l^D(t)\big)\theta_l(0)}.
\end{equation}
$\tilde{\kappa}^D(t)$ represents the aggregate reduction in household demand caused by the pandemic shock and is equal to $\textstyle 1 - \sum_i \theta_k(0) (1-\kappa_k^D(t))$. However, households have the choice to save all the money they are not spending ($\Delta s = 1$), or to spend all their money on goods of other industries ($\Delta s = 0$). We can thus redefine the aggregate reduction in household demand shock as,
\begin{equation}
    \tilde{\kappa}^D(t) = \Delta s \bigg(1 - \sum_k \theta_k(0) \big(1-\kappa_k^D(t)\big)\bigg),
\end{equation}
where $\Delta s$ is the household savings rate, assumed $\Delta s = 0.75$ given evidence of forced saving during the \covid{} pandemic in Belgium \citep{basselier2021}.\\

\textbf{Intermediate demand} For industry $l$ to produce one unit of output, inputs from industry $k$ are needed. The production recipe is encoded in the matrix of technical coefficients $\mathcal{A}$, where an element $\mathcal{A}_{kl} = Z_{kl}/x_l(0)$ represents the expense in inputs $k$ to produce one unit of output $l$. In the model, production and demand are not immediate, rather, each industry $l$ aims to keep a target inventory of inputs $k$, $n_l Z_{kl}(0)$, so that production can go on for $n_l$ more days (Table \ref{tab:EPNM_stock}). The stock of inputs $k$ kept by industry $l$ is denoted as $S_{kl}(t)$. The intermediate demand faced by industry $l$ from industry $k$ at time $t$ is modeled as the sum of two components,
\begin{equation}
O_{kl}^d(t) = \mathcal{A}_{kl}d_l(t-1) + \frac{1}{\tau} \big( n_l Z_{kl}(0) - S_{kl}(t-1) \big).
\end{equation}
The first term represents the attempts of industry $l$ to satisfy incoming demand under the naive assumption that demand on day $t$ will be the same as on day $t-1$. The second term represents the attempts by industry $l$ to close inventory gaps. The parameter $\tau$ governs how quickly an industry aims to close inventory gaps.\\

\textbf{Other demand (exogenous)} Industry $k$ faces demand $f_k^d(t)$ from exogenous sources not explicitly included in the model. The exogenous sources are: 1) Government and non-profit organizations, 2) gross fixed capital formation (investments), 3) exports of goods (economic activities A-E, Table \ref{tab:NACE21}), and, 4) exports of services (economic activities F-T, Table \ref{tab:NACE21}). Changes in stocks are included explicitly to model supply chains. Mathematically,
\begin{equation}
f_k^d(t) = (1-\kappa_k^F(t)) \big(f_{\text{gov/npo},k}^d(t) +  f_{\text{inv},k}^d(t) + f_{\text{exp, goods},k}^d(t) + f_{\text{exp, services},k}^d(t) \big).
\end{equation}
We omit a mathematical description of $\kappa_k^F(t)$ for the sake of brevity. As their demand is small compared to investments and exports, governments and non-profits will not affect the model's dynamics by much, and hence, we assume their demand remains fixed throughout the pandemic. During the second quarter of 2020, Belgium faced an investment shock of 16.2~\%  while Sweden faced an investment shock of 6.9~\%. By the third quarter of 2020, investments in both countries had fully recovered \citep{oecd2023}. Hence, we model the time course of investment shocks as follows: 1) In March 2020, we gradually apply the shocks, 2) starting May 1st, 2020, we gradually ease out the shocks by September 1st, 2020. Because shocks to the exports of goods also recovered by September 2020, we model shocks to exports of goods identically. In April 2020, the observed shock to exports of goods was equal to 25.0~\% in Belgium \citep{nbb2023b} and 14.0\% in Sweden \citep{scb2023g}. Shocks to the exports of services recovered much slower, we assume the shocks to exports of services are equal to 21.0\%, and demand recovers only by September 2021 \citep{oecd2022}. \\

\subsubsection{Supply}\label{section:supply} 

Every industry aims to satisfy the incoming demand by producing the required output. However, production under the imposed pandemic shocks is subject to two constraints.\\

\textbf{Labor supply constraints} Productive capacity is assumed to linearly depend on the available amount of labor and hence,
\begin{equation}\label{eq:labor_supply_constraints_1}
    x_k^{\text{cap}}(t) = \frac{l_k(t)}{l_k(0)} x_k(0).
\end{equation}
Recall that under economic restrictions, labor supply is shocked and the maximum amount of available labor is reduced to,
\begin{equation}\label{eq:labor_supply_constraints_2}
    l_k^{\text{max}}(t) = \big(1 - \kappa_k^S(t)\big) l_k(0).
\end{equation}
However, as explained in Section \ref{section:hiring_firing}, industries are allowed to fire workers if productive capacity is greater than demand, and thus, the output can be constrained further by a shortage of labor. By combining Eq. \eqref{eq:labor_supply_constraints_1} and Eq. \eqref{eq:labor_supply_constraints_2}, we obtain,
\begin{equation}
    x_k^{\text{cap}}(t) \leq \big(1 - \kappa_k^S(t)\big) x_k(0).
\end{equation}

\textbf{Input bottlenecks} The productive capacity of an industry can be constrained if an insufficient supply of inputs is in stock. The productive capacity of an industry can be constrained in several ways, referred to as a \textit{production function} ($\mathcal{F}$). In a classical Leontief production function, every input encoded in the recipe matrix $A_{kl}$ is considered critical to production. However, this assumption is unrealistic, Pichler et al. \citep{pichler2022} have clearly demonstrated the added value of relaxing the classical Leontief approach to what is referred to as a \textit{Partially Binding Leontief} production function (PBL). The approach is based on a survey assessing the criticality of inputs to different industries (conducted by IHS Markit Analysts, consult \citep{pichler2022}). Industries were asked to label each input as \textit{critical}, \textit{important}, or \textit{non-critical}, corresponding to a numerical value of 1, 0.5, and 0 respectively. In this work, we use a \textit{half-critical} PBL production function, in which the depletion of \textit{critical} inputs halts production completely while the depleting \textit{important} inputs reduce production with 50~\%, consistent with the label 0.5 in the survey. Mathematically, we may write,
\begin{equation}
    x_j^{\text{inp}}(t) = \min_{ \{l \in \mathcal{C}_k, m \in \mathcal{I}_k\}} \bigg\{ \frac{S_{lk}(t)}{\mathcal{A}_{lk}}, \frac{1}{2} \bigg( \frac{S_{mk}(t)}{\mathcal{A}_{mk}} + x_k^{\text{cap}}(0) \bigg)  \bigg\}.
\end{equation}

\subsubsection{Realised output and rationing}\label{section:rationing} As each industry aims to maximally satisfy incoming demand under its production constraints, the realized output of sector $k$ at time $t$ is,
\begin{equation}\label{eq:realised_output}
    x_k(t) = \min \{ x_k^{\text{cap}}(t),\ x_k^{\text{inp}}(t),\ d_k(t)\},
\end{equation}
so the output is constrained by the smallest of three values: the labor-constrained productive capacity $x_k^{\text{cap}}(t)$, the input-constrained productive capacity $x_k^{\text{inp}}(t)$ and total demand $d_k(t)$. If productive capacity was lower than total demand, industries ration their output equally across customers (\textit{strict proportional rationing}), as follows,
\begin{equation}
    c_k(t) =  c_k^d(t) \bigg( \frac{x_k(t)}{d_k(t)} \bigg),
\end{equation}
\begin{equation}
    f_k(t) = f_k^d(t) \bigg( \frac{x_k(t)}{d_k(t)} \bigg),
\end{equation}
\begin{equation}
    O_{kl}(t) = O_{kl}^d(t) \bigg( \frac{x_k(t)}{d_k(t)} \bigg).
\end{equation}\\
Alternative rationing schemes were deemed unsatisfactory and are not discussed here \citep{pichler2021,alleman2023c}.

\subsubsection{Inventory adjustment}\label{section:inventory_updating}
After the realized output has been rationed among the customers, inventories can be updated,
\begin{equation}
    S_{kl}(t) = \max_{k,l} \{ S_{kl}(t-1) + O_{kl}(t) - \mathcal{A}_{kl} x_l(t),\ 0\}.
\end{equation}
The new stocks of input $k$ in industry $l$ are thus equal to the intermediate inputs $k$ received minus the inputs $k$ consumed in the production of $l$ outputs. The maximum operator prevents stocks from assuming negative values.

\subsubsection{Hiring and firing}\label{section:hiring_firing}
Firms will adjust their labor force depending on what constraint was limiting in Eq. \eqref{eq:realised_output}. If the supply of labor, $x_k^{\text{cap}}(t)$, was limiting then industry $k$ will attempt to hire as many workers as needed to make the supply of labor not limiting. Opposed, if either input constraints $x_k^{\text{inp}}(t)$ or total demand $d_k(t)$ were limiting, industry $k$ will attempt to lay off workers until labor supply constraints become limiting,
\begin{equation}
    \Delta l_k(t) = \frac{l_k(0)}{x_k(0)} \big[ \min \{ x_k^{\text{inp}}(t), d_k(t) \} - x_k^{\text{cap}(t)} \big].
\end{equation}
However, the process of adjusting the labor force is not an instantaneous one and takes time, which is accounted for as follows,
\begin{equation}\label{eq:hiring_firing}
    l_k(t) =
    \begin{cases}    
        l_k(t-1) + 1/\iota_H \Delta l_k(t), & \text{if } \Delta l_k(t) \geq 0, \\
        l_k(t-1) + 1/\iota_F \Delta l_k(t), & \text{if } \Delta l_k(t) < 0, \\
    \end{cases}
\end{equation}
where $\iota_H$ is the average time needed to hire a new employee and $\iota_F$ is the average time needed to lay off an employee. Both parameters were included in the model calibration (Appendix \ref{app:calibration}).

\section{Model limitations and assumptions}\label{app:assumptions}

Here, we list the most important assumptions and simplifications underpinning our model. While we believe these are unlikely to alter our work's conclusions, we choose to explicitly mention them below as good scientific practice.

\begin{enumerate}
    \item In this work, ``voluntary" behavioral changes are a black box encompassing all behavioral changes that result from awareness to sars-cov-2, including sanitary recommendations, fear, prosocial behavior, and social pressure \citep{bavel2020}. Awareness can be spread by individuals, scientific institutes, and governments. Opposed, ``forced" behavioral changes are the result of enforcement by law and are thus always induced by governments. Behavioral changes in Sweden were predominantly ``voluntary" according to our definition, although this does not imply the Swedish population's behavior should be regarded as a ``business-as-usual" mentality. Pressure to alter behavior increases as \sars{} incidence rises, blurring the line as ``voluntary" behavioral changes may eventually be perceived as ``forced".\\
    
    \item Important similarities between Belgium and Sweden are the presence of a functioning government, public health agencies, and independent news media that work ``together" (even though it often does not seem so), to raise awareness when \sars{} incidence surges. The results of our work are only transferable to countries that satisfy these criteria. We do not expect an uninformed or misinformed population to alter its behavior in the same way.\\

    \item Opposed to the collective feedback model resulting in voluntary behavioral changes, there is likely an inversely working behavioral cost of too frequently changing forceful interventions, resulting in declining compliance over time. The human brain, constrained by its fixed attention budget, faces significant challenges in adhering to complex rules, particularly when these rules undergo frequent changes. Similar to Ash et al. \cite{ash2022}, we plan to use a utility-based approach to account for a learning cost of new rules, where the marginal costs of additional rules increases, as well as a cost of forgetting old rules when these are substituted by new rules. Our guess is this channel, which is rooted in principles of behavioral economics and cognitive psychology, may be at least as important as the more benevolent collective memory channel.\\
    
    \item We use social contact data obtained from Beraud et al. \citep{beraud2015} for France to model social contacts in Belgium and Sweden. Gauging if these social contact data are representative for Belgium and Sweden is difficult. There seems to be substantial variation in the number of contacts per country (f.i. BE 11.6/FI 10.7/DE 7.6/IT 18.0) \citep{mossong2008}. However, the absolute magnitude of the contacts is less relevant than the mixing patterns, as we assume for both Sweden and Belgium that $R_0=3$. Because the mixing patterns found by Mossong et al. \citep{mossong2008} seem consistent across countries, and consistent with the French study by Beraud et al \citep{beraud2015}, we expect our use of French contact data, which was motivated by the absence of information on economic occupation of correspondents in other contact studies, is not likely to have influenced the results of this work.\\
    
    \item Cross-border mobility, which may be relevant to transmission in Skåne (Sweden) and Belgium in general, is not included in the model. The influence of importing \sars{} cases is highest when \sars{} incidence is low during the summer of 2020. We expect the lack of cross-border mobility may partially overlap with the estimated parameters of the seasonal forcing on the transmission rate.\\
    
    \item Our model only makes use of recurrent commuter mobility, as such, mobility for leisure and other purposes is neglected. Commuting mobility represents the lion’s share of overall mobility, and mobility in Europe is very local, even though there is a small share of citizens traveling frequently over longer distances \citep{fiorello2016}. Further, both Belgium and Sweden imposed travel restrictions which may have lowered the influence of non-commuter travel further.\\
    
    \item Implementing seasonality using a cosine function is a high-level mathematical abstraction of several factors such as, but not limited to, the effects of humidity and temperature on viral survival in the environment, and behavioral changes such as spending more time indoors.
\end{enumerate}

\section{Model calibration}\label{app:calibration}

\subsection{Methods}

\textbf{Parameters} Our aim is to infer the distribution of 12 model parameters using the affine-invariant ensemble sampler for Markov Chain Monte Carlo (MCMC) \citep{goodman2010,emcee2013} (Table \ref{tab:overview_L2_priors}).\\

\begin{table}[b!]
    \centering
    \caption{Initial estimates of the model parameters ($\theta_k$). Parameters used to inform the L2 regularised normal prior probabilities (Eq. \eqref{eq:P2prior}), where $\lambda_k$ are the L2 regularisation weights, $\mu_k$ are the desired parameter values, and $\sigma_k$ are the assumed standard deviations on the desired parameter values.}
    {\renewcommand{\arraystretch}{1.10}
    \begin{tabular}{p{1.3cm}p{0.75cm}p{4.7cm}p{0.7cm}p{0.5cm}p{0.5cm}p{0.5cm}}
        \toprule
        \textbf{Par.} & \textbf{Eq.} & \textbf{Meaning} & $\theta_k$ &  $\lambda_k$ & $\mu_k$ & $\sigma_k$ \\ \midrule
        $\nu~(\text{d}.)$ & \eqref{eq:EMA_hospital_load} & Mean lifetime of the exponential moving average hospital load & 18.0 & 10 & 22.0 & 1.0 \\
        $\xi_\text{{eff}}~(-)$ & \eqref{eq:gompertz} & Contact effectivity in the absence of hospitalizations & 0.40 & 10 & 0.45 & 0.01 \\
        $\pi_{\text{eff}}~(-)$ & \eqref{eq:gompertz} & Steepness of the contact effectivity & 0.060 & 25 & 0.0 & 0.015 \\
        $\pi_\text{{work}}~(-)$ & \eqref{eq:gompertz} & Steepness of the work contact reduction & 0.035 & 25 & 0.035 & 0.004 \\
        $\pi_\text{{leisure}}~(-)$ & \eqref{eq:gompertz} & Steepness of leisure contact reduction & 0.060 & 15 & 0.060 & 0.006 \\
        $\mu~(-)$ & \eqref{eq:EMA_spatial_average} & Spatial connectivity of the hospital load awareness network & 0.72 & 10 & 1.0 & 0.1 \\
        $A_\text{{BE}}~(-)$ & \eqref{eq:seasonality} & Amplitude of cosine wave governing seasonal variations on the transmission rate & 0.16 & 20 & 0.18 & 0.03 \\
        $\Delta t_\text{{BE}}~(\text{d}.)$ & \eqref{eq:seasonality} & Temporal shift of cosine wave governing seasonal variations on the transmission rate (compared to July 14th) & -14.0 & 15 & 0.0 & 3.5 \\
        $A_\text{{SWE}}~(-)$ & \eqref{eq:seasonality} & Amplitude of cosine wave governing seasonal variations on the transmission rate & 0.23 & 20 & 0.22 & 0.03 \\
        $\Delta t_\text{{SWE}}~(\text{d}.)$ & \eqref{eq:seasonality} & Temporal shift of cosine wave governing seasonal variations on the transmission rate (compared to July 14th) & 14.0 & 15 & 0.0 & 3.5 \\
        $\iota_\text{H}~(\text{d}.)$ & \eqref{eq:hiring_firing} & Average time to hire labor & 7.0 & 10 & 7.0 & 2.0 \\
        $\iota_\text{F}~(\text{d}.)$ & \eqref{eq:hiring_firing} & Average time to fire labor & 7.0 & 10 & 7.0 & 2.0 \\
      \bottomrule
    \end{tabular}
    }
    \label{tab:overview_L2_priors}
\end{table}

\textbf{Available data} For both countries, time series on the evolution of the hospitalization incidence, gross domestic product, and employment must be matched with the simulations for the model states $H_\text{{in}}$ (daily hospitalization incidence), $x$ (gross productivity) and $l$ (labor compensation). For Belgium, the daily hospitalization incidence per province was obtained from Sciensano \citep{sciensano2023} while monthly data on synthetic GDP were obtained from the business surveys of the Belgian National Bank \citep{nbb2023a}. As a proxy for labor compensation, biweekly surveys performed by the Economic Risk Management Group of the Belgian National Bank \citep{ermg2021} on the fraction of furloughed workers were used. For Sweden, the weekly hospitalization incidence per county was obtained from Socialstyrelsen \citep{socialstyrelsen2023}. Monthly data on synthetic GDP were obtained from the Central Bureau of Statistics \citep{scb2023b}. As a proxy for labor compensation, the monthly number of hours worked, obtained from the Central Bureau of Statistics \citep{scb2023a}, was used. All economic data used were seasonally adjusted and smoothed by the publisher.\\

\textbf{Time course of model inputs} To match the model to the available timeseries of data, we must, as realistically as possible, implement the government-mandated policies imposed during the 2020 \covid{} pandemic in Belgium and Sweden. As schematically shown in Fig. \ref{fig:epinomic_flowchart}, mandated policies are governed by three parameters in our model: $\bm{A}(t)$, which represents the degree to which the economic activity $k$ is allowed in spatial patch $g$ at time $t$, $\bm{E}(t)$, the degree to which telework is mandated for economic activity $k$ in spatial patch $g$ at time $t$, and, $\bm{F}(t)$, the degree to which leisurely interactions in the private sphere are prohibited in spatial patch $g$ at time $t$. \\

Forced behavioral changes were much more important during the 2020 \covid{} pandemic in Belgium than in Sweden, with the Swedish government mostly relying on voluntary recommendations. The only exceptions were: 1) Mandatory distance learning for the last two years of secondary education and tertiary education (ages 17 and above). 2) A ban on large gatherings, first of more than 500 people, and later of more than 50 people. 3) A ban on national and international travel \citep{ludvigsson2020}. The policies for Sweden are implemented as follows, $A^g_{\text{P85}}(t) = 0.2$, and $A^g_{\text{R90-92}}(t) = 0.2$ for all spatial patches $g$, from March 11th 2020 onwards. National and international travel restrictions are not explicitly included in the model. No restrictions on leisurely contacts in the private sphere were imposed, and hence $\bm{F}(t) = 0$. Telework was strongly recommended by the Swedish government, but not imposed at the risk of fines as in Belgium, and hence, $\bm{E}(t) = 0$. We note that in Sweden, forced behavioral changes were induced at the beginning of 2021 in light of the Alpha variant, but, as no variants of concern and vaccinations were incorporated into the disease transmission model, comparing the model results to the actual epidemiological situation is only valid until the beginning of 2021. So, these restrictions fall outside the temporal range considered in this work. For the sake of brevity, we omit a detailed overview of \covid{} policies in Belgium here \citep{alleman2021,alleman2023a,rollier2023}. The time course for $\bm{A}(t)$ is listed in Table \ref{tab:policies_BE_A}, while the time course of $\bm{E}(t)$ and $\bm{F}(t)$ are tabulated in Table \ref{tab:policies_BE_EF}.\\

\textbf{Bayesian framework} We attempt to maximise the parameter's log posterior probability given the data, defined as \citep{hartig2011},
\begin{equation}\label{eqn:posterior}
\log p (\bm{\tilde{x}}(\bm{\theta}) \mid \bm{x}) = \log \left[ \frac{p(\bm{x} \mid \bm{\tilde{x}}(\bm{\theta})) p(\bm{\theta})}{p(\bm{x})} \right],
\end{equation}
where $p (\bm{\tilde{x}}(\bm{\theta}) \mid \bm{x})$ is the posterior probability of the unknown parameters in light of the observations, $p(\bm{x} \mid \bm{\tilde{x}}(\bm{\theta}))$ is the likelihood of the observations in light of the parameters, $ p(\bm{\theta})$ is the prior probability of the parameters, and $p(\bm{x})$ is the probability of the data, which is constant since it doesn't depend on the unknown parameters $\bm{\theta}$. In this way, Eq.~\eqref{eqn:posterior} can be simplified to,
\begin{equation}
\log p (\bm{\tilde{x}}(\bm{\theta}) \mid \bm{x}) \propto \log p(\bm{x} \mid \bm{\tilde{x}}(\bm{\theta})) + \log p(\bm{\theta}),
\end{equation}
or, upon expansion of the vectors,
\begin{equation}
\log p (\bm{\tilde{x}}(\bm{\theta}) \mid \bm{x}) \propto \sum_n \log p(x_n \mid \tilde{x}_n(\bm{\theta})) + \sum_k \log p(\theta_k),
\end{equation}
where $n$ sums over the time series used (36) and $k$ over the number of calibrated parameters (12). We used a locally optimized estimate of the parameter's values to inform an L2 regularised normal prior for every parameter. Mathematically, this prior is equal to,
\begin{equation}\label{eq:P2prior}
p(\theta_k) = \lambda_k \left[ \frac{1}{\sigma_k \sqrt{2\pi}} e^{-\dfrac{(\theta_k - \mu_k)^2}{2 \sigma_k^2}}~\right],
\end{equation}
where $\theta_k$ is an estimate of the model parameter $k$, $\lambda_k$ are the L2 regularisation weights, $\mu_k$ is the expected value of parameter $k$, $\sigma_k$ is the standard deviations on the expected value of parameter $k$ (Table~\ref{tab:overview_L2_priors}). L2 regularised regression is also referred to as ``ridge regression" \citep{hoerl1970}.\\

The likelihood function used differed between the datasets and countries. Hospital incidence data in Belgium is available every day and for every province. A quadratic relationship between the mean and variance of these data was previously established \citep{alleman2023a}, and thus the data resulted from a negative binomial observation process. Hospital incidence data in Sweden was available every week and for every county. The daily surveillance data were (most likely) generated by a similar negative binomial observation process as the Belgian data, but have been aggregated to the weekly temporal level by Socialstyrelsen \citep{socialstyrelsen2023}, significantly reducing the noise from the original observation process. For the sake of simplicity, we assume the data generation process of the aggregated weekly data follows a Poisson distribution and thus a Poisson likelihood function was used \citep{alleman2021}. For all economic data, no measure for the observation error was available, hence we assumed these data were the result of a Gaussian process with a standard deviation of 2~\%.\\

\textbf{Initial condition and calibration procedure} An initial condition that results in an adequate representation of the spatial spread of \sars{} in Belgium and Sweden is needed. To this end, the model was initialized on February 1st, 2020 with a basic reproduction number of $R_0 = 3$. The initial number of exposed individuals in every spatial patch on February 1st, 2020 was then identified. As the spread of \sars{} during February 2020 depends on the calibrated parameters, we employed an iterative calibration procedure.\\

First, a reasonable parameter estimate that captured the overall trends in the datasets was formulated. Because all parameters have a physical meaning, formulating a meaningful estimate was straightforward. Keeping the estimate for all 12 parameters fixed, we used the simplex method by Nelder and Mead \citep{nelder1965} to calibrate the initial number\footnote{The initial number of exposed individuals are distributed over the 17 age groups according to the relative fraction of the total number of social contacts.} of exposed individuals in every spatial patch between February 1st, 2020, and May 1st, 2020 (encompassing the first 2020 \covid{} wave in both countries). The values of the model's states on March 1st, 2020, were then used as the initial condition for the calibration of the 12 parameters between March 1st, 2020, and January 1st, 2021. We used the simplex method \citep{nelder1965} to locally optimize our initial estimate. This procedure of iteratively optimizing the initial number of exposed individuals in every spatial patch on a smaller subset of data and then subsequently optimizing the 12 model parameters using all data was repeated until convergence, which occurred after one iteration (initial estimates are shown in Table \ref{tab:overview_L2_priors}). For Sweden, the epidemic appears to have been seeded in the Stockholm metropolitan area and to a much lesser extent in J\"onk\"oping, which is consistent with the first confirmed case being found in J\"onk\"oping on Jan. 31st, 2020 \citep{krisinformation2020}. For Belgium, the largest number of initial infections was found in Limburg and Hainaut provinces \citep{rollier2023}. During the optimization, the 12 model parameters were calibrated to data from both countries simultaneously as fitting them separately would not yield a valid basis for comparisons.\\

The Markov chains used to sample the posterior probability were initialized by perturbating the obtained initial estimates uniformly by 5~\%. After discarding the initial burn-in, the sampler was run for 50 times the chain's integrated autocorrelation time. The resulting chains were thinned by a factor equal to half the chain with the maximum integrated autocorrelation time.

\clearpage
\begin{table}[!ht]
    \raggedright
    \caption{Extent to which telework was mandated, $\bm{E}(t)$, and the extent to which leisure contacts in the private sphere were prohibited, $\bm{F}(t)$, by the Belgian government during the 2020 \covid{} pandemic, on a scale from 0 to 1.}
    \begin{tabular}{>{\raggedright\arraybackslash}p{0.8cm}>{\centering\arraybackslash}p{0.6cm}>{\centering\arraybackslash}p{0.6cm}>{\centering\arraybackslash}p{0.6cm}>{\centering\arraybackslash}p{0.6cm}>{\centering\arraybackslash}p{0.6cm}>{\centering\arraybackslash}p{0.6cm}>{\centering\arraybackslash}p{0.6cm}>{\centering\arraybackslash}p{0.6cm}>{\centering\arraybackslash}p{0.6cm}>{\centering\arraybackslash}p{0.6cm}}
        \toprule
        & \textbf{Date} & & & & & & & & &\\
        \midrule
        \textbf{Par.} & 03-15 &  05-04 & 05-18 & 06-08 & 07-01 & 08-03 & 08-24 & 10-19 & 11-02 & 11-27 \\
        \midrule
        $\bm{E}(t)$ & 1 & 1 & 1 & 0 & 0 & 1 & 0 & 1 & 1 & 1 \\
        $\bm{F}(t)$ & 1 & 1 & 1 & 0 & 0 & $0^1$ & 0 & 1 & 1 & 1 \\
        \bottomrule
    \end{tabular}
\vspace{0.2cm}
\raggedright{\footnotesize{$^1$Except in Antwerp province, where $F(t) = 1$.}}
\label{tab:policies_BE_EF}    
\end{table}

\begin{table}[!ht]
    \footnotesize
    \centering
    \caption{Extent to which an economic activity (Table \ref{tab:NACE64}) was prohibited by the Belgian government during the 2020 \covid{} pandemic, $\bm{A}(t)$, on a scale between 0 and 1.}
    \begin{tabular}{>{\raggedright\arraybackslash}p{1.0cm}>{\centering\arraybackslash}p{0.6cm}>{\centering\arraybackslash}p{0.6cm}>{\centering\arraybackslash}p{0.6cm}>{\centering\arraybackslash}p{0.6cm}>{\centering\arraybackslash}p{0.6cm}>{\centering\arraybackslash}p{0.6cm}>{\centering\arraybackslash}p{0.6cm}>{\centering\arraybackslash}p{0.6cm}>{\centering\arraybackslash}p{0.6cm}>{\centering\arraybackslash}p{0.6cm}}
        \toprule
        & \textbf{Date} & & & & & & & & &\\
        \midrule
        \textbf{Sect.} & 03-15 &  05-04 & 05-18 & 06-08 & 07-01 & 08-03 & 08-24 & 10-19 & 11-02 & 11-27 \\
        \midrule
        A01 & 1 & 0 & 0 & 0 & 0 & 0 & 0 & 0 & 0 & 0 \\ 
        A02 & 1 & 0 & 0 & 0 & 0 & 0 & 0 & 0 & 0 & 0 \\ 
        A03 & 1 & 0 & 0 & 0 & 0 & 0 & 0 & 0 & 0 & 0 \\ 
        B05-09 & 1 & 0 & 0 & 0 & 0 & 0 & 0 & 0 & 0 & 0 \\ 
        C10-12 & 1 & 0 & 0 & 0 & 0 & 0 & 0 & 0 & 0 & 0 \\ 
        C13-15 & 1 & 0 & 0 & 0 & 0 & 0 & 0 & 0 & 0 & 0 \\ 
        C16 & 1 & 0 & 0 & 0 & 0 & 0 & 0 & 0 & 0 & 0 \\ 
        C17 & 1 & 0 & 0 & 0 & 0 & 0 & 0 & 0 & 0 & 0 \\ 
        C18 & 1 & 0 & 0 & 0 & 0 & 0 & 0 & 0 & 0 & 0 \\ 
        C19 & 1 & 0 & 0 & 0 & 0 & 0 & 0 & 0 & 0 & 0 \\ 
        C20 & 1 & 0 & 0 & 0 & 0 & 0 & 0 & 0 & 0 & 0 \\ 
        C21 & 1 & 0 & 0 & 0 & 0 & 0 & 0 & 0 & 0 & 0 \\ 
        C22 & 1 & 0 & 0 & 0 & 0 & 0 & 0 & 0 & 0 & 0 \\ 
        C23 & 1 & 0 & 0 & 0 & 0 & 0 & 0 & 0 & 0 & 0 \\ 
        C24 & 1 & 0 & 0 & 0 & 0 & 0 & 0 & 0 & 0 & 0 \\ 
        C25 & 1 & 0 & 0 & 0 & 0 & 0 & 0 & 0 & 0 & 0 \\ 
        C26 & 1 & 0 & 0 & 0 & 0 & 0 & 0 & 0 & 0 & 0 \\ 
        C27 & 1 & 0 & 0 & 0 & 0 & 0 & 0 & 0 & 0 & 0 \\ 
        C28 & 1 & 0 & 0 & 0 & 0 & 0 & 0 & 0 & 0 & 0 \\ 
        C29 & 1 & 0 & 0 & 0 & 0 & 0 & 0 & 0 & 0 & 0 \\ 
        C30 & 1 & 0 & 0 & 0 & 0 & 0 & 0 & 0 & 0 & 0 \\ 
        C31-32 & 1 & 0 & 0 & 0 & 0 & 0 & 0 & 0 & 0 & 0 \\ 
        C33 & 1 & 0 & 0 & 0 & 0 & 0 & 0 & 0 & 0 & 0 \\ 
        D35 & 0 & 0 & 0 & 0 & 0 & 0 & 0 & 0 & 0 & 0 \\ 
        E36 & 0 & 0 & 0 & 0 & 0 & 0 & 0 & 0 & 0 & 0 \\ 
        E37-39 & 0 & 0 & 0 & 0 & 0 & 0 & 0 & 0 & 0 & 0 \\ 
        F41-43 & 1 & 0 & 0 & 0 & 0 & 0 & 0 & 0 & 0 & 0 \\ 
        G45 & 1 & 0 & 0 & 0 & 0 & 0 & 0 & 0 & 0 & 0 \\ 
        G46 & 1 & 0 & 0 & 0 & 0 & 0 & 0 & 0 & 0 & 0 \\ 
        G47 & 1 & 1 & 0 & 0 & 0 & 0 & 0 & 0 & 1 & 0 \\ 
        H49 & 1 & 1 & 1 & 0 & 0 & 0 & 0 & 0 & 0 & 0 \\ 
        H50 & 1 & 1 & 1 & 0 & 0 & 0 & 0 & 0 & 0 & 0 \\ 
        H51 & 1 & 1 & 1 & 0 & 0 & 0 & 0 & 0 & 0 & 0 \\ 
        H52 & 1 & 0 & 0 & 0 & 0 & 0 & 0 & 0 & 0 & 0 \\ 
        H53 & 1 & 0 & 0 & 0 & 0 & 0 & 0 & 0 & 0 & 0 \\ 
        I55-56 & 1 & 1 & 1 & 0 & 0 & 1 & 0 & 1 & 1 & 1 \\ 
        J58 & 1 & 0 & 0 & 0 & 0 & 0 & 0 & 0 & 0 & 0 \\ 
        J59-60 & 1 & 0 & 0 & 0 & 0 & 0 & 0 & 0 & 0 & 0 \\ 
        J61 & 1 & 0 & 0 & 0 & 0 & 0 & 0 & 0 & 0 & 0 \\ 
        J62-63 & 1 & 0 & 0 & 0 & 0 & 0 & 0 & 0 & 0 & 0 \\ 
        K64 & 1 & 0 & 0 & 0 & 0 & 0 & 0 & 0 & 0 & 0 \\ 
        K65 & 1 & 0 & 0 & 0 & 0 & 0 & 0 & 0 & 0 & 0 \\ 
        K66 & 1 & 0 & 0 & 0 & 0 & 0 & 0 & 0 & 0 & 0 \\ 
        L68 & 1 & 0 & 0 & 0 & 0 & 0 & 0 & 0 & 0 & 0 \\ 
        M69-70 & 1 & 0 & 0 & 0 & 0 & 0 & 0 & 0 & 0 & 0 \\ 
        M71 & 1 & 0 & 0 & 0 & 0 & 0 & 0 & 0 & 0 & 0 \\ 
        M72 & 1 & 0 & 0 & 0 & 0 & 0 & 0 & 0 & 0 & 0 \\ 
        M73 & 1 & 0 & 0 & 0 & 0 & 0 & 0 & 0 & 0 & 0 \\ 
        M74-75 & 1 & 0 & 0 & 0 & 0 & 0 & 0 & 0 & 0 & 0 \\ 
        N77 & 1 & 1 & 1 & 1 & 0.5 & 1 & 0.5 & 1 & 1 & 0 \\ 
        N78 & 1 & 0 & 0 & 0 & 0 & 0 & 0 & 0 & 0 & 0 \\ 
        N79 & 1 & 1 & 1 & 1 & 0.5 & 1 & 0.5 & 1 & 1 & 0 \\ 
        N80-82 & 1 & 0 & 0 & 0 & 0 & 0 & 0 & 0 & 0 & 0 \\ 
        O84 & 1 & 0 & 0 & 0 & 0 & 0 & 0 & 0 & 0 & 0 \\ 
        P85 & 1 & 1 & 0.8 & 0 & 0 & 0 & 0 & 0 & 0 & 0 \\ 
        Q86 & 1 & 0 & 0 & 0 & 0 & 0 & 0 & 0 & 0 & 0 \\ 
        Q87-88 & 1 & 0 & 0 & 0 & 0 & 0 & 0 & 0 & 0 & 0 \\ 
        R90-92 & 1 & 1 & 1 & 1 & 0.5 & 1 & 0.5 & 1 & 1 & 1 \\ 
        R93 & 1 & 1 & 1 & 1 & 0.5 & 1 & 0.5 & 1 & 1 & 1 \\ 
        S94 & 1 & 1 & 1 & 1 & 0.5 & 1 & 0.5 & 0 & 1 & 1 \\ 
        S95 & 1 & 0 & 0 & 0 & 0 & 1 & 0 & 0 & 0 & 0 \\ 
        S96 & 1 & 1 & 0 & 0 & 0 & 1 & 0 & 0 & 1 & 0 \\ 
        T97-98 & 1 & 1 & 0 & 0 & 0 & 1 & 0 & 0 & 1 & 0 \\ 
        \bottomrule
    \end{tabular}
\label{tab:policies_BE_A}    
\end{table}

\clearpage
\subsection{Results}

\begin{figure}[h!]
    \centering
    \includegraphics[width=\linewidth]{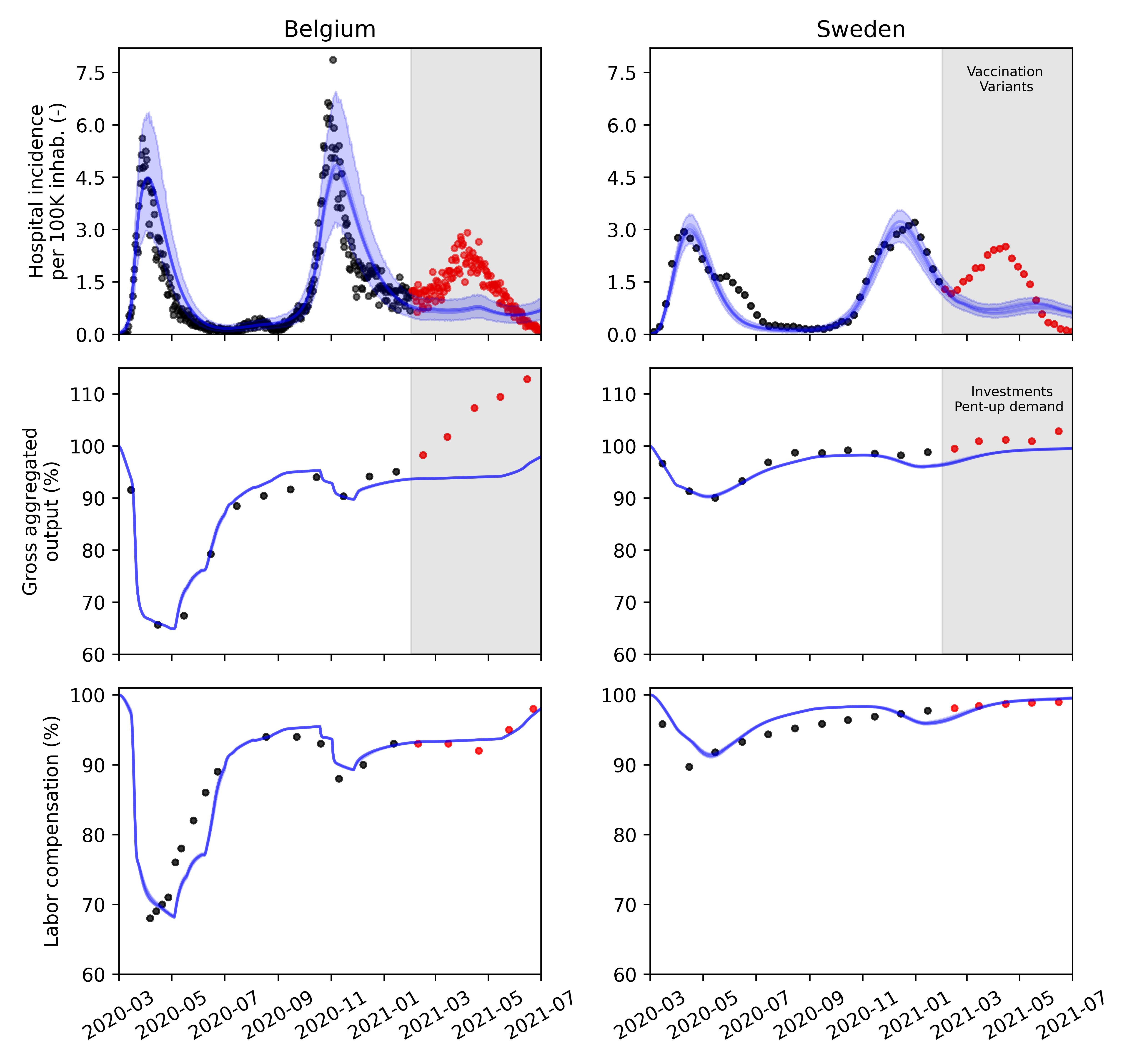}
    \caption{Comparison between 36 model realizations simulated using the posterior parameter distributions obtained from the model calibration and (normalized) data on hospital incidence (top), percentage gross aggregated output retained during the pandemic (middle), and percentage labor compensation retained during the pandemic (bottom). Data used in the calibration are colored black, while data not used in the calibration are colored red.} 
    \label{fig:calibration_epinomic_national}
\end{figure}

\begin{figure}[h!]
    \raggedleft
    \includegraphics[scale=0.45]{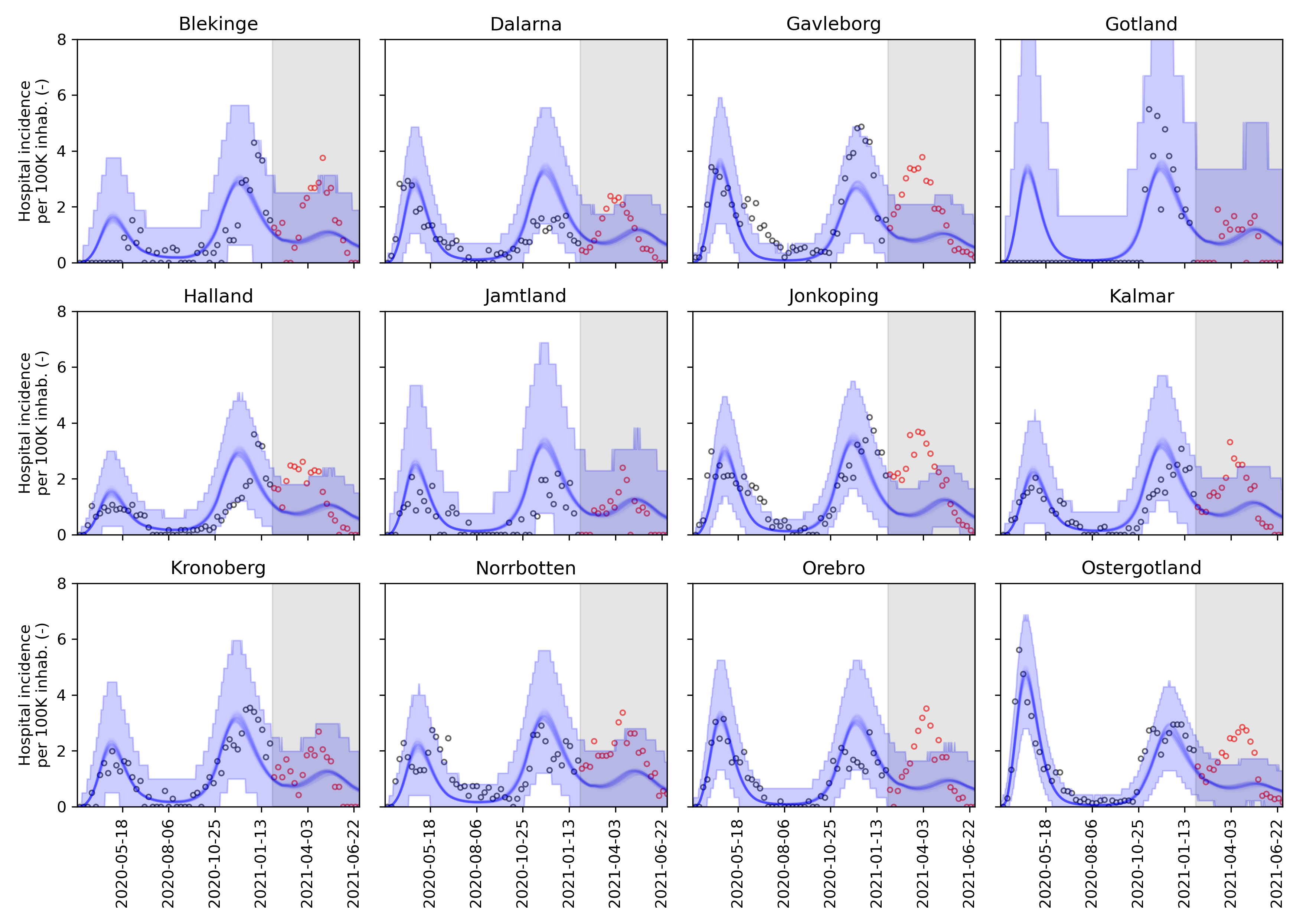}
    \caption{Simulated county-level hospitalization incidence for Sweden. Based on 36 model realizations of the hospital incidence per \num{100000} inhabitants between March 1st 2020 and July 1st 2021 (solid lines) with a binomial 95~\% confidence region (transparent band). Black crosses signify data were used in the calibration while red crosses signify data were not used. From February 2021 onward, the appearance of Variants of Concern and vaccines influences the disease's dynamics.} 
    \label{fig:calibration_SWE_1}
\end{figure}

\begin{figure}[h!]
    \raggedleft
    \includegraphics[scale=0.45]{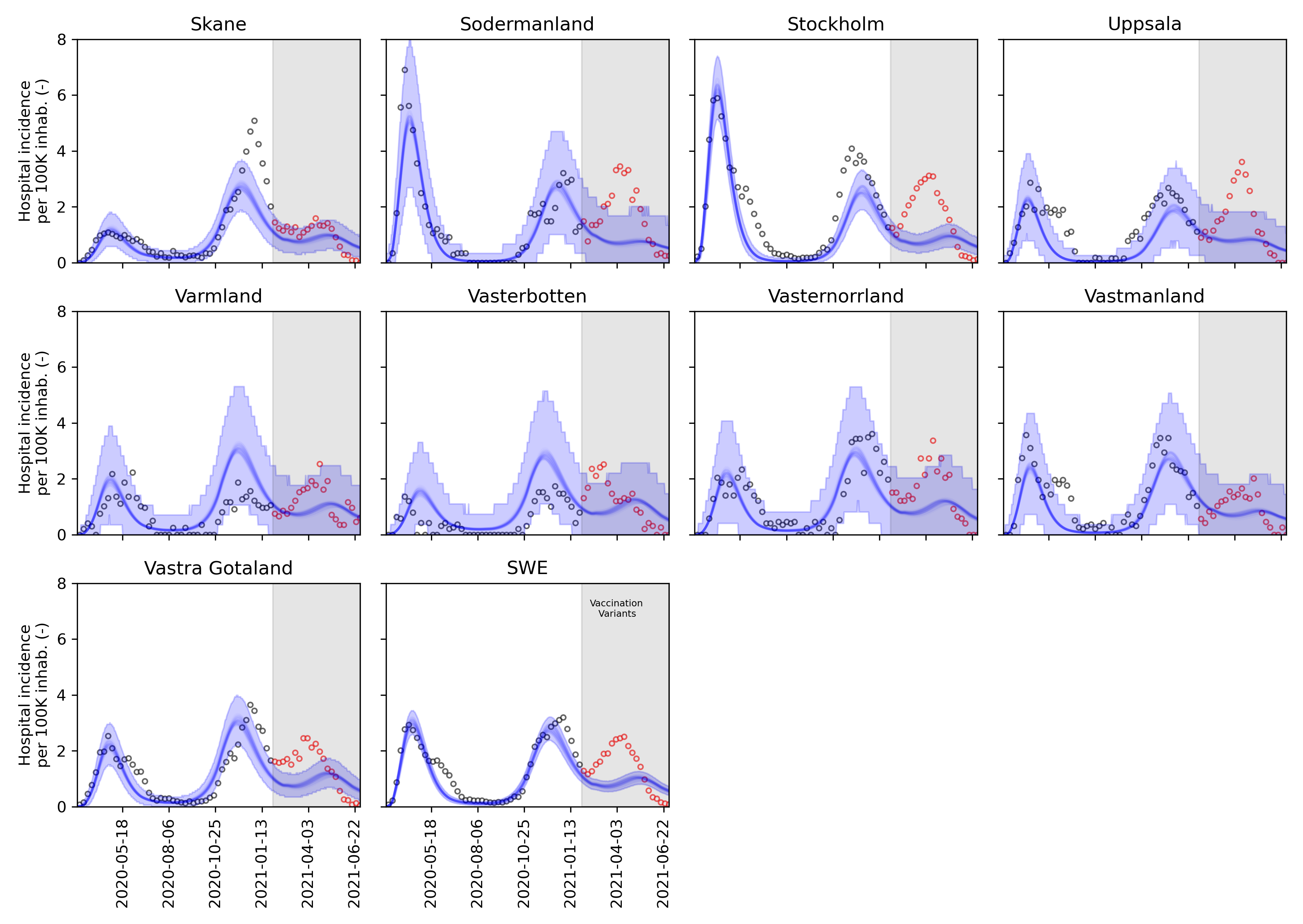}
    \caption{Simulated county-level hospitalization incidence for Sweden. Based on 36 model realizations of the hospital incidence per \num{100000} inhabitants between March 1st 2020 and July 1st 2021 (solid lines) with a binomial 95~\% confidence region (transparent band). Black crosses signify data were used in the calibration while red crosses signify data were not used. From February 2021 onward, the appearance of Variants of Concern and vaccines influences the disease's dynamics.} 
    \label{fig:calibration_SWE_2}
\end{figure}

\begin{figure}[h!]
    \raggedleft
    \includegraphics[scale=0.45]{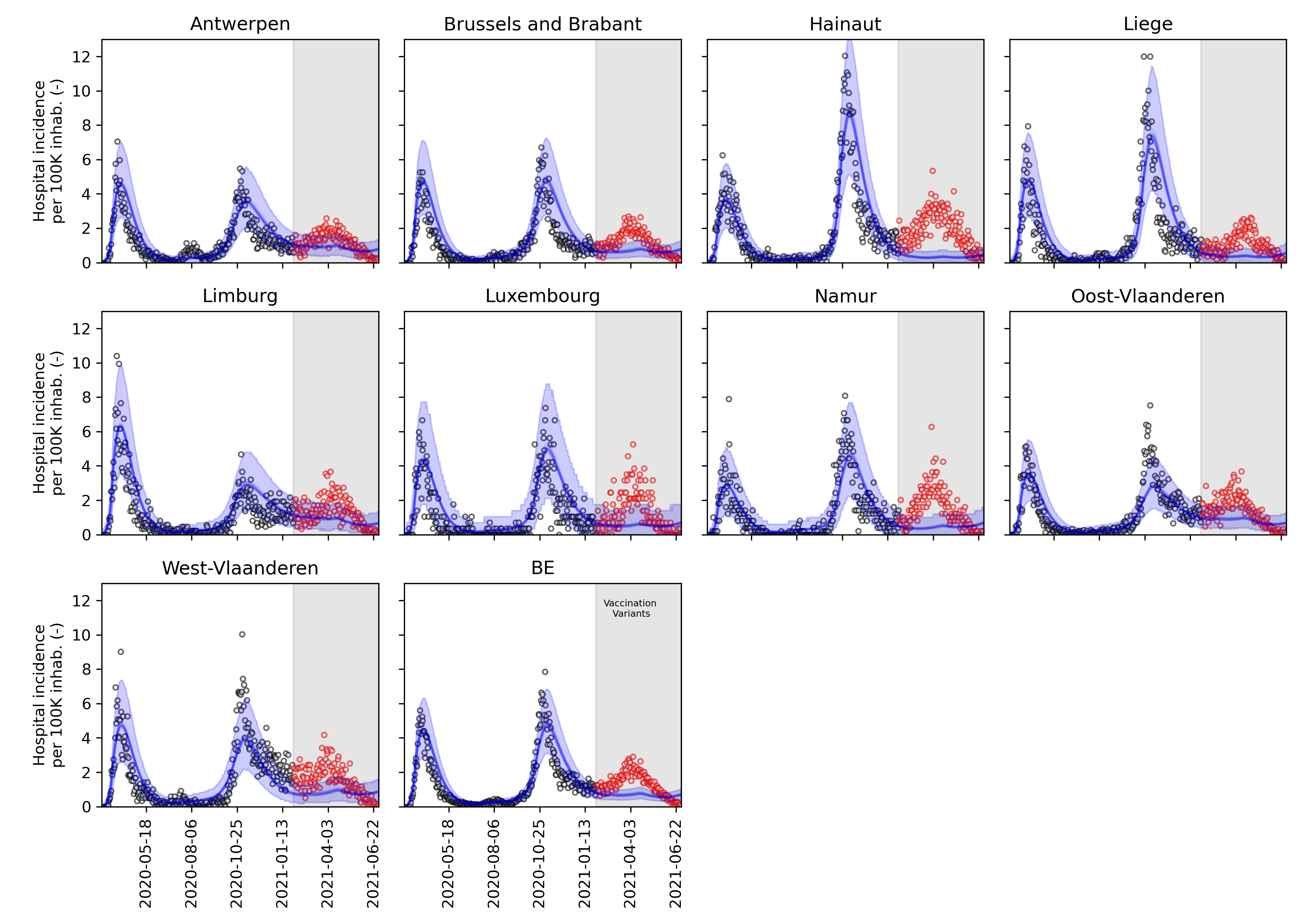}
    \caption{Simulated provincial-level hospitalization incidence for Belgium. Based on 36 model realizations of the hospital incidence per \num{100000} inhabitants between March 1st 2020 and July 1st 2021 (solid lines) with a binomial 95~\% confidence region (transparent band). Black crosses signify data were used in the calibration while red crosses signify data were not used. From February 2021 onward, the appearance of Variants of Concern and vaccines influences the disease's dynamics.} 
    \label{fig:calibration_BE}
\end{figure}

In Fig. \ref{fig:calibration_epinomic_national}, a comparison between the model realizations simulated using the posterior parameter distributions obtained from the model calibration and the empirical data on hospital incidence (top), percentage gross aggregated output retained during the pandemic (middle), and percentage labor compensation retained during the pandemic (bottom) is shown. From an epidemiological point-of-view, the model adequately captures the recurrent nature of \covid{} surges and the peak height of the epidemic in both countries. During the third \covid{} surge driven by the Alpha variant, which was not included in our model, a minor surge in hospital incidence is observed in the simulations for both countries. Extending the model to include variants of concern and vaccinations is straightforward and would likely yield a more satisfactory simulation for the remainder of 2021 \citep{alleman2023a}.  Further, our model is able to adequately capture the spatiotemporal spread of \sars{} in Sweden and Belgium at the regional level (Figs. \ref{fig:calibration_SWE_1}, \ref{fig:calibration_SWE_2}, and \ref{fig:calibration_BE}).\\

In Table \ref{tab:calibration_comparison_model_data}, we compare the observed and modeled reduction of gross aggregated output and labor compensation at the quarterly and yearly temporal levels. During 2020, the model provides an accurate description of gross aggregated output and labor compensation. In Sweden, gross aggregated output declined by 3.1~\%, whereas the model forecasts a decline of 4.5~\% (relative error 45~\%), and labor compensation declined by 4.3~\%, whereas the model predicted a decline of 3.7~\% (relative error 19~\%). In Belgium, gross aggregated output faced a much bigger decline of 13.4~\%, whereas the model predicted a decline of 14.1~\% (relative error 5.2~\%), and labor compensation declined by 12.2~\%, whereas the model predicted a decline of 12.8~\% (relative error 4.9~\%). For both Sweden and Belgium, the model's economic projections were thus slightly too pessimistic. However, the model is generally accurate, mostly for Belgium but also for Sweden. Although the relative predictive error for Sweden might seem large, the observations are close to zero, and hence the relative error is inflated. At the quarterly level, the relative prediction error is lowest in both countries during the first \covid{} epidemic in 2020 (Q2; 2.3~\% and 5.4~\% for gross agg. output) while it is highest between the \covid{} surges (Q3; 63.2~\% and 23.5~\% for gross agg. output). For Belgium, both gross output and labor compensation are adequately captured by the model (Fig. \ref{fig:calibration_epinomic_national}). Labor compensation in Belgium started recovering roughly one week before the first sectors were reopened on May 4th, 2020, resulting in a mismatch between the observations and model projection. However, the model does not explicitly account for preparations needed for a re-opening of economic activity. For Sweden, gross output is captured sufficiently accurately but the decline of employment during the first \sars{} epidemic is underestimated, and its predicted recovery too quick.\\
 
After February 1st, 2021, which was chosen as the end date of the calibration, drawing comparisons between the epinomic model trajectories and datasets becomes difficult for epistemic reasons. Epidemiologically speaking, the introduction of the Alpha variant and the start of the nationwide vaccination campaign in 2021 will influence the model's dynamics. Economically, we expect that investments, such as the EU stimulus package \citep{EU2021}, and pent-up demand after forced savings by households in 2020 \citep{basselier2021} become key drivers of economic recovery. The values of the model's parameters found during the calibration procedure, which seem physically meaningful to us, are listed in Table \ref{tab:calibration_results}. Our model's ability to capture the long-term epidemiological and economic trends in two countries using one calibrated set of parameters justifies its further use in modeling counterfactual scenarios (Section \ref{section:scenarios}).\\

\begin{table}[!h]
    \centering
    \caption{Values of the model parameters resulting in the maximization of the posterior probability of the model parameters in light of the data (Eq. \ref{eqn:posterior}).}
    {\renewcommand{\arraystretch}{1.35}
    \begin{tabular}{p{1.5cm}p{6.5cm}p{1cm}p{1.8cm}}
        \toprule
        \textbf{Par.} & \textbf{Meaning} & \textbf{Value} & \textbf{95~\% CI} \\ \midrule
        $\nu~(\text{d}.)$ & Mean lifetime of the exponential moving average hospital load & 20.8 & [19.7, 22.4] \\
        $\xi_\text{{eff}}~(-)$ & Contact effectivity in the absence of hospitalizations & 0.39 & [0.38, 0.40] \\
        $\pi_{\text{eff}}~(-)$ & Steepness of the contact effectivity & 0.070 & [0.067, 0.073] \\
        $\pi_\text{{work}}~(-)$ & Steepness of the work contact reduction & 0.032 & [0.029, 0.035] \\
        $\pi_\text{{leisure}}~(-)$ & Steepness of leisure contact reduction & 0.055 & [0.051, 0.061] \\
        $\mu~(-)$ & Spatial connectivity of the hospital load awareness network & 0.76 & [0.64, 0.86]\\
        $A_\text{{BE}}~(-)$ & Amplitude of cosine wave governing seasonal variations on the transmission rate & 15.8 & [14.9, 16.7] \\
        $\Delta t_\text{{BE}}~(\text{d}.)$ & Temporal shift of cosine wave governing seasonal variations on the transmission rate (compared to July 14th) & -15.8 & [-17.8, -13.3] \\
        $A_\text{{SWE}}~(-)$ & Amplitude of cosine wave governing seasonal variations on the transmission rate & 24.3 & [22.8, 25.7] \\
        $\Delta t_\text{{SWE}}~(\text{d}.)$ & Temporal shift of cosine wave governing seasonal variations on the transmission rate (compared to July 14th) & 7.7 & [5.3, 10.9] \\
        $\iota_\text{H}~(\text{d}.)$ & Average time to hire labor & 7.0 & [5.5, 8.7] \\
        $\iota_\text{F}~(\text{d}.)$ & Average time to fire labor & 6.1 & [4.2, 8.1]  \\
      \bottomrule
    \end{tabular}
    }
    \label{tab:calibration_results}
\end{table}

\clearpage

\clearpage
\begin{table}[!h]
    \centering
    \caption{Comparison between the observed and simulated reductions of gross aggregated output and labor compensation during the 2020 \covid{} pandemic in Belgium and Sweden. The mean absolute error (MAE) is used as a measure of model accuracy.}
    {\renewcommand{\arraystretch}{1.35}
    \begin{tabular}{p{1.2cm}p{1cm}p{1cm}p{1cm}p{1cm}p{1cm}p{1cm}}
        \toprule
        \multicolumn{7}{l}{\textbf{Sweden}} \\
        \midrule
        & \multicolumn{3}{l}{\textbf{Gross agg. output (\%)}} & \multicolumn{3}{l}{\textbf{Labor compensation (\%)}} \\
         & \textbf{Obs.} & \textbf{Sim.} & \textbf{MAE} & \textbf{Obs.} & \textbf{Sim.} & \textbf{MAE} \\
        \midrule
        2020 Q1 & -0.7 & - & - & -1.0 & - & - \\
        2020 Q2 & -8.5 & -8.3 & 0.2 & -8.4 & -6.8 & 1.6 \\
        2020 Q3 & -1.9 & -3.1 & 1.2 & -4.9 & -2.8 & 2.1 \\
        2020 Q4 & -1.3 & -2.3 & 1.0 & -3.1 & -2.2 & 0.9 \\ \midrule
        2020 & -3.1 & -4.5 & 1.4 & -4.3 & -3.7 & 0.7 \\
        \midrule
        \multicolumn{7}{l}{\textbf{Belgium}} \\
        \midrule
        2020 Q1 & -4.2 & - & - & - & - & - \\
        2020 Q2 & -29.2 & -27.6 & 1.6 & -23.4 & -25.0 & 1.6 \\
        2020 Q3 & -9.8 & -7.5 & 2.3 & -6.0 & -6.3 & 0.3 \\
        2020 Q4 & -7.2 & -7.6 & 0.4 & -9.6 & -7.8 & 1.8 \\ \midrule
        2020 & -13.4 & -14.1 & 0.7 & -12.2 & -12.8 & 0.6 \\        
      \bottomrule
    \end{tabular}
    }
    \label{tab:calibration_comparison_model_data}
\end{table}

\clearpage

\clearpage

\clearpage

\clearpage
\pagebreak
\section{Sensitivity analysis}\label{app:sensitivity}

We present a sensitivity analysis of the collective memory feedback model's parameters: $\nu$, $\mu$, $\xi_{\text{eff}}$, $\pi_{\text{eff}}$, $\pi_{\text{work}}$, and $\pi_{\text{leisure}}$, to give readers more insight in the model structure. The sensitivity analysis is one-dimensional, we vary one parameter and keep all other parameters fixed at their ``optimal" values (Table \ref{tab:calibration_results}). For every parameter, we choose three values, of which one is equal or close to the actual value used in this work, while one value is significantly lower and one value is significantly higher. We do not incorporate seasonal forcing, holidays, and shocks to investments and exports. Additionally, we do not trigger any government measures and assume the simulation starts with an $R_0=3$, and general awareness to \sars{} already triggered. We assume the first infected individuals are located in Stockholm County and Brussels in Sweden and Belgium respectively.\\

\begin{figure}[h!]
    \raggedleft
    \includegraphics[width=1\textwidth]{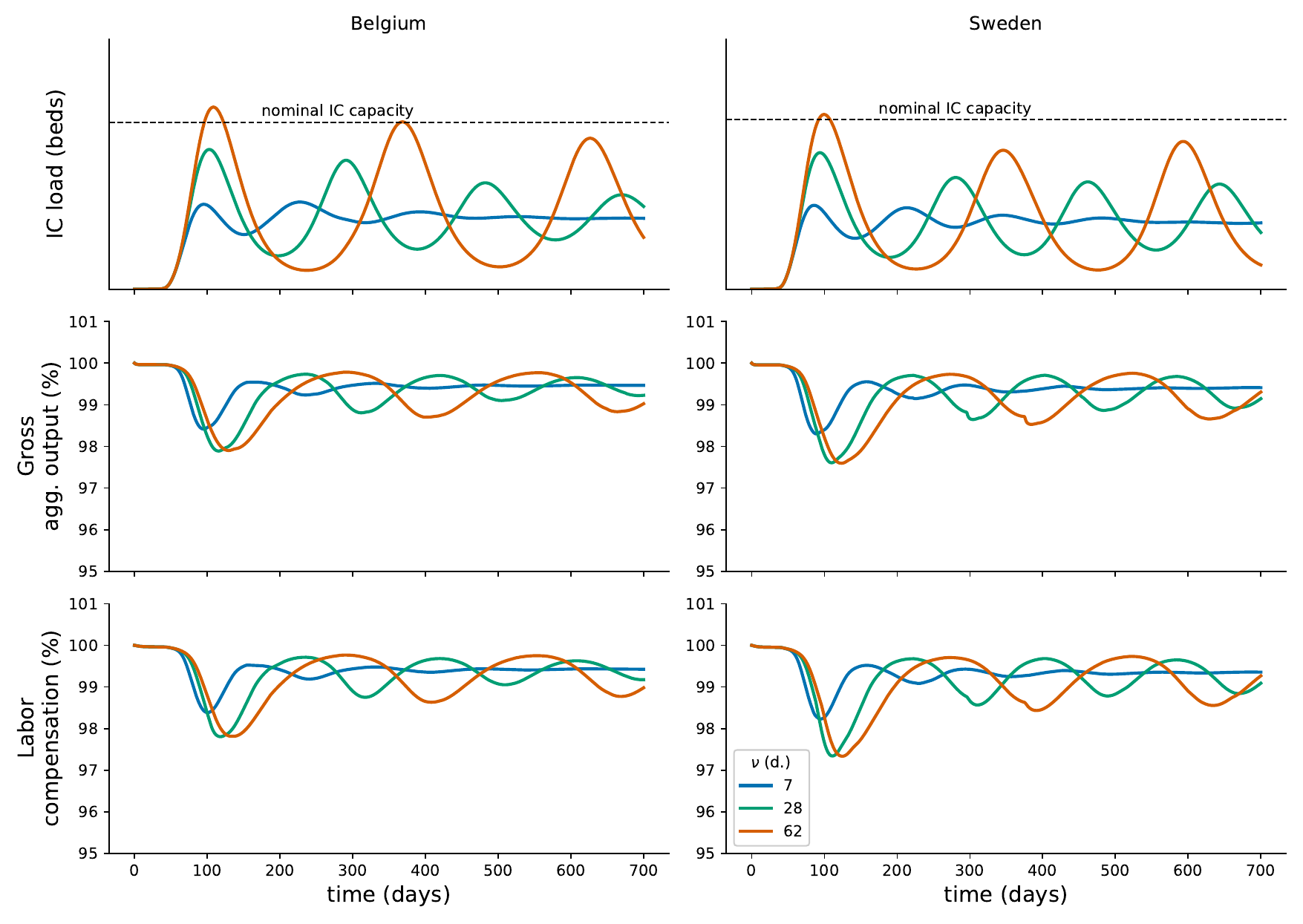}
    \caption{Impact of varying $\nu$, the mean lifetime of the EMA hospital load, on IC load (top), gross aggregated output (middle) and labor compensation (bottom).} 
    \label{fig:sensitivity_nu}
\end{figure}

\textbf{Hospital load memory mean lifetime or ``forgetfullness" ($\bm{\nu}$)} The impact of varying $\nu$ (Eq. \eqref{eq:EMA_hospital_load}) is shown in Fig. \ref{fig:sensitivity_nu}. Larger values of $\nu$ result in slower voluntary behavioral changes, as the history of the hospital load is retained longer in the population's collective memory. Increasing $\nu$ results in low-frequency, high-amplitude oscillations while lowering $\nu$ results in high-frequency, low-amplitude oscillations. For $\nu=7~\text{days}$, we observe a dynamic equilibrium at $\approx~50\%$ of the nominal IC bed capacity. Changing $\nu$ does not seem to alter the dynamic equilibrium of the system, only the frequency and amplitude of the oscillations. For a long mean lifetime of $\nu=62~\text{days}$, the nominal IC load is almost reached in subsequent \covid{} waves. Keeping the mean lifetime short minimizes the system’s oscillations, rendering it more easy to control. However, translating this shorter mean lifetime into practical advice is challenging.\\

\begin{figure}[h!]
    \raggedleft
    \includegraphics[width=1\textwidth]{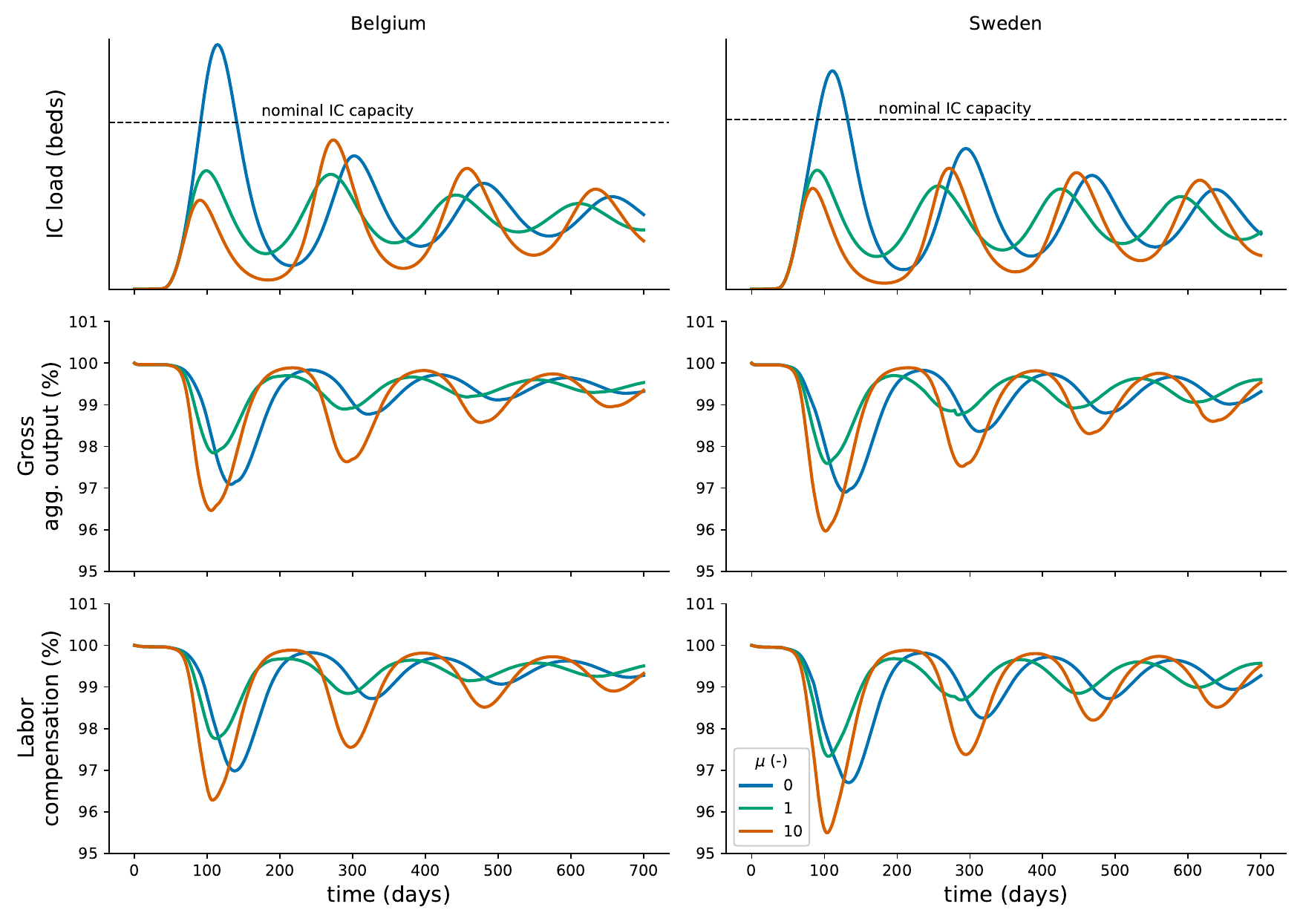}
    \caption{Impact of varying $\mu$, which governs the degree to which localized epidemics can induce behavioral changes in other spatial patches, on IC load (top), gross aggregated output (middle) and labor compensation (bottom).} 
    \label{fig:sensitivity_mu}
\end{figure}

\textbf{Spatial spread of awareness to localised epidemics ($\bm{\mu}$)} The impact of varying $\mu$ (Eq. \eqref{eq:EMA_spatial_average}) is shown in Fig. \ref{fig:sensitivity_mu}. Increasing $\mu$ results in smaller outbreaks of \covid{} (lower number of IC beds occupied), both during the initial outbreak and in subsequent waves. However, increasing $\mu$ results in more economic damage as individuals in any spatial patch now alter their consumption patterns based on the hospital load in the spatial patch with the highest hospital load, instead of the hospital load on their own spatial patch. The Swedish economy is more sensitive to the spatial spread of awareness than Belgium's. We conclude that keeping the population well informed on the spatial incidence of \covid{} cases may result in less severe \covid{} epidemics at the cost of greater economic damage due to induced fear of infection.\\

\begin{figure}[h!]
    \raggedleft
    \includegraphics[width=0.95\textwidth]{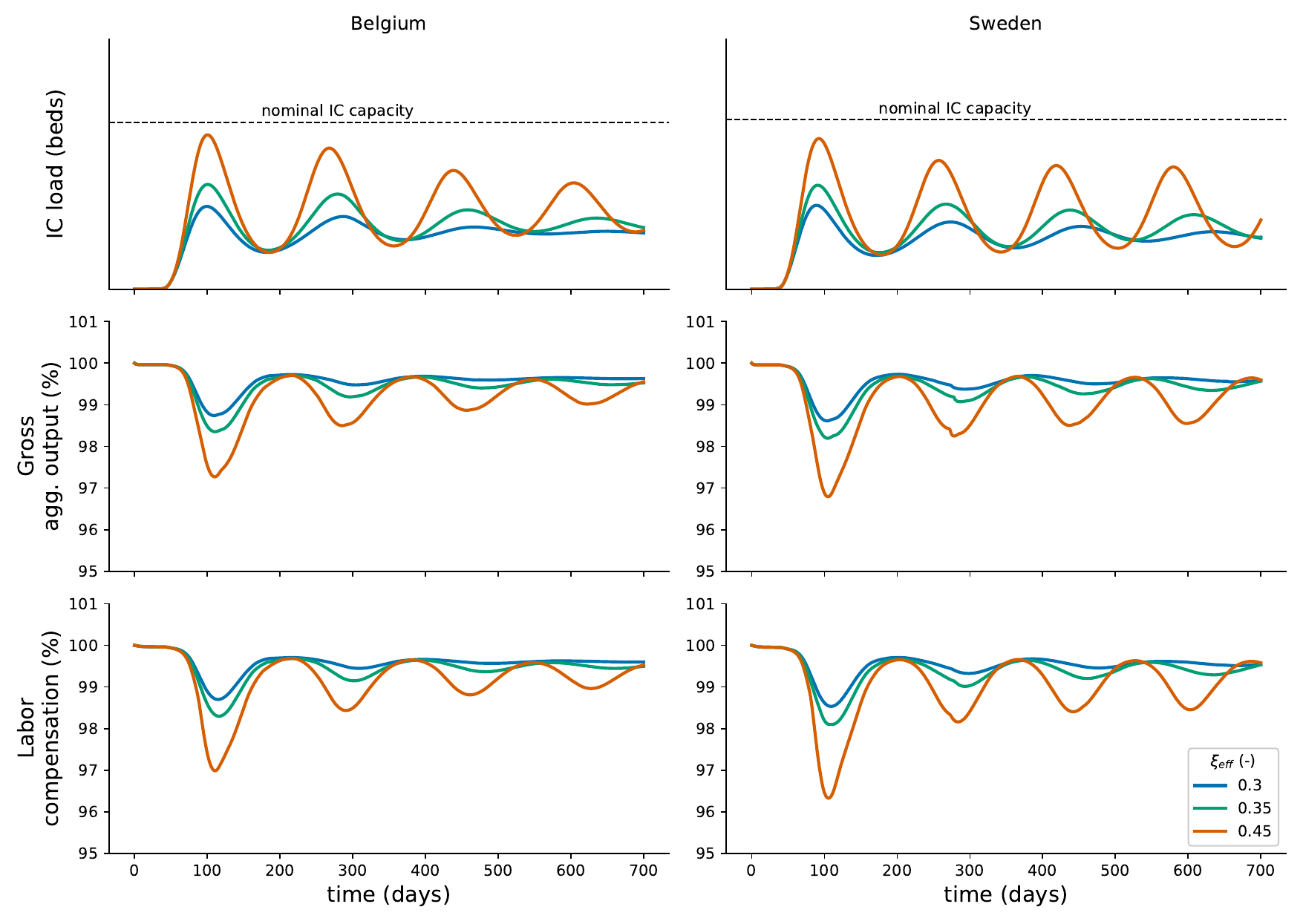}
    \caption{Impact of varying $\xi_{\text{eff}}$, governing the contact effectivity in the absence of hospitalizations, on IC load (top), gross aggregated output (middle) and labor compensation (bottom).} 
    \label{fig:sensitivity_xi_eff}
\end{figure}

\textbf{Contact effectivity in the absence of hospitalizations ($\bm{\xi_{\text{eff}}}$)} The impact of varying $\xi_{\text{eff}}$ is shown in Fig. \ref{fig:sensitivity_xi_eff}. After triggering ``general awareness" to \sars{}, the use of preventive measures such as face masks, and testing and tracing at the beginning of the  \covid{} pandemic will lower the effective reproduction number below $R_0=3$ for the remainder of the pandemic. Lower values of $\xi_{\text{eff}}$ result in a lower effective reproduction number during the pandemic. This in turn results in a lower dynamic equilibrium IC load and a lower amplitude of the oscillations. We conclude that keeping the population aware of the dangers posed by \covid{}, in combination with preventive measures such as face masks and testing \& tracing, may result in fewer IC beds occupied, fewer oscillations, and less economic damage overall. \\

\begin{figure}[h!]
    \raggedleft
    \includegraphics[width=0.95\textwidth]{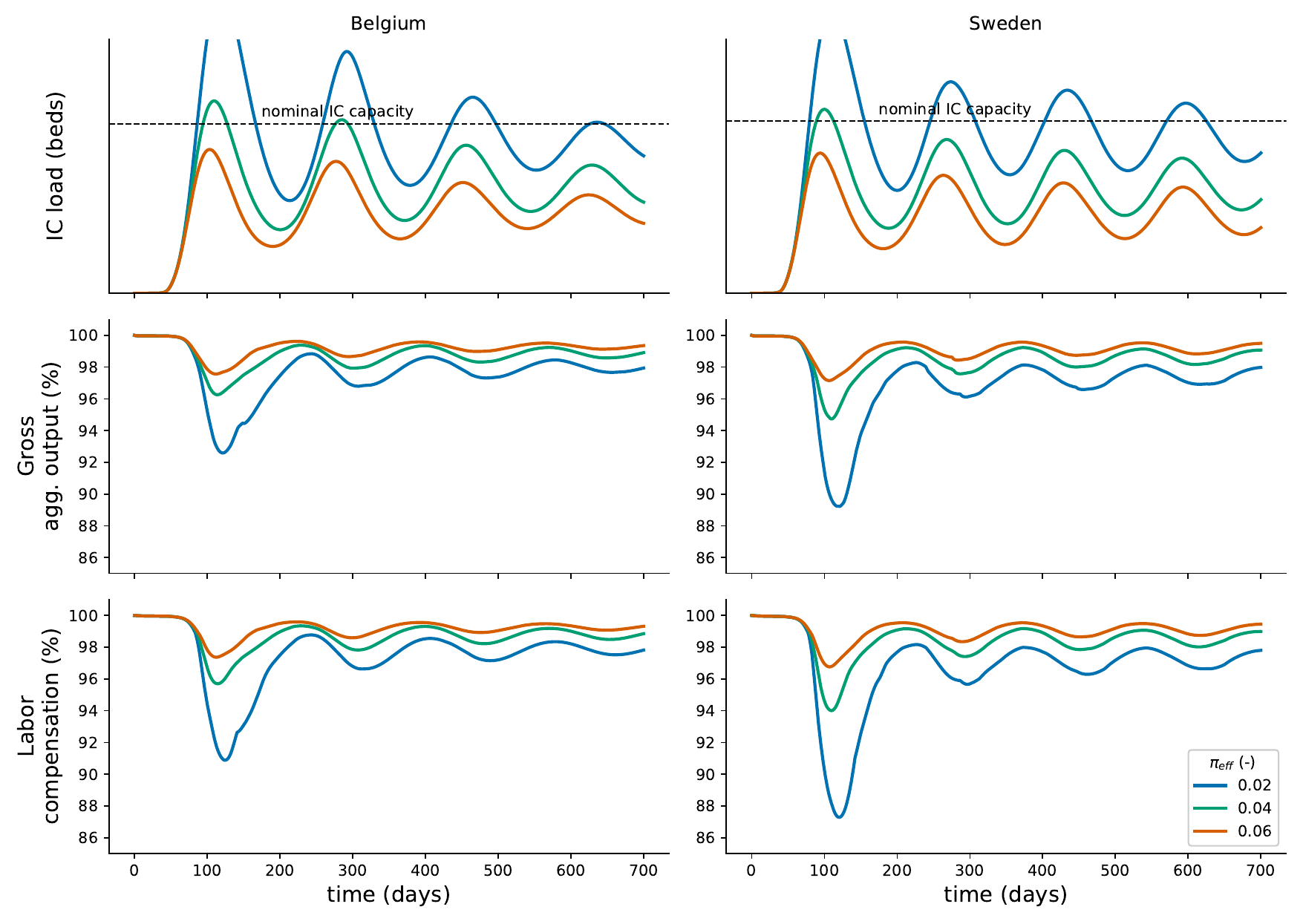}
    \caption{Impact of varying $\pi_{\text{eff}}$, governing the rate at which individuals lower their contact effectivity through the use of preventive measures and general prudence, as the hospital load increases, on IC load (top), gross aggregated output (middle) and labor compensation (bottom).} 
    \label{fig:sensitivity_pi_eff}
\end{figure}

\textbf{Sensitivity of contact effectivity to hospital load ($\bm{\pi_{\text{eff}}}$)} The impact of varying $\pi_{\text{eff}}$ is shown in Fig. \ref{fig:sensitivity_pi_eff}. Smaller values of $\pi_{\text{eff}}$ imply individuals lower the effectivity of their contacts slower when the hospital load increases. Smaller values of $\pi_{\text{eff}}$ result in a lower dynamic equilibrium IC load but do not alter the frequency or amplitude of oscillations. Faster adoption of preventive measures, such as preventive measures and testing, tracing and quarantine may result in fewer IC beds occupied, fewer oscillations, and less economic damage. In addition, these results indicate that the adoption speed of preventive measures matters, and government forcing may be beneficial both to public health and the economy.\\

\begin{figure}[h!]
    \raggedleft
    \includegraphics[width=0.95\textwidth]{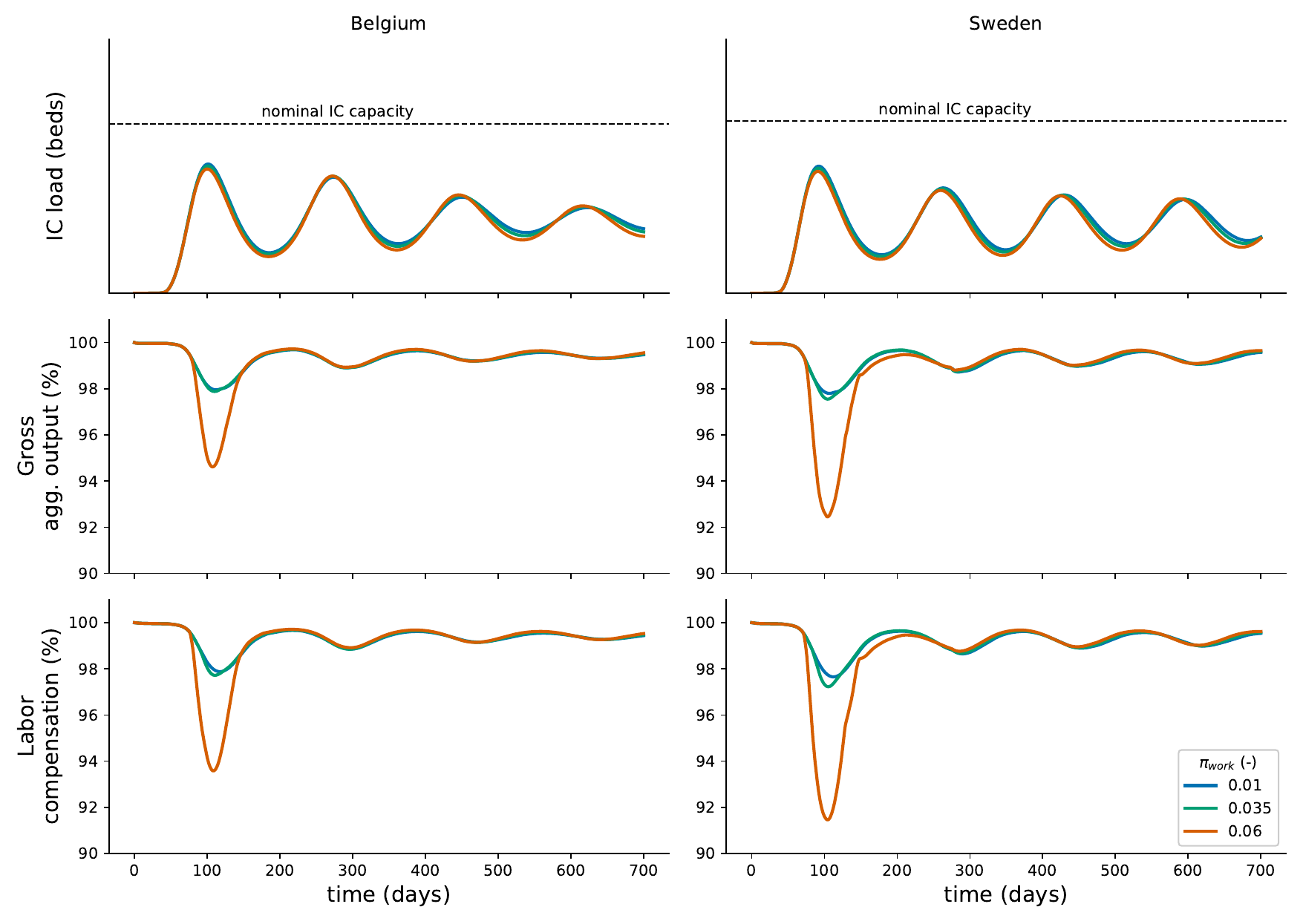}
    \caption{Impact of varying $\pi_{\text{work}}$, governing the rate at which individuals lower their workplace contacts as the hospital load increases, on IC load (top), gross aggregated output (middle) and labor compensation (bottom).} 
    \label{fig:sensitivity_pi_work}
\end{figure}

\textbf{Sensitivity of voluntary reduction in workplace contacts to hospital load ($\bm{\pi_{\text{work}}}$)} The impact of varying $\pi_{\text{work}}$ is shown in Fig. \ref{fig:sensitivity_pi_work}. The impact of $\pi_{\text{work}}$ on the spread of the epidemic is small, while the impact on the economy is large. Larger values of $\pi_{\text{work}}$ result in people staying away from the workplace at lower hospital loads, inducing a larger labor supply shock. We conclude that keeping fear of infection in the workplace low, for instance, through the use of preventive measures, can drastically reduce the induced labor supply shock during a pandemic. \\

% The impact of $\pi_{\text{work}}$ on the spread of \sars{} is small due to the simultaneous lowering of the contact effectivity. The authors have exploited all means to omit the contact effectivity from the model but this never resulted in satisfactory results.

\begin{figure}[t!]
    \raggedleft
    \includegraphics[width=0.95\textwidth]{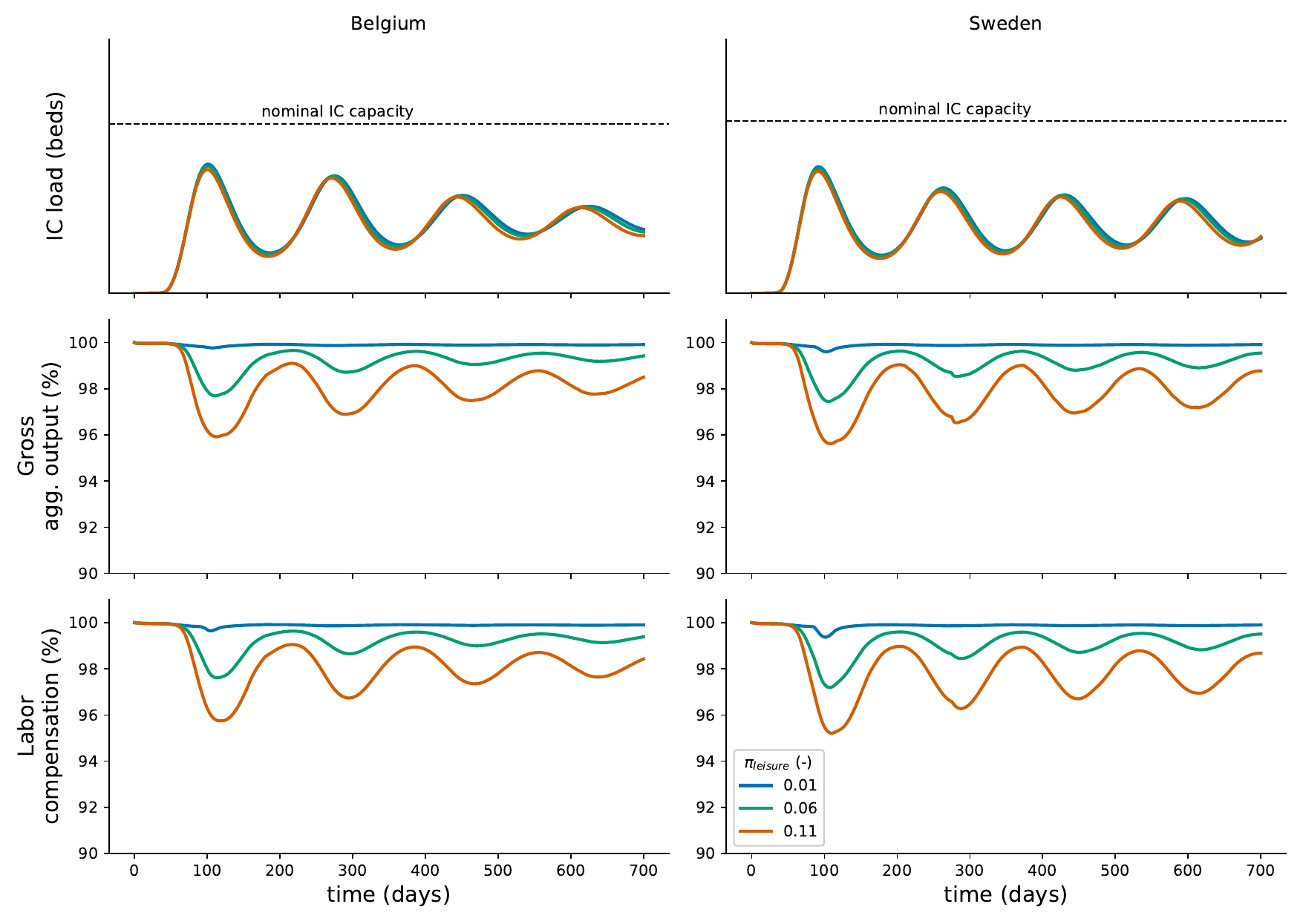}
    \caption{Impact of varying $\pi_{\text{leisure}}$, governing the rate at which individuals lower their leisure contacts as the hospital load increases, on IC load (top), gross aggregated output (middle) and labor compensation (bottom).} 
    \label{fig:sensitivity_pi_leisure}
\end{figure}

\textbf{Sensitivity of voluntary reduction in leisure contacts to hospital load ($\bm{\pi_{\text{leisure}}}$)} The impact of varying $\pi_{\text{leisure}}$ is shown in Fig. \ref{fig:sensitivity_pi_leisure}. Similar to $\pi_{\text{work}}$. Larger values of $\pi_{\text{leisure}}$ result in people avoiding leisure contacts at lower hospital loads, inducing a larger household demand shock. $\pi_{\text{leisure}}$ has a smaller impact on the economy than $\pi_{\text{work}}$. From both Fig. \ref{fig:sensitivity_pi_work} and \ref{fig:sensitivity_pi_leisure}, we conclude that making sure employees feel safe in the workplace must, under logistical constraints, be prioritized over making sure people feel safe during leisure activities if the economic damage is to be minimized. This finding ignores the impact on the population's general well-being. \\ 

\clearpage
\pagebreak
\section{Supplementary results}\label{app:supplementary_results}

\textbf{Scenario 1: Responding late necessitates restrictions with higher economic damages}

\begin{table}[!h]
    \centering
    \caption{Simulated reduction of gross aggregated output, labor compensation, and cumulative number of IC patients during the second quarter of 2020 (April, May, June). Policies are defined in Table \ref{tab:scenarios_policies_BE}, policy P1 imposed on 2020-03-15 is factual (underlined).}
    {\renewcommand{\arraystretch}{1.35}
    \begin{tabular}{p{1.5cm}p{1cm}p{1cm}p{1cm}p{1cm}p{1cm}p{1cm}}
        \toprule
        & \multicolumn{5}{l}{\textbf{Reduction in gross aggregated output (\%)}} \\
         & P1 & P2 & P3 & P4a & P4b \\
        \midrule
        2020-03-03 & -24.4 & -20.8 & -14.5 & -12.2 & -12.6 \\
        2020-03-06 & -24.4 & -20.6 & -14.5 & -12.4 & -12.9 \\
        2020-03-09 & -24.1 & -20.4 & -14.6 & -12.8 & -13.3 \\
        2020-03-12 & -24.1 & -20.2 & -14.7 & -13.4 & -14.0 \\
        2020-03-15 & \underline{-24.0} & -19.9 & -15.1 & -14.9 & -15.3 \\
        2020-03-18 & -24.7 & -20.1 & -16.7 & -17.2 & -17.9 \\ \midrule     
        & \multicolumn{5}{l}{\textbf{Reduction in labor compensation (\%)}} \\
         & P1 & P2 & P3 & P4a & P4b \\
        \midrule
        2020-03-03 & -21.0 & -18.5 & -10.7 & -8.2 & -8.6 \\
        2020-03-06 & -20.9 & -18.3 & -10.7 & -8.4 & -8.9 \\
        2020-03-09 & -20.6 & -18.0 & -10.7 & -8.8 & -9.3 \\
        2020-03-12 & -20.5 & -17.7 & -10.8 & -9.4 & -10.0 \\
        2020-03-15 & \underline{-20.2} & -17.3 & -11.2 & -11.2 & -12.0 \\
        2020-03-18 & -21.1 & -17.5 & -13.6 & -14.8 & -15.6 \\ \midrule      
        & \multicolumn{5}{l}{\textbf{Cumulative number of IC patients (-)}} \\
         & P1 & P2 & P3 & P4a & P4b \\
        \midrule
        2020-03-03 & 333 & 1914 & 2251 & 3430 & 4439 \\
        2020-03-06 & 612 & 2074 & 2448 & 3677 & 4687 \\
        2020-03-09 & 1097 & 2297 & 2657 & 4109 & 5221 \\
        2020-03-12 & 1913 & 2828 & 3221 & 4931 & 5872 \\
        2020-03-15 & \underline{3416} & 4066 & 4495 & 6487 & 7559 \\
        2020-03-18 & 6154 & 6739 & 7180 & 9206 & 9956 \\ 
      \bottomrule
    \end{tabular}
    }
    \label{tab:hypothetical_scenario_results}
\end{table}

\clearpage
\pagebreak

\begin{landscape}
\thispagestyle{empty}
\begin{figure}[h!]
    \centering
    \includegraphics[width=0.95\linewidth]{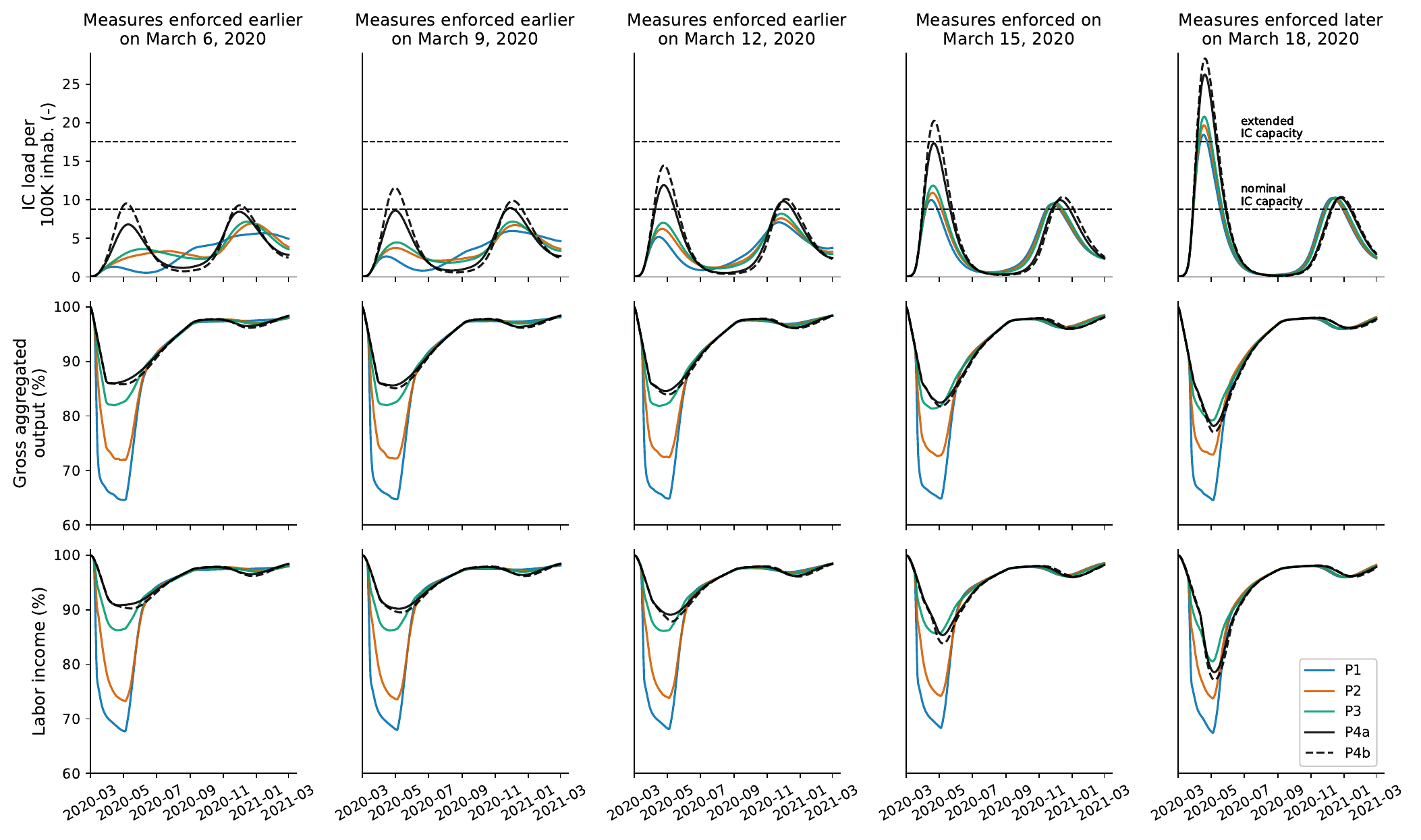}
    \caption{Simulated IC load per \num{100000} inhabitants, percentage gross aggregated output and labor income as compared to before the \covid{} pandemic, for Belgium. Policies range from a strict lockdown as implemented in Belgium (P1) to voluntary recommendations as implemented in Sweden (P4b) (Table \ref{tab:scenarios_policies_BE}). Policies are released gradually over a two-month period, starting on May 5, 2020. Our model forecasts a second \covid{} surge after summer, which occurred in reality (Fig. \ref{fig:data_calibration}).} 
    \label{fig:hypothetical_scenarios_BE_full}
\end{figure}
\end{landscape}

\textbf{Scenario 2: Prolonging lockdown after an initial surge can strain the healthcare system and increase economic damages}

\begin{figure}[h!]
    \centering
    \includegraphics[width=1\linewidth]{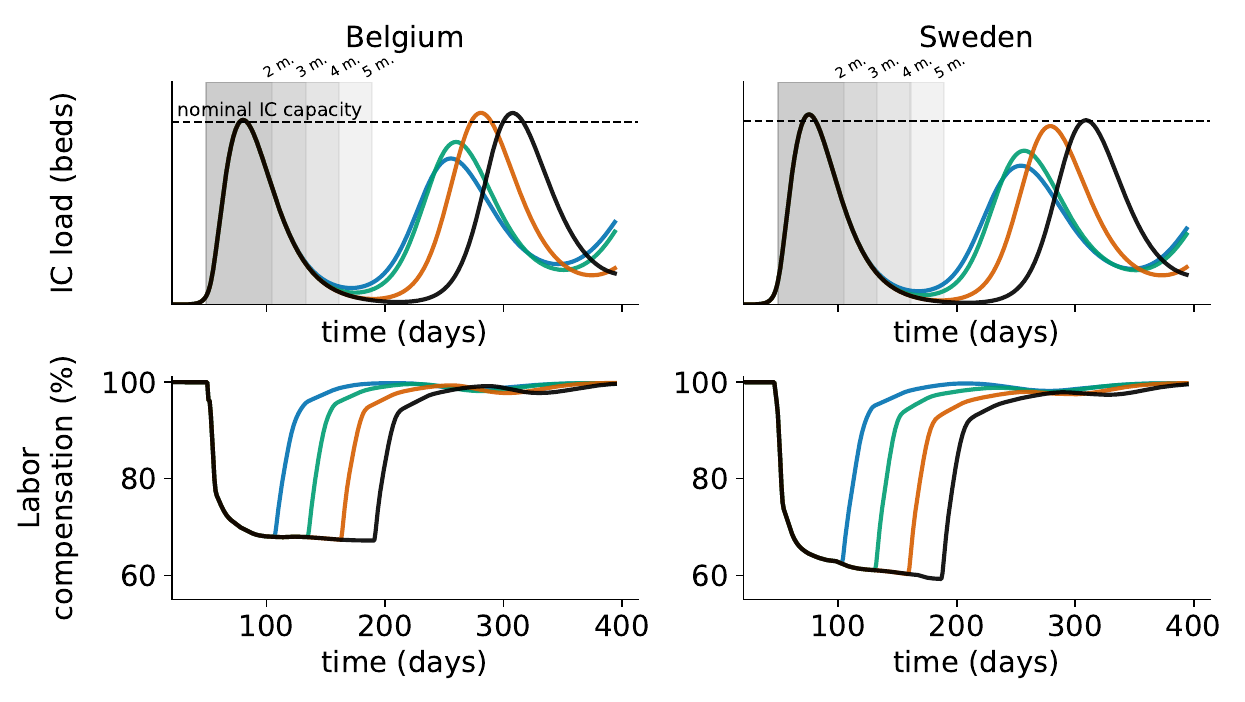}
    \caption{Simulated IC load and reduction of labor compensation in Belgium and Sweden during one year. Both governments impose a lockdown (policy P1, Table \ref{tab:scenarios_policies_BE}) to counter the initial epidemic. The lockdown is then maintained for 2 (blue), 3 (green), 4 (orange) and 5 months (black). After releasing the lockdown, no measures are in place.} 
    \label{fig:prevention_paradox_full}
\end{figure}

\begin{table}[!h]
    \centering
    \caption{Simulated reduction of gross aggregated output, labor compensation, and cumulative number of IC patients. Prolonging lockdown results in a (small) decline in the number of IC patients, but a large increase in economic damages.}
    {\renewcommand{\arraystretch}{1.35}
    \begin{tabular}{p{1.2cm}p{1.5cm}p{1.5cm}p{1.5cm}p{1.5cm}p{1.5cm}}
        \toprule
        & \multicolumn{4}{l}{\textbf{Reduction in gross aggregated output (\%)}} \\
        & 2 months & 3 months & 4 months & 5 months \\
        \midrule
        Sweden & -6.4 & -9.2 & -12.1 & -15.0  \\
        Belgium & -5.9 & -8.3 & -10.9 & -13.4 \\ \midrule
        & \multicolumn{4}{l}{\textbf{Reduction in labor compensation (\%)}} \\
        & 2 months & 3 months & 4 months & 5 months \\
        \midrule
        Sweden & -5.5 & -7.7 & -10.2 & -12.6  \\
        Belgium & -6.5 & -9.4 & -12.5 & -15.5 \\ \midrule   
        & \multicolumn{4}{l}{\textbf{Cumulative number of IC patients (-)}} \\
        & 2 months & 3 months & 4 months & 5 months \\
        \midrule
        Sweden & 6869 & 6792 & 6560 & 6339  \\
        Belgium & 10608 & 10444 & 10183 & 9747 \\ 
      \bottomrule
    \end{tabular}
    }
    \label{tab:prevention_paradox_results}
\end{table}

\textbf{Scenario 4: Sweden may be more resilient to the spread of \sars{} than Belgium but its economy is not - Why Sweden may have been fortunate} \\ 

\begin{figure}[h!]
    \raggedleft
    \includegraphics[width=1\textwidth]{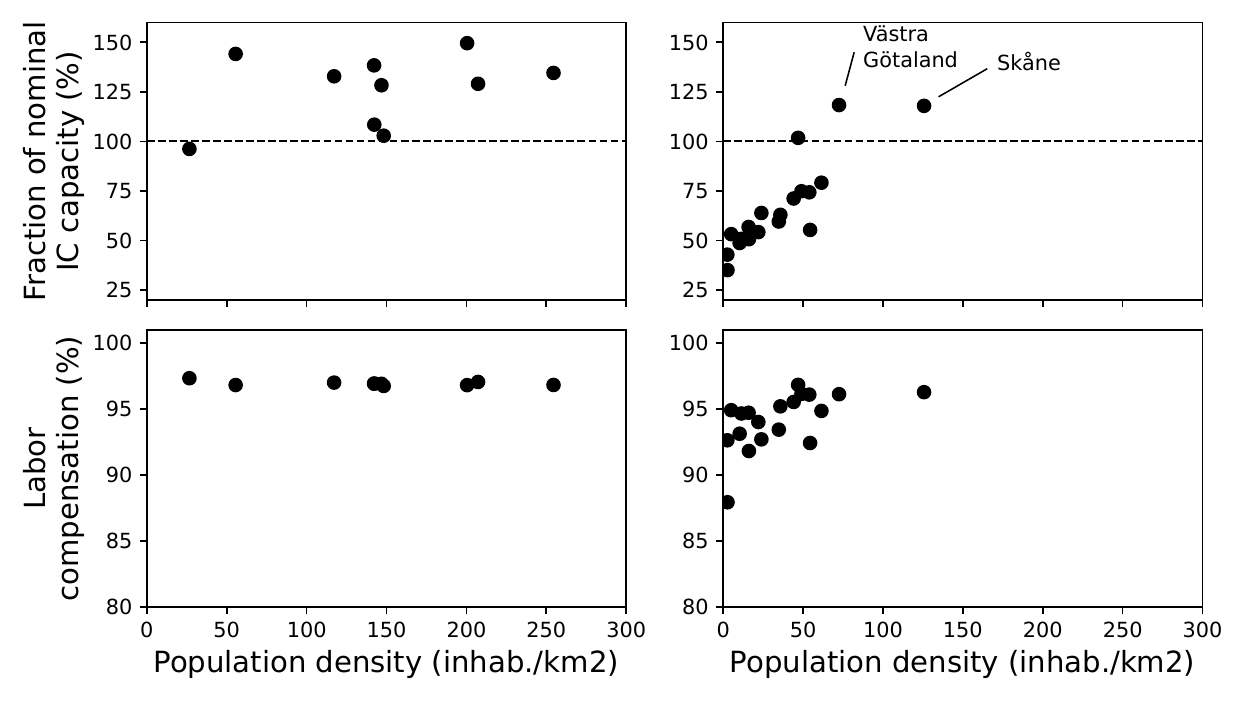}
    \caption{The relationship between the population density of a Swedish county (right) or Belgian province (left) in which the second infected individual is seeded and the maximum impact of the resulting \sars{} epidemic on the occupied number of IC beds and the economy.}
    \label{fig:hypothetical_spatial_spread_1}
\end{figure}

Figure \ref{fig:hypothetical_spatial_spread_1} shows the relationship between the population density of the Swedish county (right) or Belgian province (left) in which the second infected individual is placed and the maximum impact of the resulting \sars{} epidemic on the occupied number of IC beds and the economy. In Belgium, we conclude no such relationship exists and thus the use of a spatially explicit model is not needed. In Sweden, seeding the epidemic in more sparsely populated areas results in fewer occupied IC beds but more economic damage. This is caused by the way awareness spreads spatially, if the epidemic is seeded in a more rural area, this will result in a local epidemic that is small in absolute magnitude but overwhelms the local HCS capacity. Images of crowded hospitals will make national media, resulting in behavioral changes in the rest of the country. The result is a smaller overall epidemic but larger economic damage.\\

%Most notable for Sweden is the effect of seeding one infected individual in Stockholm and the other in the counties of Skåne or Västra Götaland, which are Sweden's number two and three counties in terms of population density, as well as containing Sweden's number two and three largest cities, Malmö and Göteborg. In reality, the epidemic originated in the Stockholm metropolitan area (Appendix \ref{app:calibration}) and thus these scenarios are counterfactual. If the epidemic was simultaneously seeded in either county, the resulting epidemic would have surpassed the nominal IC bed capacity of 600 beds by 19 ~\%. If \sars{} had reached Sweden via the land bridge to Denmark at Malmö this would have resulted in an acute IC bed shortage during the first 2020 \covid{} epidemic. We conclude the Swedish territory, which is much larger, more sparsely populated, and less connected than Belgium's, lends itself more to a strategy of voluntary recommendations. However, Swedish policymakers should be more wary of pursuing such a strategy during future epidemics. If \sars{} had simultaneously reached Sweden from Denmark at Malmö, a counterfactual scenario that is not unimaginable at all, there would have been an acute IC bed shortage. Instead, based on the counterfactual scenarios for Belgium (Fig. \ref{fig:hypothetical_scenarios_BE}), mandating work-at-home (policy P4a) would result in less \sars{} circulation with no additional economic damage, thus being the least disruptive alternative policy.\\ 

\end{appendices}

\end{document}